\documentclass[11pt]{article}	
\pdfoutput=1 
\usepackage{color} 
\usepackage{amsmath}
\usepackage{amsfonts}
\usepackage{amssymb}
\usepackage{enumerate}
\usepackage{float}
\usepackage{caption}
\usepackage{graphicx}
\usepackage{array,booktabs}
\usepackage{collref}				
\usepackage{framed}				
\usepackage{placeins}
\usepackage{jheppub}		
\usepackage{pstricks}
\usepackage{bbm}
\usepackage{latexsym}

\newcommand{\centeron}[2]{{\setbox0=\hbox{#1}\setbox1=\hbox{#2}\ifdim
\wd1>\wd0\kern.5\wd1\kern-.5\wd0\fi \copy0
\kern-.5\wd0\kern-.5\wd1\copy1\ifdim\wd0>\wd1
                                   \kern.5\wd0\kern-.5\wd1\fi}}
\newcommand{\ltap}{\>\centeron{\raise.35ex\hbox{$<$}}
                           {\lower.65ex\hbox{$\sim$}}\>}
\newcommand{\gtap}{\>\centeron{\raise.35ex\hbox{$>$}}
                           {\lower.65ex\hbox{$\sim$}}\>}

\newcommand{\lsim}{\mathrel{\ltap}}

\newcommand\ZZ{\hbox{\zfont Z\kern-.4emZ}}
\font\zfont = cmss10 

\newcommand{\fref}[1]{Fig.~\ref{f.#1}}
\newcommand{\eref}[1]{eq.~(\ref{e.#1})}

\newcommand{\aref}[1]{Appendix~\ref{a.#1}}
\newcommand{\sref}[1]{Section~\ref{s.#1}}
\newcommand{\ssref}[1]{Section~\ref{ss.#1}}
\newcommand{\sssref}[1]{Section~\ref{sss.#1}}
\newcommand{\cref}[1]{Chapter~\ref{c.#1}}
\newcommand{\tref}[1]{Table~\ref{t.#1}}

\newcommand{\ba}{\begin{array}}
\newcommand{\ea}{\end{array}}

\newcommand{\beq}{\begin{eqnarray}}
\newcommand{\eeq}{\end{eqnarray}}
\newcommand{\beqs}{\begin{eqnarray*}}
\newcommand{\eeqs}{\end{eqnarray*}}

\newcommand{\bal}{\begin{align}} 
\newcommand{\eal}{\end{align}}

\def\bi{\begin{itemize}}
\def\ei{\end{itemize}}
\def\ben{\begin{enumerate}}
\def\een{\end{enumerate}}
\def\bc{\begin{center}}
\def\ec{\end{center}}
\def\bt{\begin{table}}
\def\et{\end{table}}
\def\btb{\begin{tabular}}
\def\etb{\end{tabular}}

\def\ev{\, {\rm  eV}}
\def\gev{\, {\rm GeV}}
\def\kev{\, {\rm keV}}
\def\mev{\, {\rm MeV}}
\def\tev{\, {\rm TeV}}

\def\mass2{mass${}^2$}

\def\gram{\mathrm{g}}
\def\cmmthree{\mathrm{cm}^{-3}}

\def\cmtwo{\mathrm{cm}^2}
\def\invsecond{\mathrm{s}^{-1}}

\usepackage{tikz}
\usetikzlibrary{arrows,shapes}
\usetikzlibrary{trees}
\usetikzlibrary{matrix,arrows} 				
\usetikzlibrary{positioning}				
\usetikzlibrary{calc,through}				
\usetikzlibrary{decorations.pathreplacing}  
\usepackage{pgffor}							

\usetikzlibrary{decorations.pathmorphing}	
\usetikzlibrary{decorations.markings}
\usetikzlibrary{snakes}

\tikzset{
    vector/.style={decorate, decoration={snake}, draw},
	provector/.style={decorate, decoration={snake,amplitude=2.5pt}, draw},
	antivector/.style={decorate, decoration={snake,amplitude=-2.5pt}, draw},
    fermion/.style={draw, postaction={decorate},
        decoration={markings,mark=at position .55 with {\arrow[draw]{>}}}},
    fermionbar/.style={draw, postaction={decorate},
        decoration={markings,mark=at position .55 with {\arrow[draw=black]{<}}}},
    fermionnoarrow/.style={draw},
    gluon/.style={decorate, draw,decoration={coil,amplitude=4pt, segment length=6pt}, line width=1},
    scalar/.style={dashed,draw, postaction={decorate},
        decoration={markings,mark=at position .55 with {\arrow[draw]{>}}}},
    scalarbar/.style={dashed,draw, postaction={decorate},
        decoration={markings,mark=at position .55 with {\arrow[draw]{<}}}},
    scalarnoarrow/.style={dash pattern = on 6 pt off 3 pt,draw},
    electron/.style={draw, postaction={decorate},
        decoration={markings,mark=at position .55 with {\arrow[draw]{>}}}},
	bigvector/.style={decorate, decoration={snake,amplitude=4pt}, draw},
	vectorscalar/.style={loosely dotted,draw, postaction={decorate}},
}

\title{The Double-Dark Portal}
\author{David Curtin$^1$,}
\author{Yuhsin Tsai$^2$}
\affiliation{$^1$C. N. Yang Institute for Theoretical Physics, Stony Brook University, 
 Stony Brook, NY 11794, U.S.A.\\$^2$Physics Department, University of California Davis, Davis, California 95616, U.S.A.}

\abstract{
In most models of the dark sector, dark matter is charged under some new symmetry to make it stable. We explore the possibility that not just dark matter, but also the force carrier connecting it to the visible sector is charged under this symmetry. This dark mediator then acts as a Double-Dark Portal. 
We realize this setup in the \emph{dark mediator Dark matter} model (dmDM), featuring a fermionic DM candidate $\chi$ with Yukawa couplings to light scalars $\phi_i$. The scalars couple to SM quarks via the operator $\bar q q \phi_i^* \phi_j/\Lambda_{ij}$.
This can lead to large direct detection signals via the $2\rightarrow3$ process $\chi N \rightarrow \chi N \phi$  if one of the scalars has mass $ \lesssim 10 \kev$. 
For dark matter Yukawa couplings $y_\chi \sim 10^{-3} - 10^{-2}$, dmDM features a thermal relic dark matter candidate while also implementing the SIDM scenario for ameliorating inconsistencies between dwarf galaxy simulations and observations.
We undertake the first systematic survey of constraints on light scalars coupled to the SM via the above operator.
The strongest constraints are derived from a detailed examination of the light mediator's effects on stellar astrophysics. 
LHC experiments and cosmological considerations also yield important bounds.
Observations of neutron star cooling exclude the minimal model with one dark mediator, but a scenario with two dark mediators remains viable and can give strong direct detection signals. 
We explore the direct detection consequences of this scenario and find that a heavy $\mathcal{O}(100 \gev)$ dmDM candidate fakes different $\mathcal{O}(10 \gev)$ WIMPs at different experiments.  Large regions of dmDM parameter space are accessible above the irreducible neutrino background.
}

\begin{document} 
\maketitle


\setcounter{page}{3}

\section{Introduction}
\label{s.intro}

The existence of dark matter (DM) is firmly established by a myriad of astrophysical and cosmological observations \cite{Ade:2013zuv}.  Nevertheless, the exact characteristics of dark matter particles remain almost completely mysterious. Weakly Interacting Massive Particles (WIMPs) are the most popular DM candidate since they arise in supersymmetry and can naturally occur with the correct relic abundance \cite{Jungman:1995df}, but many other scenarios are possible.

Direct detection via DM-nucleus scattering \cite{Goodman:1984dc} has made tremendous strides, with experiments like LUX \cite{Akerib:2013tjd}, Super-CDMS \cite{Agnese:2013rvf} and XENON100 \cite{Aprile:2012nq}   achieving sensitivities to WIMP-nucleon scattering cross sections of $\sigma^\mathrm{SI}_n \sim 10^{-45} \ \cmtwo$ for a $\mathcal{O}(100 \gev$) WIMP. There have also been several anomalies in the $\mathcal{O}(10 \gev)$ mass range \cite{Bernabei:2013cfa, Aalseth:2012if, Angloher:2011uu, Agnese:2013rvf} that seem to conflict with each other, as well as with various exclusion bounds by the above experiments when assuming a WIMP-like scattering. It remains possible that some or all of these hints will be explained by something other than dark matter, especially given how challenging these measurements and their background suppression is in that mass range. Even so, past and current anomalies naturally stimulate a great deal of work by the theory community in an attempt to reconcile conflicting experimental results. The myriad of plausible  models demonstrates the necessity to explore as many different dark matter scenarios as possible, lest a crucial signal be overlooked. 

In most models of the dark sector, dark matter is charged under some new symmetry to make it stable. However, in light of the complex structure of the Standard Model (SM) there is no particularly strong reason to assume the dark sector to be so simple. We explore the possibility that not just dark matter, but also the force carrier connecting it to the visible sector is charged under this symmetry. This dark mediator then acts as a Double-Dark Portal. 

In \cite{Curtin:2013qsa} we introduced a model to realize this scenario: \emph{Dark Mediator Dark Matter} (dmDM). It features a fermionic dark matter candidate $\chi$ with Yukawa couplings to one or more light scalars $\phi_i$. These scalars carry dark charge and can only couple to the SM in pairs, realized as a nonrenormalizable coupling to quarks, $\bar q q \phi \phi/\Lambda$. For sufficiently light $\phi$ this can lead to direct detection via a $2\to3$ nuclear scattering process, shown in \fref{feynmandiagram}.

Bounds from direct detection experiments are usually analyzed assuming a contact operator interaction $\bar \chi \chi \bar q q/\tilde \Lambda^2$. The shape of the resulting nuclear recoil spectrum is entirely determined by the nuclear form factor and dark matter velocity distribution. Many past models feature different nuclear recoil spectra. Examples include the introduction of a mass splitting  \cite{TuckerSmith:2001hy, Graham:2010ca, Essig:2010ye}; considering matrix elements $|\mathcal{M}|^2$ with additional velocity- or momentum transfer suppressions (for a complete list see e.g. \cite{MarchRussell:2012hi}), especially at low DM masses close to a GeV \cite{Chang:2009yt}; light scalar or `dark photon' mediators (see e.g. \cite{Essig:2013lka, Essig:2010ye}) which give large enhancements at low nuclear recoil; various forms of composite dark matter \cite{Alves:2009nf, Kribs:2009fy, Lisanti:2009am, Cline:2012bz, Feldstein:2009tr} which may introduce additional form factors; and DM-nucleus scattering with intermediate bound states \cite{Bai:2009cd} which enhances scattering in a narrow range of DM velocities. 
Notably missing from this list are alternative process topologies for DM-nucleus scattering. 
This omission is remedied by the dmDM scenario, which generates a functionally unique recoil suppression and overall cross section dependence on DM and nucleus mass. Direct detection constraints on dmDM are explored in this paper in detail, and we show that a $\sim 100 \gev$ dmDM candidate fakes different $\mathcal{O}(10 \gev)$ standard WIMPs at different experiments. 

Dark Mediator Dark Matter has important consequences outside of direct detection. Coupling dark matter to a light scalar can ameliorate inconsistencies between simulations and observations of dwarf galaxies \cite{Carlson:1992fn,Spergel:1999mh,Tulin:2013teo} while being compatible with a thermal relic. Perhaps more drastic however is the unique pair-wise coupling of light scalars to SM quarks. 

We conduct the first systematic survey to constrain operators of the form $\bar q q \phi_i \phi_j^*/
\Lambda_{ij}$ where $\phi_i$ is a very light scalar, checking a large variety of cosmological, astrophysical and collider bounds. 
The heaviest stable dark mediator has to be lighter than $\sim \ev$ to avoid overclosing the universe. This makes emission during direct detection plausible. The most stringent bounds on its coupling come from observations of neutron star cooling, which require $\Lambda \gtrsim 10^8 \tev$ for a single dark mediator. However, all constraints are easily circumvented in a model with two mediators, which can generate a strong direct detection signal. The constraints we derive are important outside of the dmDM context as well, applying to any light scalars with the above coupling to the SM. 

The pairwise dark mediator coupling to quarks is not gauge invariant above the electroweak breaking scale, necessitating a UV completion. We present one such possibility featuring dark vector quarks, leading to discoverable TeV scale LHC signatures.

This paper is organized as follows. In \sref{dmdm} we define the dark mediator Dark Matter model and outline how dmDM could be realized in a UV-complete theory with its own set of LHC signatures. \sref{yXbounds} summarizes bounds on the dark matter Yukawa coupling to dark mediators. 
In \sref{stellarconstraints} we derive stellar astrophysics bounds on dark mediators coupled to SM quarks, which give the most powerful constraints on our scenario. Cosmology and LHC experiments also yield important bounds, which are discussed in \sref{constraints}.  A realistic model of dmDM, which avoids all constraints, is defined in \sref{phisummary}. \sref{directdetection} reviews the direct detection phenomenology of dmDM, and we conclude in \sref{conclusion}. Some technical details and additional calculations are presented in the Appendices.

\vspace{3mm}
\section{Dark Mediator Dark Matter}
\label{s.dmdm}
In this section we define the Dark Mediator Dark Matter model and discuss a possible UV-completion involving heavy vector-like quarks that could be discoverable at the LHC.

\subsection{Model Definition}
Given its apparently long lifetime, most models of DM include some symmetry under which the DM candidate is charged to make it stable. An interesting possibility is that not only the DM candidate, but also the mediator connecting it to the visible sector is charged under this dark symmetry. Such a `dark mediator' 
$\phi$ could only couple to the SM fields in pairs, at leading order.

There are several possibilities for writing down a dark-mediator model. However, if the mediator couples via additional derivatives or through loops, direct detection is  suppressed below observable levels. This limits the choice of dark mediator couplings to the simple construction introduced in \cite{Curtin:2013qsa}, which we repeat here.

\begin{figure}
\begin{center}
\includegraphics[width=8cm]{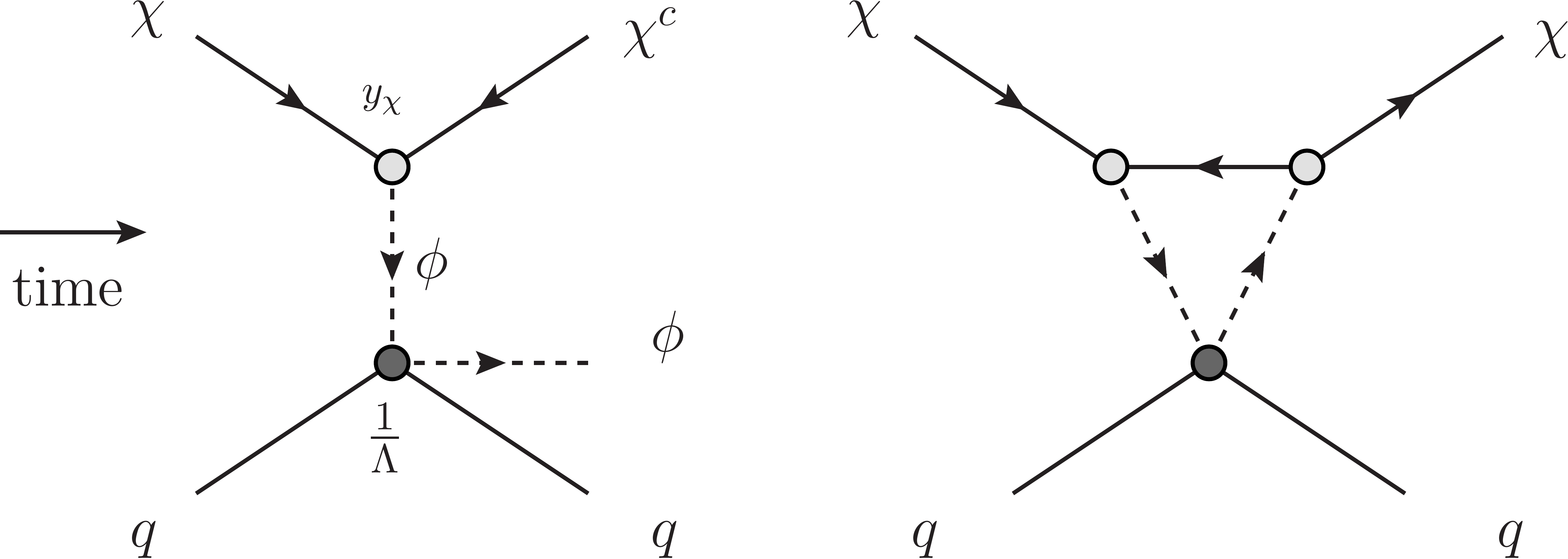}
\end{center}
\caption{
The quark-level Feynman diagrams responsible for DM-nucleus scattering in \emph{Dark Mediator Dark Matter} (dmDM). Left: the $2\rightarrow3$ process at tree-level. Right: the loop-induced $2\rightarrow2$ process. The arrows indicate flow of dark charge.
}
\label{f.feynmandiagram}
\end{figure}

Consider real or complex SM singlet scalars $\phi_i$ coupled to quarks,
along with Yukawa couplings to a Dirac fermion DM $\chi$.
The relevant terms in the effective Lagrangian are 
\begin{equation}
\label{e.dmdm}
\mathcal{L}_\mathrm{DM}  \supset 
\displaystyle{\sum_{i,j}^{n_{\phi}}}\, \frac{1}{\Lambda_{ij}}  \bar q\,q \,\phi_i \phi_j^* + \displaystyle{\sum_{i}^{n_{\phi}}}\left ( y^i_\chi \overline{\chi^c}\chi \phi_i + h.c. \right) 
+ \sum_{i,j,k,l} \lambda_{ijkl} \phi_i \phi_j^* \phi_k \phi_l^* + \cdots,
\end{equation}
where $\ldots$ stands for $\phi, \chi$ mass terms, as well as the rest of the dark sector, which may be more complicated than this minimal setup.  This interaction structure can be enforced by a $\mathbb{Z}_4$ symmetry.  The first two terms dictate the dark sector's interaction with the SM, while the quartics are only important in the early universe (see \sref{constraints}).\footnote{The $\mathbb{Z}_4$ symmetry also allows higgs portal couplings of the form $|H|^2 \phi_i \phi_j^*$, but they will have a very subdominant effect on phenomenology compared to the first term in \eref{dmdm}.} 

The leading order process for DM-nucleus scattering is $\chi N \to \bar \chi N \phi$ if $m_\phi \lesssim \mathcal{O}(10 \kev)$. However, an elastic scattering $\chi N \to \chi N$ is always present at loop-level since it satisfies all possible symmetries, see \fref{feynmandiagram}. This low-energy $2\to2$ loop process is equivalent to the operator
\begin{equation}
\label{e.2to2operator}
\frac{\,y_{\chi}^2}{2\,\pi^2}\,\frac{1}{\Lambda\,q} \ (\bar{\chi}\,\chi\,\bar{N}\,N),
\end{equation}
(for $n_\phi = 1$) in the massless $\phi$ limit, where $q=\sqrt{2\,m_N\,E_r}$ is the momentum transfer in the scattering.\footnote{Note that in this limit, the process has an IR pole similar to tree-level $t$-channel exchange, hence the $q^{-1}$ dependence.} Effectively, this is identical to a standard WIMP with a $\bar \chi \chi \bar N N$ contact operator, but with an additional $1/E_r$ suppression in the cross section. This gives a similar phenomenology as a light mediator being exchanged at tree-level with derivative coupling.

The main new features of this model for direct detection  in \sref{directdetection} are captured by the minimal case with a single mediator $n_\phi = 1$. However, the actual number of dark mediators is important for interpreting indirect constraints in Sections \ref{s.yXbounds}, \ref{s.stellarconstraints} and \ref{s.constraints}. It also affects the relative importance of the two nuclear scattering processes.  When $n_{\phi}=1$, the $2\rightarrow3$ process will dominate direct detection for Yukawa coupling $y_\chi$ below some threshold as long as $m_\phi \lesssim \kev$. If $n_\phi = 2$, however, the dominant scalar-DM coupling could be $\bar q q \phi_1 \phi_2^*/\Lambda_{12}$. In that case, the $2\to2$ operator above is $\propto y_\chi^{\phi_1} y_\chi^{\phi_2}$ and can be suppressed without reducing the $2\to3$ rate by taking $y_\chi^{\phi_1} \ll y_\chi^{\phi_2}$. Both processes will be considered for direct detection in \sref{directdetection}.

The effect of strong differences between proton and neutron coupling to DM have been explored by \cite{Feng:2011vu}. To concentrate on the kinematics we shall therefore assume the operator $\bar q q \phi \phi^*/\Lambda$ is flavor-blind in the quark mass basis.

We point out that depending on the UV completion of the model, a leptonic coupling via $\bar \ell \ell \phi \phi^*$ is also possible. We do not consider it here, since direct detection would be very difficult, but indirect constraints, in particular from white dwarf cooling, could be sensitive to such a scenario.

\subsection{A possible UV-completion}
\label{ss.uvcompletion}

\begin{table}[t]
\begin{center}
\begin{tabular*}{0.45\textwidth}{@{\extracolsep{\fill}}c|cccc}
	\hline
	\\[-7pt]
	 $\quad$  & $SU(3)_c$ & $SU(2)_L$ & $U(1)_Y$ & $\mathbb{Z}_4$ \\[2pt]
	\hline\hline
	\\[-6pt]
	$\bar Q$ & $\bar 3$ & $\bar 2$ & $-1/6$ & $0$  \\[2pt]
	\\[-6pt]
	$u$ & $3$ & $1$ & $2/3$ & $0$ \\[2pt]
	\\[-6pt]
	$d$ & $3$ & $1$ & $-1/3$ & $0$ \\[2pt]
	\\[-6pt]
	$H$ &$1$ & $ 2$ & $1/2$ & $0$ \\[2pt]
	\hline
	\\[-6pt]
	$\phi$ &$1$& $1$ & $0$ &  $\pi$\\[2pt]
	\hline
	\\[-6pt]
	$\psi_{Q_{1,2}}$ &$3$  & $ 2$ & $1/6$ &  $\pi$\\[2pt]
	\\[-6pt]
	$\psi_{u_{1,2}}$ &$3$  & $1$ & $2/3$ &  $\pi$\\[2pt]
	\\[-6pt]
	$\psi_{d_{1,2}}$ &$3$  & $1$ & $-1/3$ &  $\pi$\\[2pt]
	\hline
	\\[-6pt]
	$\chi$ & $1$ & $1$ & $\!\!\!\!0$ & $\pi/2$\\[2pt]
	\hline
\end{tabular*}
\caption{Particle content of the dark vector quark UV completion of dmDM: complex scalar $\phi$, Dirac fermions $\psi$ (with index $1,\,2$ for the two Weyl fermion components) and $\chi$. $\tilde{H}=i\,\sigma^2H^*$.}\label{t.particlecontent}
\end{center}
\end{table}

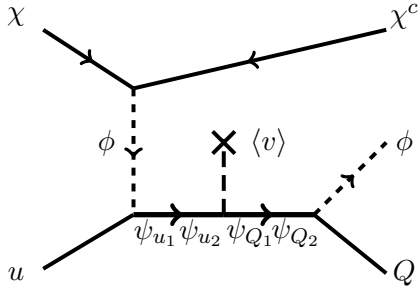
\begin{figure}
\begin{center}
		\begin{tikzpicture}[line width=1.5 pt, scale=1.2] 
			\draw[fermion] (-1.5,1.25) -- (-0.5,0.6);
			\draw[fermion] (2.3,1.25) -- (-0.5,0.6);
			\draw[scalar] (-0.5,0.6) -- (-0.5,-0.8);
			\draw[fermionnoarrow] (-1.5,-1.4) -- (-0.5,-0.8);
                        \draw[fermion][line width=2pt] (-0.5,-0.8) -- (0.5,-0.8);
                        \draw[scalarnoarrow] (0.5,-0.8) -- (0.5,0.04);
                        \draw[fermion][line width=2pt] (0.5,-0.8) -- (1.5,-0.8);
                        \draw[fermionnoarrow] (1.5,-0.8) -- (2.3,-1.45);
                        \draw[scalar] (1.5,-0.8) -- (2.3,0);			%
			\begin{scope}[rotate=90]
				\begin{scope}[shift={(0,-.5)}] 
					\clip (0,0) circle (.175cm);
					\draw[fermionnoarrow] (-1,1) -- (1,-1);
					\draw[fermionnoarrow] (1,1) -- (-1,-1);
				\end{scope}	
			\end{scope}
						           \node at (-0.8,0) {$\phi$};
						           \node at (2.5,0) {$\phi$};
						           \node at (-1.8,-1.45) {$u$};
           \node at (-1.8,1.4) {$\chi$};
           \node at (1,0) {$\langle v\rangle$};
            \node at (0.8,-1) {$\psi_{Q_1}$};
            \node at (1.3,-1) {$\psi_{Q_2}$};
           \node at (-0.25,-1) {$\psi_{u_1}$};
            \node at (0.25,-1) {$\psi_{u_2}$};

           \node at (2.5,-1.45) {$Q$};
           \node at (2.5,1.4) {$\chi^c$};
                       		\end{tikzpicture}
	\end{center}            
\caption{The $2\to 3$ direct detection scattering process within the UV completion of dmDM. When treating Higgs vev as a mass insertion, the propagator of heavy Dirac quark is dominated by the chirality-flipping piece, $\frac{M_Q}{p^2-M_Q^2}$, at low energy. This gives the suppression scale in \eref{effcoupling}.
}
\label{f.2to3process}
\end{figure}

Above the electroweak symmetry breaking scale the $\bar q q \phi \phi^*/\Lambda$ operator is realized as $\bar Q_L H q_R \phi\phi^*/M^2$. This is suggestive of a particular UV completion involving heavy vector-like fermions coupling to $\phi$ and SM quarks via Yukawa couplings. The minimal particle content to realize dmDM is therefore a light scalar mediator $\phi$, heavy vector-like quarks $\psi_{Q,\,q}$ in the same gauge representations as the SM $Q_L, u_R, d_R$ respectively, and a Dirac fermion dark matter candidate $\chi$. Their charges are shown in \tref{particlecontent}. The Lagrangian\footnote{We show the $n_\phi = 1$ complex scalar case, generalization to more or real dark mediators are trivial.} contains Yukawa couplings 
\begin{eqnarray}
\nonumber \mathcal{L} &\subset& y_Q\,\phi^*\,\bar Q \,{\psi_{Q_2}}+ y_{h}\left(\bar{\psi}_{Q_{1,2}} H\psi_{d_{2,1}}+\bar{\psi}_{Q_{1,2}}\tilde{H}\psi_{u_{2,1}}\right)\\
&& + y_q\left(\phi\,\bar{\psi}_{d_1}\,d+\phi\,\bar{\psi}_{u_1}\,u\right)+ h.c.,
\label{e.HQyukawa}
\end{eqnarray}
where the index $1,\,2$ represents the chirality component of Dirac fermion $\psi$'s. $\psi_{Q,u,d}$ have Dirac masses
\begin{equation}
M_Q\,\bar \psi_{Q}\, {\psi}_{Q}+M_u\,\bar \psi_{u}\,{\psi}_{u}+M_d\,\bar\psi_{d}\,{\psi}_{d}.
\end{equation}
The DM mass and its coupling to $\phi$ are given by
\begin{equation}\label{eq:dmcoupling}
m_{\chi}\,\bar \chi\,\chi + y_{\chi}\, \overline{\chi^c}\,\chi\,\phi + h.c.\,\,.
\end{equation}
We assume all the couplings are flavor universal and 
$M_Q=M_q$, $y_Q = y_q$
for simplicity. 

The direct detection $2\rightarrow3$ scattering process is shown in \fref{2to3process}. 
When the momentum transfer through heavy quarks is much smaller than $M_Q$, we can integrate out the lower part of the diagram to generate the dimension 6 operator
\begin{equation}\label{eq:phiphiQq}
\frac{y_Q^2 y_{h}}{M_Q^2} \left(\bar{Q}\,H\,d+\bar{Q}\,\tilde{H}\,u\right)\phi \phi^*.
\end{equation}
Below the scale of electroweak symmetry breaking this becomes the operator of \eref{dmdm} with 
\begin{equation}\label{e.effcoupling}
\Lambda = \frac{M_Q^2}{y_Q^2 y_h v}
\end{equation}
where $v = 246 \gev$ is the SM Higgs VEV.

As we will show, $M_Q$ could easily be TeV scale, allowing for discovery of these heavy vector-like quarks at the LHC. As long as the LHC with $\sqrt{s} = 8 \tev$ has not produced them on shell they are not trivially excluded despite being new colored states that couple to the Higgs. Since they do not receive their mass primarily from the Higgs vev, their contribution to the $h\gamma\gamma$ loop coupling is strongly suppressed. As we discuss in \sref{constraints}, the collider constraints on additional vector-like quark generations can be satisfied for $M_Q \gtrsim \tev$. The quark Yukawa couplings do receive a flavor-universal correction which may lead to the light quark Yukawa couplings being tuned to the order of $0.1\%$, 
but like the origin of the light scalar $\phi$ we put these naturalness issues aside to concentrate on the phenomenology of dmDM.

\vspace{3mm} 
\section{Constraining the DM Yukawa Coupling}
\label{s.yXbounds}

The dark matter Yukawa coupling $y_\chi \overline {\chi^c} \chi \phi$ can be constrained by various astrophysical and cosmological observations, the most important of which we summarize here. For simplicity these bounds are formulated for $n_\phi = 1$, but can also be applied directly to $n_\phi > 1$ scenarios if one Yukawa coupling dominates.

The dark matter relic density $\Omega_\mathrm{CDM} = 0.1196 \pm 0.0031$  has been accurately measured by the Planck Satellite \cite{Ade:2013zuv}. Under the assumptions of a simple thermal relic this fixes $y_\chi$ to a \emph{specific value} (which depends on $m_\chi$). The lowest-order annihilation cross section for the process $\chi\,\bar{\chi}\to\phi\phi^*$ is
\begin{equation}
\sigma_{\chi\,\bar{\chi}\to\phi\phi^*}=\frac{y_{\chi}^4}{64\pi m_{\chi}^2}\label{eq:XsecRelic},
\end{equation}
assuming no sizable $\phi^3$ couplings. Performing the standard WIMP freeze-out calculation \cite{Kolb:1990vq} we find that the $\phi \phi^* \leftrightarrow \bar \chi \chi$ process freezes out at the usual $T \sim m_\chi/20$. Requiring that $\Omega_\chi = \Omega_\mathrm{CDM}$ gives
\begin{equation}
\label{e.DMrelicyX}
y_\chi \approx 0.0027 \sqrt{\frac{m_\chi}{\gev}}.
\end{equation}
This is generically very small, of order $0.01$ for $\sim 10 \gev$ DM, and is compared to the other $y_\chi$ bounds in \fref{yXboundplot} (magenta line). We emphasize that this constraint will be shifted if $\chi$ is non-thermally produced. 
Although  DM interaction is mediated by light scalars, the Sommerfeld enhancement, which is proportional to \cite{Feng:2010zp}
\begin{equation}
S\simeq\frac{\pi\alpha_{\chi}/v}{1-e^{-\pi\alpha_{\chi}/v}},
\end{equation}
is negligible due to the small Yukawa coupling $y_{\chi}^2$, as well as the relatively large velocity $v\simeq 0.3$ during freeze-out.

\begin{figure}
\begin{center}
\includegraphics[width=8cm]{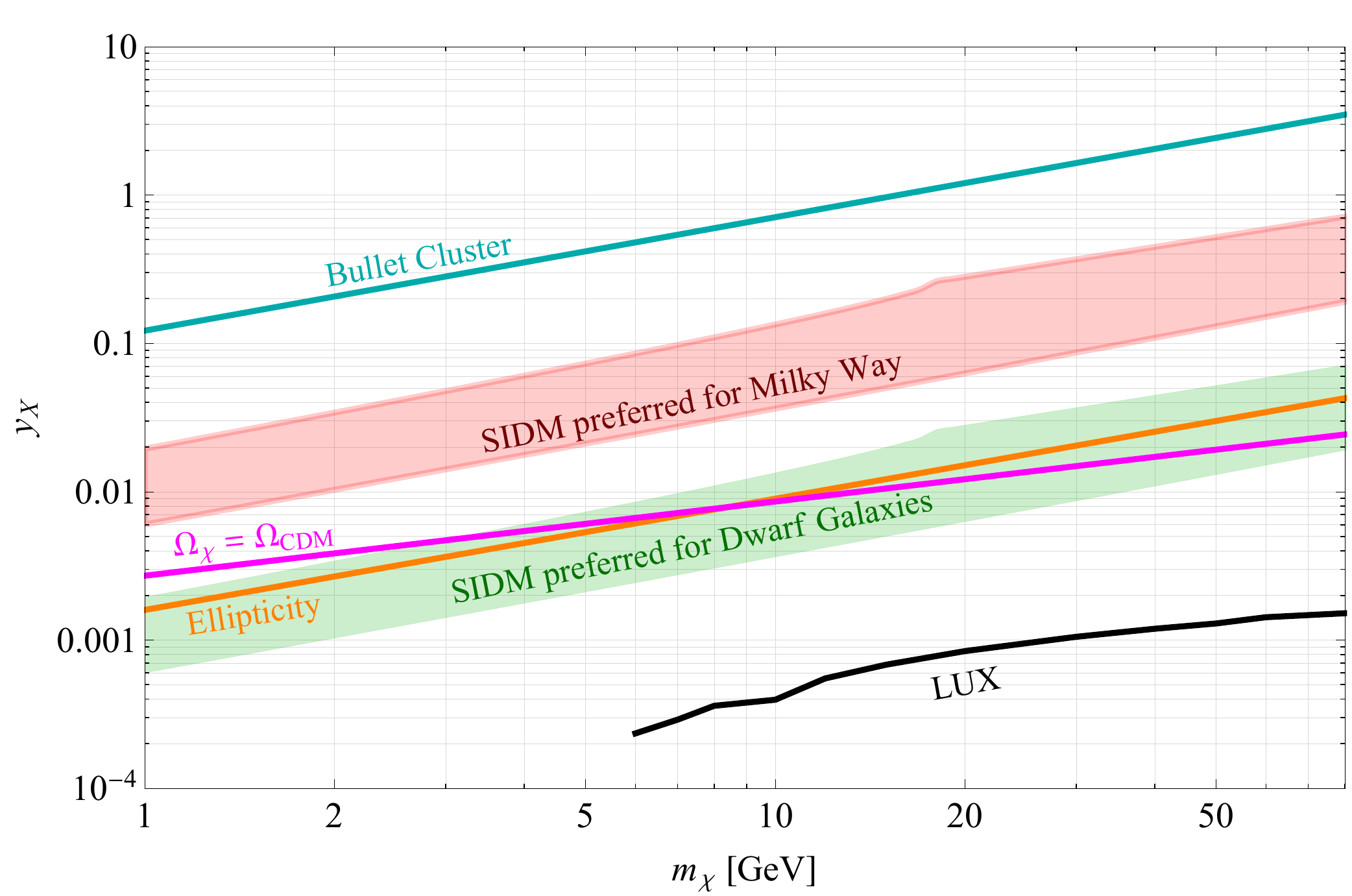}
\end{center}
\caption{
Bounds on  the Yukawa coupling $y_\chi \overline {\chi^c} \chi \phi$ for $n_\phi = 1$. (These bounds can also be applied directly to $n_\phi > 1$ scenarios if one Yukawa coupling dominates)
Magenta: required value of $y_\chi$ for $\chi$ to be a thermal relic. Cyan and Orange: upper bounds on $y_\chi$ from bullet cluster and ellipticity observations. The green shaded region implements the SIDM solution to the core-cusp and too-big-to-fail problems of dwarf galaxies \cite{Carlson:1992fn,Spergel:1999mh,Tulin:2013teo}, while the pink region can modify the halo of milky way size galaxies. See text for details. Black curve: $2\to3$ dominated direct detection requires $y_\chi$ to lie below this curve if $n_\phi = 1$, see \ssref{ddconstraints}. 
}
\label{f.yXboundplot}
\end{figure}

An \emph{upper bound} on the dark matter self-interaction may be obtained from observations of the Bullet Cluster and galactic ellipticities. This was done by the authors of  \cite{Feng:2009mn} for a massless mediator. We can apply those bounds directly to our model as long as $m_\phi$ is much smaller than the momentum transfer of a characteristic DM-DM collision  ($q \gtrsim \mev$ for $m_\chi \gtrsim \gev$). The bullet cluster bound
\begin{equation}
\label{e.bulletclusterbound}
y_\chi \lesssim 0.13 \left(\frac{m_\chi}{\gev}\right)^{3/4}
\end{equation}
is considered quite reliable, but concerns have been raised about the ellipticity bound, the strength of  which may have been overestimated \cite{Peter:2012jh}. Both upper bounds are shown in \fref{yXboundplot} (cyan and orange lines).

 
Rather than merely requiring the light mediator to not spoil well-understood aspects of galaxy formation and interaction, one could go one step further and use the dark matter self-interaction to address existing inconsistencies between prediction and observation. 
Current $N$-body simulations of Cold Dark Matter halos predict an overabundance of dwarf spheroidals, as well as dwarf galaxy halos that are more cusped than observed. These inconsistencies are called the too-big-to-fail and core-cusp problems. It has been shown that the disagreement between simulations and observation can be ameliorated by introducing a sizable dark matter self-interaction, dubbed the Selft Interacting Dark Matter (SIDM) scenario \cite{Carlson:1992fn,Spergel:1999mh,Tulin:2013teo}.

The presence of a light scalar in the $m_\phi \lesssim \mev$ mass range allows dmDM to act as a realization of SIDM. To derive the \emph{preferred range of $y_\chi$} we follow the procedure in \cite{Tulin:2013teo}. 

The small ratio between the potential energy of $\phi$ mediation and the kinetic energy of DM in galactic halos, $2\alpha_{\chi}m_{\phi}/(m_{\chi}v^2)\ll 1$, shows that DM self-interaction should be described in the classical limit. The transfer cross section for DM scattering, 
\begin{equation}\label{eq:sigmaTdmDM}
\frac{\sigma_T}{m_{\chi}} \simeq \frac{y_{\chi}^4}{\pi\,m_{\chi}^3\,v^4}\,\ln\left(\frac{4\pi\,m_{\chi}\,v^2}{2\,y_{\chi}^2\,m_{\phi}}\right),
\end{equation}
is just the total cross section weighted by fractional longitudinal momentum transfer. A value of
\begin{equation}
\label{e.SIDMxsec}
\frac{\sigma_T}{m_{\chi}} = 0.5-30\,\rm{cm}^{2}/\rm{g}
\end{equation}
could reconcile the inconsistencies between $N$-body simulations and observations. The required coupling depends on the ambient dark matter velocity, which is $\sim 30$ km/s for dwarf galaxies and $\sim 300$ km/s in larger milky way size galaxies. \fref{yXboundplot} shows the preferred bands of $y_\chi$ to achieve the cross section \eref{SIDMxsec} in these two systems. In this plot, $m_\phi = \mev$, but the change for $m_\phi = \ev$ is not substantial.\footnote{The heavier $\phi$ is chosen to evade neutron star bounds, see \sssref{neutronstarcooling}. $2\to3$ direct detection with emission of a $\lesssim \kev$ dark scalar can still occur in an $n_\phi = 2$ model, see \sref{phisummary}.} As we can see, the dmDM model with a thermal relic DM does provide a potential solution to the core-cups and too-big-to-fail problem of dwarf galaxies. 
 
 
Finally, as we discuss in \ssref{ddconstraints}, there is an \emph{upper bound} on $y_\chi$ for the $2\to3$ process to dominate direct detection when $n_\phi = 1$.  If $y_\chi$ is larger, direct detection proceeds via the $2\to2$ loop process. This is shown for the LUX experiment as the black line in \fref{yXboundplot}. (The corresponding upper bound for other experiments is somewhat weaker.) Note that this boundary between the two direct detection regimes is arbitrarily shifted for $n_\phi = 2$.


 In summary, \fref{yXboundplot} shows both the \emph{preferred values} of $y_\chi$ for a thermal relic and to resolve inconsistencies between observations and simulations for dwarf galaxies and the milky way; it also shows the \emph{upper bounds} on $y_\chi$ to satisfy bullet cluster and self-interaction bounds, and to ensure $2\to3$ dominated direct detection. Roughly speaking, the most relevant values of $y_\chi$ are $\sim 10^{-3} - 10^{-2}$.

\vspace{3mm} 
\section{Constraining the Dark Mediator $\phi$ through Stellar Astrophysics}
\label{s.stellarconstraints}

A light dark mediator like $\phi$ coupling to the SM via
\begin{equation}
\label{e.qqphiphi}
\frac{1}{\Lambda}\bar q q \phi \phi^*
\end{equation}
is produced in the early universe, as well as stellar cores and high energy colliders.

In this section we compute $m_\phi$-dependent bounds on $\Lambda$ from stellar astrophysics. The light scalar $\phi$ is produced in stellar cores if $m_\phi \lesssim T$. This can affect the length of the neutrino burst in supernovae explosions, radiative heat transfer and energy loss in the sun, and the cooling of stellar relics. We assume $n_\phi = 1$, but the constraints are easily applied to the more general case. 

The derivation of these bounds differs from the corresponding calculations  for axions, since light scalars couple more strongly at low energy due to the scaling of the operator \eref{qqphiphi}. In the regime where respective bound can be set, $\phi$ fully thermalizes in the sun and white dwarfs. 
By far the strongest constraints are obtained from observations of neutron star cooling: $\Lambda \gtrsim 10^8 \tev$ for $m_\phi \lesssim 100 \kev$, which excludes this scenario for direct detection completely. However, in \sref{phisummary} we construct  $n_\phi = 2$ scenario with one eV and one MeV dark mediator that evades all constraints while allowing for sizable direct detection signals.

It is useful to keep in mind the range of $\Lambda$ relevant for direct detection.  As discussed in \sref{yXbounds}, the preferred range for the dominant DM Yukawa coupling is $y_\chi \sim 10^{-3} - 10^{-2}$. Direct detection bounds on dmDM were computed in \cite{Curtin:2013qsa} and are reviewed in \sref{directdetection}. For dmDM to be detectable above the irreducible neutrino background, $\Lambda \lesssim 10^4 \tev$ in the relevant dark mediator coupling to quarks.

\subsection{$\phi$ interaction and production cross sections}

\begin{figure*}
\begin{center}
\includegraphics[width=11.4cm]{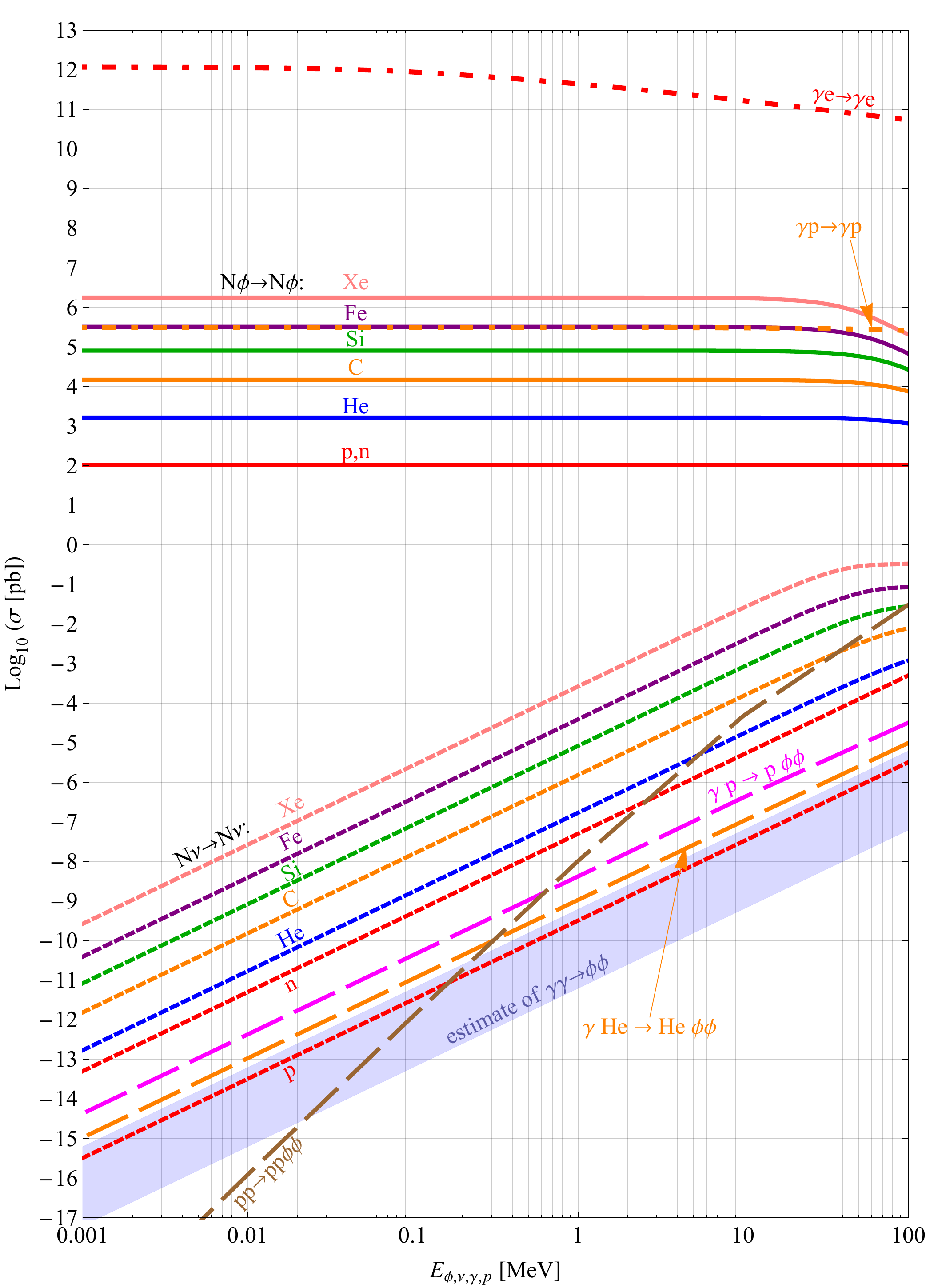}
\end{center}
\caption{
Low-energy scattering and production cross sections  for computing bounds on the dmDM model, compared to some relevant SM processes. Solid lines: coherent scattering of $\phi$ off a stationary nucleus via the operator $\bar q q \phi \phi^*/\Lambda$. Dashed lines: coherent scattering of a neutrino off a stationary nucleus via $Z$-exchange (see also \cite{Brice:2013fwa}). Long-Dash-Dotted lines: Compton scattering of a photon off a stationary electron or proton. Long-Dashed lines: $p p \rightarrow p p \phi \phi^*$, $\gamma p \rightarrow p \phi \phi^*$  and $\gamma \mathrm{He} \rightarrow \mathrm{He} \  \phi \phi^*$ where one initial proton is stationary. The blue band represents a naive dimensional analysis estimate \eref{gagaphiphiestimate} of $\gamma \gamma \rightarrow \phi \phi^*$ (or the reverse annihilation process $\phi \phi^* \to \gamma \gamma$), taking $\mathcal{B}^2 = 1 - 100$. In all cross sections involving the $ \bar q q \phi \phi^*/\Lambda$ operator we used $\Lambda = 10 \tev$. }
\label{f.crosssectionphilowE}
\end{figure*}

Computing stellar astrophysics and cosmological bounds requires an understanding of the $\phi$-nucleus \emph{scattering} cross sections at sub-GeV energies. This is easily computed analytically using standard methods for DM scattering and is shown in \fref{crosssectionphilowE} for $\Lambda = 10 \tev$.  For illustration we also compare these cross sections to some relevant SM scattering processes, $\nu N \rightarrow \nu N$ and Compton scattering.  Note the different energy scaling of these cross sections, with $\sigma(\phi N \rightarrow \phi N)$ being independent of energy for $E_\phi \lesssim 100 \mev$.

At tree-level, $\phi$ only couples hadronically. Therefore, the most relevant \emph{production} processes for $\phi$ in stellar cores are
\begin{equation}\label{e.phiproduction}
N \gamma \rightarrow N \phi \phi^* \ , \ \  \ \ 
p p \rightarrow p p \phi \phi^* \ , \ \  \ \ 
\gamma \gamma \rightarrow \phi \phi^*
\end{equation}
Again we are only concerned with sub-GeV energy scales. We can model the first two processes, shown in \fref{crosssectionphilowE}, in \texttt{MadGraph5} by treating the proton as a fundamental fermion and multiplying the cross section by a quark-nucleon matrix element factor, see \eref{Bsqfactor}. The Helm form factor \eref{Helm} is also included for nuclei. The one-pion exchange approximation was employed for the first process \cite{steigman}, and the obtained cross section should be seen as an $\mathcal{O}(1)$ estimate. The first process can occur off any nucleus, with $N = p, \ \mathrm{He}$ shown in \fref{crosssectionphilowE} (the cross sections for N = He, C, O are nearly identical), which is relevant in the Sun and white dwarfs. The second process proceeds identically for protons and neutrons and is relevant in neutron stars, with additional subtleties due to neutron degeneracy discussed in \sssref{neutronstarcooling}.

The photon annihilation process $\gamma \gamma \rightarrow \phi \phi$ is difficult to calculate due to  unknown form factors connecting quarks to hadronic QCD resonances. A rough estimate of the amplitude can be obtained by treating it as a loop process mediated by constituent quarks. The same approach is used to calculate the photon meson couplings, for example in \cite{Volkov:2009mz}. With the correct power of electric charge and one mass insertion for the correct chirality, the size of operator $|\phi|^2F_{\mu\nu}F^{\mu\nu}$ is approximated as
\begin{equation}
\frac{\alpha}{4\,\pi}\frac{\mathcal{B}}{\Lambda\,m_q}\,|\phi|^2F_{\mu\nu}F^{\mu\nu},
\end{equation}
where $\mathcal{B}$ is the form factor between the $\phi$ and constituent quarks, and $m_{u,d}\simeq 263$ MeV \cite{Volkov:2009mz} is the mass of the constituent quarks within the NJL model. The resulting cross section is
\begin{eqnarray}
\label{e.gagaphiphiestimate}
\sigma_{\gamma\gamma\to\phi\phi}
&\sim&
\frac{1}{16\,\pi}\left(\frac{\alpha}{\pi\,m_q}\right)^2\left( \frac{\mathcal{B}}{\Lambda}\right)^2\,E_{\gamma}^2
\\ \nonumber
&\approx&(7 \times 10^{-14} \mathrm{pb} ) \  \mathcal{B}^2 \left(\frac{\tev}{\Lambda}\right)^2 \left(\frac{E_\gamma}{\kev}\right)^2 
\end{eqnarray}
The blue band in \fref{crosssectionphilowE} is a very rough estimate with $\mathcal{B}^2 = 1 - 100$. At our level of precision we also take this to be the cross section for the reverse annihilation process $\phi \phi \to \gamma \gamma$.

\subsection{Supernovae}
\label{sss.supernovae}
Like massless axions, production and emission of $\phi$'s can lead to rapid energy loss during a supernova explosion. This can be constrained by measuring the duration of the associated neutrino burst. There are two allowed regimes \cite{Raffelt:2006cw}. The $\phi$ are trapped in the stellar medium if they couple more strongly to the SM than neutrinos. In that case they do not affect the neutrino burst. Alternatively, if the SM copuling is 5 orders of magnitude weaker, $\phi$ production is too negligible to affect the supernova. 

Rescaling $\sigma_{\phi N \to \phi N} \propto \Lambda^{-2}$ at $E_\phi \sim 10 \mev$ from \fref{crosssectionphilowE}, we see that the former constraint is  satisfied for $\Lambda \lesssim 10^6 \tev$. Therefore, supernova roughly supply the bound 
\begin{equation}
\Lambda \gtrsim 10^{11} \tev \ \ \mathrm{or} \ \ \Lambda \lesssim 10^6 \tev
\end{equation}
on $\Lambda$ involving scalars with a mass of $m_\phi \lesssim 10 \mev$, the temperature of a supernova explosion.

\subsection{Solar Energy Loss and Radiative Heat Transfer}
\label{sss.solarcooling}

A stellar core at some temperature $T \ll \gev$ can be seen as a fixed target experiment in which slow-moving nuclei are bombarded by photons as well as relativistic electrons and, in the case of dmDM, $\phi$ scalars. The most relevant production processes for $\phi$ are shown in \eref{phiproduction}, with cross sections as a function of energy illustrated in \fref{crosssectionphilowE}. The $\phi$ production rate per second per volume via a process $X_1 X_2 \rightarrow \phi \phi^* +$ SM particles, with cross section $\sigma_\mathrm{\phi prod}$ and parent particle number densities $n_{X_i}$, is
\begin{equation}
\label{e.rphicreate}
r_\phi^\mathrm{create} = 2 n_{X_1} n_{X_2} c \  \sigma_\mathrm{\phi prod} \propto \Lambda^{-2}.
\end{equation}
(We assume $\phi$ is so light that it is always relativistic.) On the other hand, the mean free path for $\phi$ before it scatters off nuclei in the star is
\begin{equation}
\label{e.Lphi}
L_\phi = \left( \sum_i n_{N_i} \sigma_{\phi N_i \rightarrow \phi N_i} \right)^{-1} \propto \Lambda^2
\end{equation}
for $N_i = $ \{p, He\} in the case of the Sun, with additional heavier elements in white dwarfs.

Estimating \emph{energy loss} due to $\phi$ emission from the star is greatly simplified if we can make four assumptions:
\begin{enumerate}[(a)]
\item The effect of $\phi$ production is small enough so as to not significantly influence the evolution of the star, allowing us to treat it as a background source of $\phi$'s.\footnote{This is a consistent assumption when setting conservative limits.}
\item $\phi$ particles are produced predominantly at the center of the star.
\item $L_\phi$ is short enough that $\phi$ scatters many times and thermalizes before leaving the star.
\item There is negligible $\phi$ annihilation in the star.
\end{enumerate}
If these conditions are satisfied, the created $\phi$ particles diffuse outwards from the center until they reach a layer of low enough density so that the surface of the star is within $\sim$ one scattering length, at which point they escape. Each $\phi$ carries away energy $E_\phi \sim T_\phi^\mathrm{escape}$, where $T_\phi^\mathrm{escape}$ is the temperature of the `layer of last scattering'.\footnote{This is to be compared to the free-streaming case, where the energy distribution of $\phi$'s from creation processes might have to be taken into account.} In the absence of annihilation processes, the equilibrium rate for $\phi$ emission is equal to the total rate of $\phi$ production, which together with $T_\phi^\mathrm{escape}$ gives the total energy loss from $\phi$ emission.

We now perform this computation for the case of the Sun. Radial density, temperature and mass fraction profiles for the standard solar model can be found in basic astrophysics textbooks and are reproduced in \aref{radialprofiles} for reference. The radius of the sun is about $R_\mathrm{sun} \approx 3.85 \times 10^{26}$ cm, while the central density and temperature are $\rho_\mathrm{sun}(0) \approx \ 150 \ \gram \ \cmmthree$ and $T_\mathrm{sun}(0) \approx 1.5 \times 10^7 \mathrm{K} \approx 1.3 \kev$. The corresponding nucleus number densities are of order $10^{25}  \ \cmmthree$, while the density of photons obeying a Bose-Einstein distribution is $n_\gamma = \frac{2 \zeta(3)}{\pi^2} T^3 \sim 10^{22} \cmmthree$. Consulting \fref{crosssectionphilowE} and \eref{rphicreate} it is clear that $\gamma N \to N \phi \phi^*$ is the dominant production process for $T \sim \kev$. 

The $\phi$ creation rate per volume as a function of distance $R$ from the sun's center is
\begin{equation}
r_\phi^\mathrm{create}(R) = 2 c n_\gamma \sum_{N = \mathrm{p}, \mathrm{He}} n_N \tilde \sigma_{N \gamma \to N \phi \phi^*}
\label{e.rphicreatesun}
\end{equation}
where $n_\gamma$ and $\tilde \sigma$ are evaluated at temperature $T_\mathrm{sun}(R)$, and $\tilde \sigma$ is the thermally averaged cross section for a Maxwell-Bolzmann distribution (excellent approximation of Bose-Einstein in the sun) of photons hitting a stationary nucleus:
\begin{equation}
\tilde \sigma_{N \gamma \to N \phi \phi^*}(T) = \int_0^\infty d E_\phi f_\mathrm{MB}(T; E_\phi) \sigma_{N \gamma \to N \phi \phi^*} (E_\phi).
\end{equation}
The resulting $r_\phi^\mathrm{create}(R) \propto \Lambda^{-2}$ is shown in \fref{SUNrphicreateLambda10tev}. About $90\%$ of $\phi$ production takes place within $0.2$ solar radii, validating assumption (b) above. The \emph{total} rate of $\phi$ creation in the entire sun is
\begin{equation}
\mathcal{R}_\phi^\mathrm{create} \approx (1.0 \times 10^{42} \ \mathrm{s}^{-1}) \left(\frac{\tev}{\Lambda}\right)^2.
\end{equation}
For our purposes here, define $R_\mathrm{core} = 0.2 R_\mathrm{star}$. Since most of the $\phi$ creation takes place within that radius, 
\begin{equation}
\mathcal{R}_\phi^\mathrm{create} \sim r_\phi^\mathrm{create}(0) \ \times \  \frac{4}{3} \pi R_\mathrm{core}^3
\end{equation}
is satisfied up to a factor of two.

\begin{figure}
\begin{center}
\includegraphics[width=7cm]{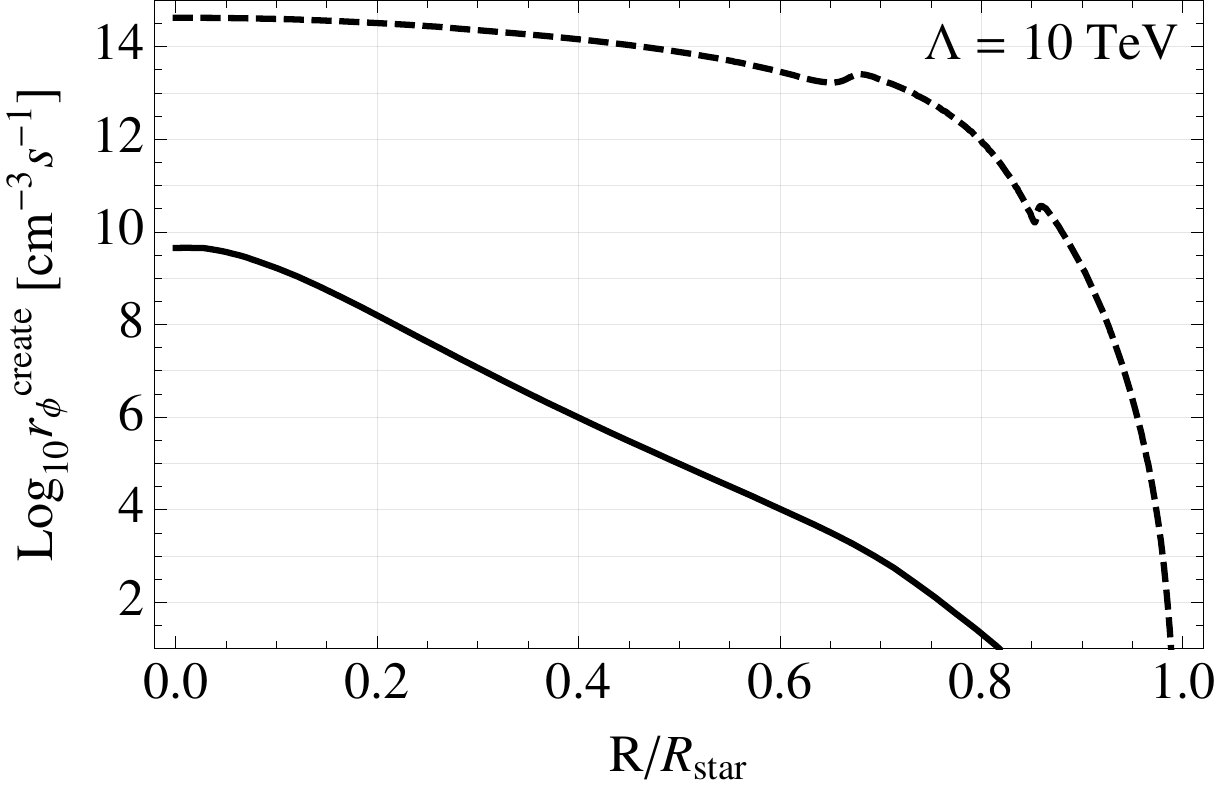}
\end{center}
\caption{
Rate of $\phi$ creation in the sun (solid) and our benchmark white dwarf with 0.1 solar luminosity  (dashed) for $\Lambda = 10 \tev$.
}
\label{f.SUNrphicreateLambda10tev}
\end{figure}

\begin{figure}
\begin{center}
\includegraphics[width=7cm]{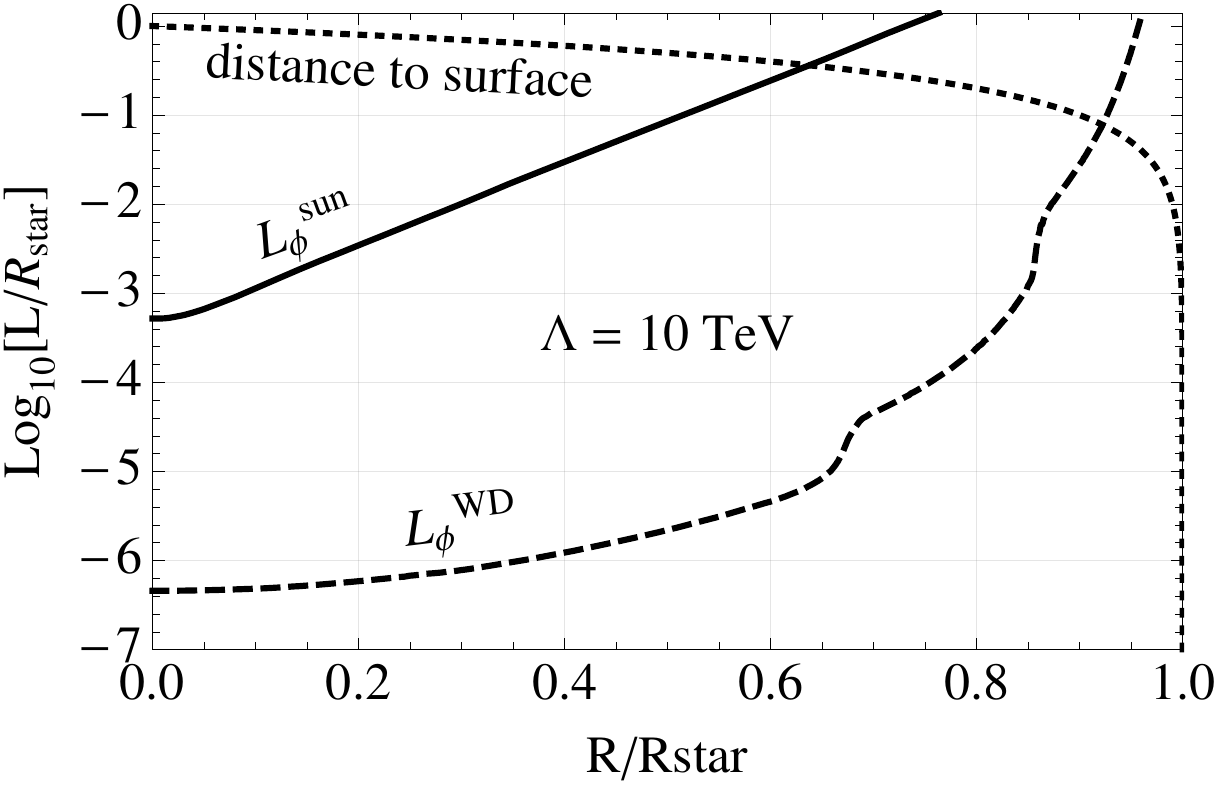}
\end{center}
\caption{
Solid (dashed) line: mean free path of $\phi$ in the sun (our benchmark white dwarf with 0.1 solar luminosity) for $\Lambda = 10 \tev$. The intersection with the dotted line marks the `layer of last scattering'.
}
\label{f.SUNLphiLambda10tev}
\end{figure}

We next compute the $\phi$ mean free path $L_\phi(R) \propto \Lambda^2$ via \eref{Lphi} using similarly averaged scattering cross sections. This is shown in \fref{SUNLphiLambda10tev} for $\Lambda = 10 \tev$. For this benchmark value assumption (c) is certainly satisfied. The `layer of last scattering' is  situated at $R \approx R_\phi^\mathrm{escape}$, where
\begin{equation}
L_\phi(R_\phi^\mathrm{escape}) = R_\mathrm{star} - R_\phi^\mathrm{escape}.
\end{equation}
This allows us to define the approximate temperature of the escaping $\phi$'s as
\begin{equation}
T_\phi^\mathrm{escape} = T_\mathrm{sun}(R_\phi^\mathrm{escape}).
\end{equation}
Both $R_\phi^\mathrm{escape}$ and $T_\phi^\mathrm{escape}$ are shown as functions of $\Lambda$ in \fref{SUNRescapeTescapetev}. Assumption (c) holds for $\Lambda \lesssim 100 \tev$. On the other hand, our calculations become unreliable around $\Lambda \sim 1 \tev$ since we then become sensitive to details of the sun's surface structure.

\begin{figure}
\begin{center}
\includegraphics[width=7cm]{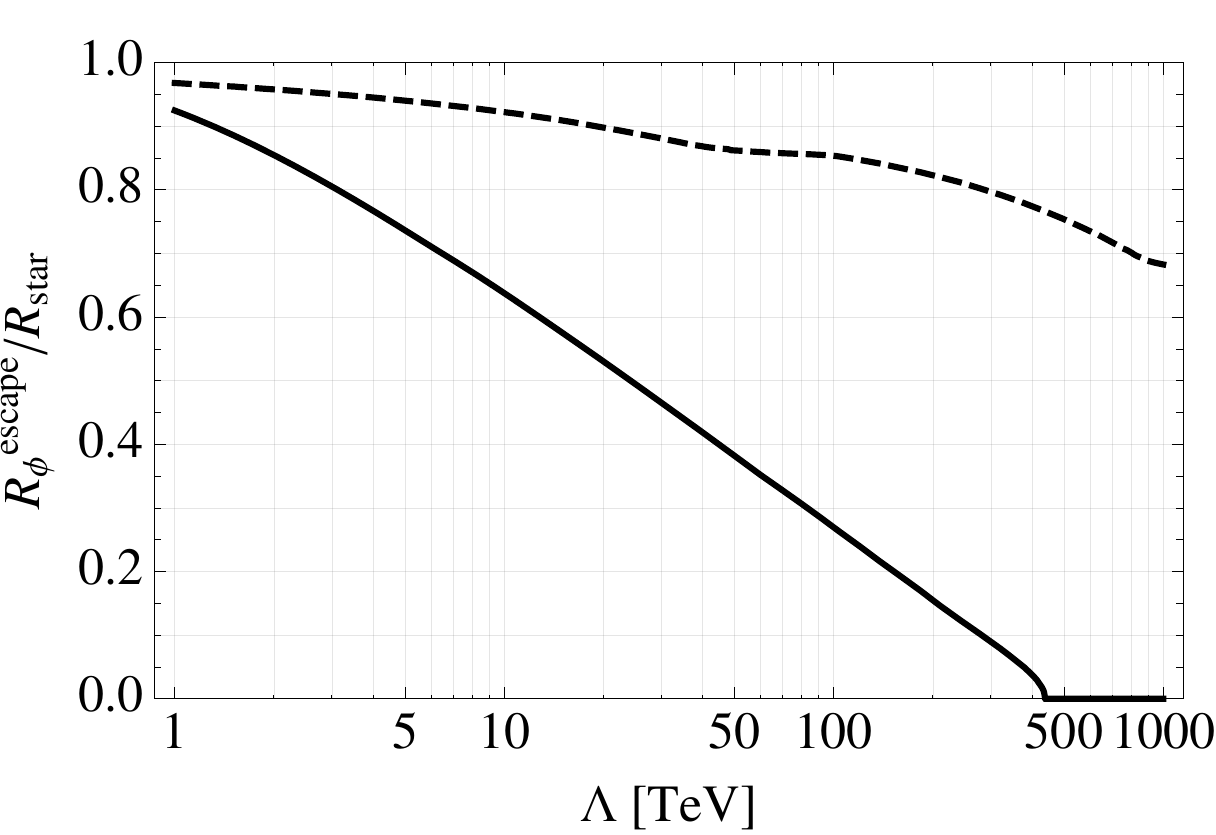}
\includegraphics[width=7cm]{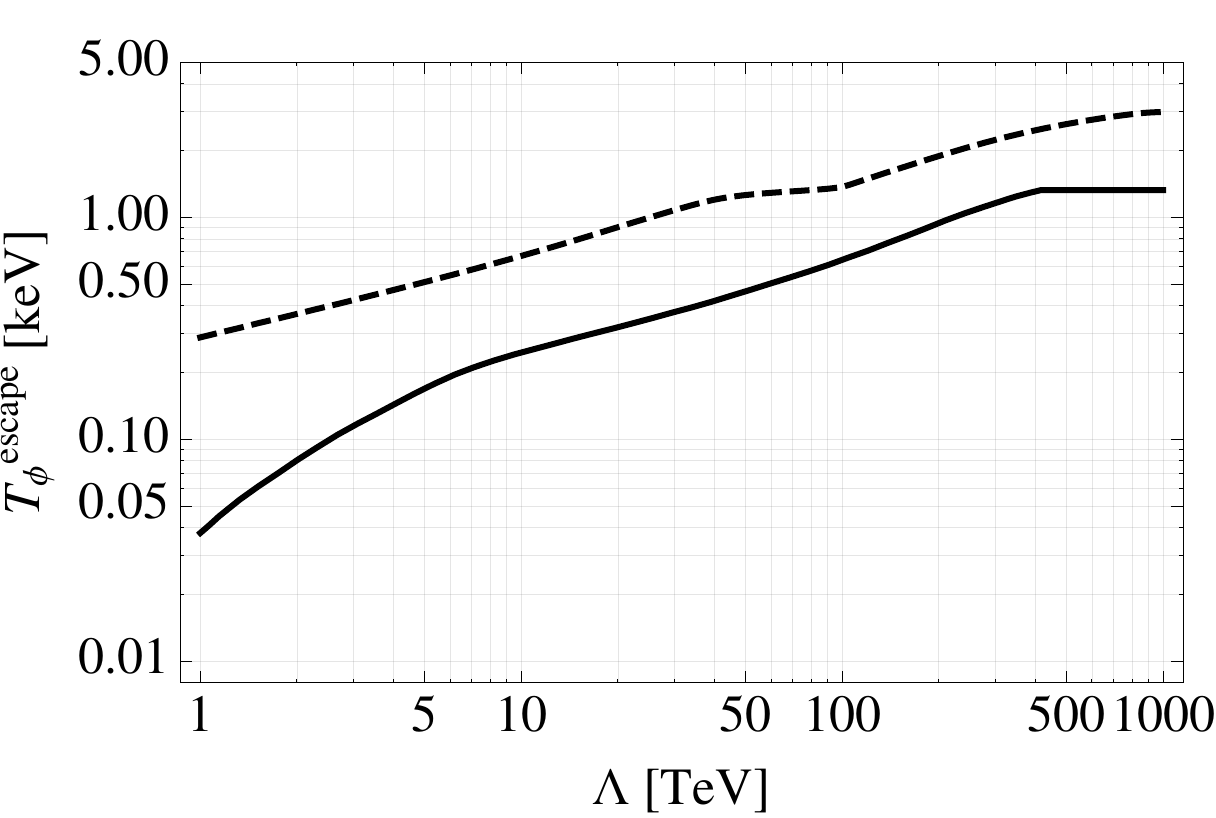}
\end{center}
\caption{
Top: $R_\phi^\mathrm{escape}$ defining the `layer of last scattering' of $\phi$'s in the Sun (solid) and our benchmark white dwarf  with 0.1 solar luminosity (dashed). The many-scattering assumption is valid in the Sun for $\Lambda \lesssim 100 \tev$. Bottom: $T_\phi^\mathrm{ecape} = T_\mathrm{star}(R_\phi^\mathrm{escape})$, the temperature of escaping $\phi$'s. (Valid in the Sun for $\Lambda \lesssim 100 \tev$).
}
\label{f.SUNRescapeTescapetev}
\end{figure}

We can now estimate the fraction of the star's luminosity in the form of $\phi$ emission, making use of the (yet to be verified) assumption (d), which gives at equilibrium:
\begin{equation}
\mathcal{R}_\phi^\mathrm{escape} = \mathcal{R}_\phi^\mathrm{create}.
\end{equation}
Therefore, the power of $\phi$ emission is
\begin{equation}
\label{e.Pphisun}
P_\phi \approx \frac{3}{2} T_\phi^\mathrm{escape} \  \mathcal{R}_\phi^\mathrm{create}.
\end{equation}
The sun's measured power output is $P_\mathrm{sun} \approx 3.85 \times 10^{26}$ Watts. The ratio $P_\phi/P_\mathrm{sun}$ as a function of $\Lambda$ is shown in \fref{SUNLambdaLimit}. The $\phi$ contribution becomes negligible\footnote{Compare to neutrino emission $P_\nu/P_\mathrm{sun} \approx 2\%$. \cite{raffelt1996}} for 
\begin{equation}
\Lambda \gtrsim 3 \tev.
\end{equation}
However, as we will see below, this does not constitute the strongest bound obtained from the sun.

We still need to verify that assumption (d) holds. Evaluating the rate of $\phi$ annihilation in the sun requires us to solve for the equilibrium number density $n_\phi(R)$. We can construct the associated diffusion equation with the information assembled here, but numerically solving it is beyond the scope of this work. However, we can make a ball-park estimate of the total equilibrium $\phi$ population by noting that the time taken for a single $\phi$ to escape is dominated by the time taken to diffuse from the dense core:
\begin{equation}
t_\phi^\mathrm{escape} \sim \frac{R_\mathrm{core}^2}{c L_\phi(0)} \approx  (2 \times 10^4 \mathrm{\ s})\left(\frac{\tev}{\Lambda}\right)^2
\end{equation}
where $L_\phi(0) \approx 5 \times 10^{-6} R_\mathrm{star} (\Lambda/\tev)^2$ (see \fref{SUNLphiLambda10tev}). This means $N_\phi$, the equilibrium total number of $\phi$'s in the sun, is approximately given by solving
\begin{equation}
\label{e.solardNphi}
\frac{dN_\phi}{dt} = \mathcal{R}^\mathrm{create}_\phi - \frac{N_\phi}{t_\phi^\mathrm{escape}} = 0,
\end{equation}
which gives
\begin{equation}
N_\phi \sim (2 \times 10^{46}) \ \left(\frac{\tev}{\Lambda}\right)^4
\end{equation}
Assuming all these $\phi$'s live in the core, the corresponding number denisty is
\begin{equation}
\label{e.nphiequilibriumsun}
 n_\phi \sim (2 \times 10^{15} \ \cmmthree) \ \left(\frac{\tev}{\Lambda}\right)^4
 \end{equation}
Consulting \fref{crosssectionphilowE} and comparing with number densities $n_\gamma \sim 10^{22} \ \cmmthree$ and $n_N \sim 10^{25} \ \cmmthree$ in the core, it is clear that the $\phi$ annihilation rate
\begin{equation}
r_\phi^\mathrm{annihilation} = 2 c n_\phi^2 \sigma_{\phi \phi \to \gamma \gamma}
\end{equation}
is completely negligible compared to the creation rate in \eref{rphicreatesun}.

\begin{figure}
\begin{center}
\includegraphics[width=7cm]{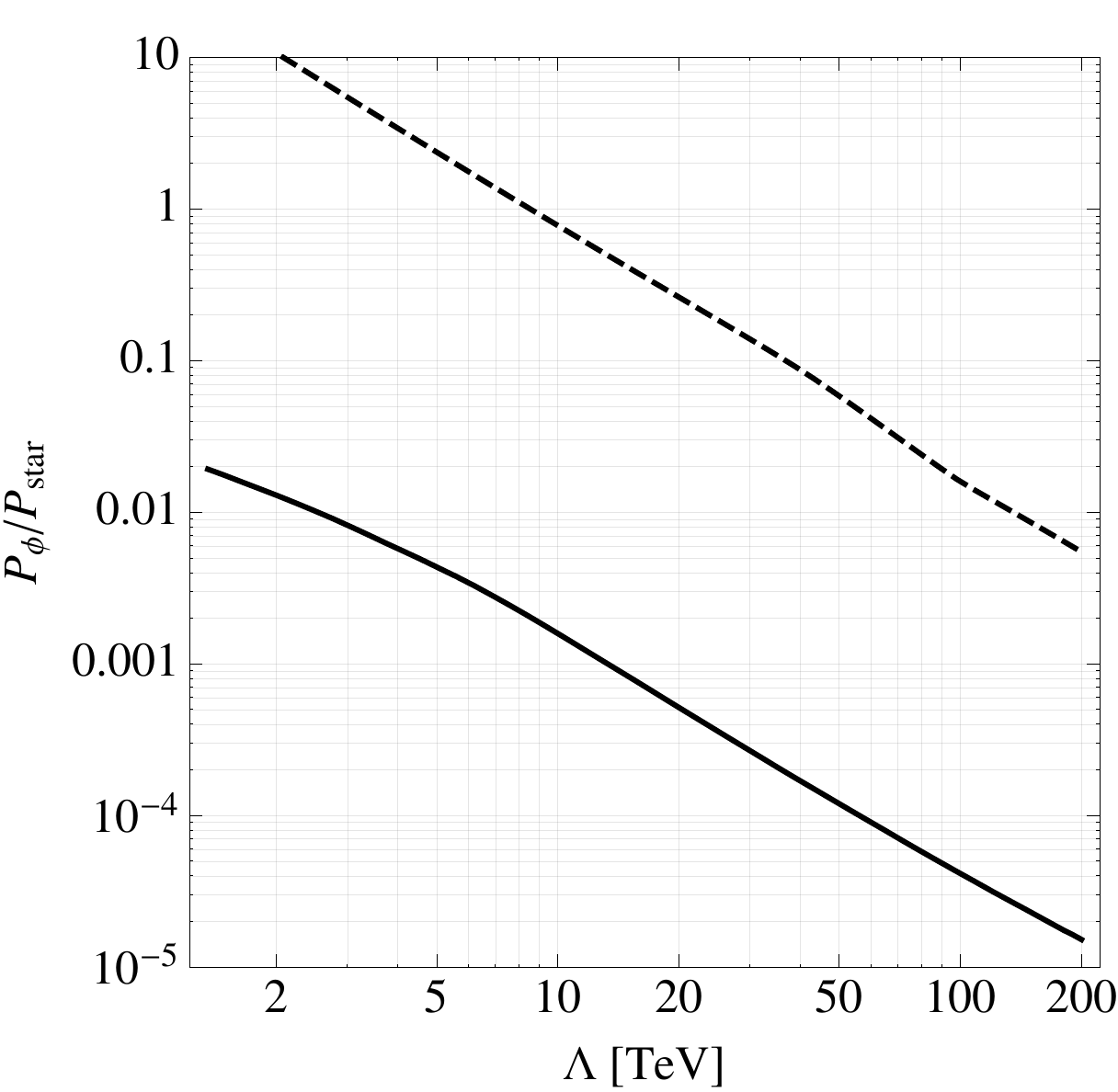}
\end{center}
\caption{
Power of emitted $\phi$ radiation as a fraction of total star power output for the Sun (solid) and our benchmark white dwarf  with 0.1 solar luminosity (dashed). 
}
\label{f.SUNLambdaLimit}
\end{figure}

To make sure the sun is not disturbed by $\phi$ production we also have to ensure that radiative heat transfer, which dominates the core and radiative zone, is relatively unaffected. The radiative heat transfer due to $\phi$ should be compared to the photon heat flux \cite{Raffelt:1988rx}:
\begin{equation}
\frac{F_\phi}{F_\gamma} \sim \frac{n_\phi L_\phi}{n_\gamma L_\gamma}.
\end{equation}
Substituting \eref{nphiequilibriumsun}, $n_\gamma(T_\mathrm{sun}(0))$, $L_\phi(0)$ as well as $L_\gamma \sim 10^{-2}$ cm (see \fref{solarprofiles} in \aref{radialprofiles}), we obtain the following heat transfer ratio in the core:
\begin{equation}
\frac{F_\phi}{F_\gamma} \sim 1 \times \left(\frac{\tev}{\Lambda}\right)^2.
\end{equation}
While this estimate is crude it does yield an important constraint,
\begin{equation}
\Lambda \gtrsim 10 \tev,
\end{equation}
which is significantly stronger than the bound from energy loss due to $\phi$ emission. 

Two final remarks are in order. Firstly, as mentioned above, \fref{SUNRescapeTescapetev} shows that assumption (c) of trapped and thermalized $\phi$'s starts breaking down when $\Lambda \gtrsim 100 \tev$. In that case $\phi$ no longer contributes to radiative heat transfer, while the lost power due to $\phi$ emission is roughly $P_\phi \sim T_\mathrm{core} \mathcal{R}_\phi^\mathrm{create} \sim (10^{-5} P_\mathrm{sun}) (100 \tev / \Lambda)^2$. The free-streaming regime in the sun therefore sets no constraints on $\Lambda$.
Secondly, we also point out that there is a sub-population of protons and photons with $E \sim 10 \mev$ produced by fusion reactions in the sun, but the total rate of fusion reactions $R_\mathrm{fusion} \approx 3.6 \times 10^{38} \ \invsecond$ is many orders of magnitude too low for this subpopulation to affect our estimates.

\subsection{White Dwarf Cooling}
\label{sss.whitedwarfcooling}

White dwarfs (WD) represent the evolutionary endpoint of stars up to several solar masses. They are supported by electron degeneracy pressure, which largely decouples their hydrostatic and thermal properties and results in a strong relationship between their mass and radius. Since white dwarfs do not support fusion processes in their cores they simply cool down after they are formed, with observable luminosities ranging from 0.5 to $\sim 10^{-4} \mathcal{L}_\mathrm{sun}$, corresponding to core temperatures of around $10$ to $0.1 \kev$ (about $10^8$ and $10^6$ K) \cite{raffelt1996}. 

Their relative simplicity makes white dwarfs suitable for constraining new physics with light particles  (see e.g. \cite{Dreiner:2013tja, raffelt1996}). Unlike the Sun, where we have a single well-studied star to compare predictions to, white dwarf cooling is constrained by the \emph{White Dwarf Luminosity Function} (WDLF), which is the number of observed WDs at different luminosities, see \fref{WDLF}. For reasonable assumptions about the star formation rate, the shape of this WDLF curve is given entirely by the WD cooling rate \cite{raffelt1996}. 

The large central density of white dwarfs $\rho_{WD} \sim 10^{6} \ \mathrm{g} \ \cmmthree$ means $\phi$'s can be copiously produced, but also thermalize completely before diffusing out of the star. This makes their emission somewhat sensitive to details of WD stellar structure, unlike e.g. free-streaming axions. Comprehensively studying the constraints on the $\frac{1}{\Lambda} q \bar q \phi \phi^*$ operator set by WD cooling would therefore require modeling a representative WD population, which is beyond the scope of this work. 

Fortunately, the WD population in our stellar neighborhood is strongly peaked around $0.5 - 0.7$ solar masses \cite{raffelt1996, DeGennaro:2007yw}. This means we can obtain a preliminary estimate of the bound on $\Lambda$ by studying a single star in this representative mass range.

Our benchmark dwarf (about 0.5 solar masses) started its life as a roughly one solar mass main sequence star that was evolved forward in time using the stellar evolution code \texttt{MESA} \cite{mesa}.\footnote{We thank Max Katz, who performed the simulation for us.} Most of the observational data in the WDLF is for luminosities $\lesssim 0.1 \mathcal{L}_\mathrm{sun}$, which corresponds to a bolometric magnitude $M_\mathrm{bol}> 7$. Photon cooling, well-described by Mestel's Law \cite{Mestel}, dominates for such cool white dwarfs. We therefore compute $\phi$-cooling in our benchmark dwarf for $\mathcal{L}_\mathrm{WD} < 0.1 \mathcal{L}_\mathrm{sun}$. 

Since the degenerate electron gas in WD cores is an excellent conductor of heat, radiative heat transfer is unimportant. We therefore only compute the total power loss due to $\phi$ emission, in an identical manner to the previous subsection. As we will see, assumptions (a) - (d) are satisfied throughout as long as $\Lambda$ is large enough. Radial profiles of density, composition and temperature produced by \texttt{MESA} for our benchmark dwarf are shown in \aref{radialprofiles}.

The $\phi$ creation rate per volume is shown in \fref{SUNrphicreateLambda10tev} for $\Lambda = 10 \tev$ when the white dwarf has 0.1 solar luminosity. Due to the similar temperature but larger density, it is 5 orders of magnitude higher than in the sun. The $p p \to p p \phi \phi$ process is still strongly temperature-suppressed. Figs. \ref{f.SUNLphiLambda10tev} and \ref{f.SUNRescapeTescapetev} show that $\phi$'s do not escape until they are very close to the white dwarf surface. The resulting power loss is shown in \fref{SUNLambdaLimit}, and $\phi$ emissivities are compared to known photon and neutrino emissivities in \fref{WDcoolingluminosities}. To a reasonable approximation, 
\begin{equation}
\label{e.WDephi}
\epsilon_\phi \approx 1.5 \times 10^{-2} \left( \frac{\tev}{\Lambda}\right)^2 \left( \frac{T}{10^7 K}\right)^{11/5}  \mathrm{erg} \  \mathrm{s}^{-1} \mathrm{g}^{-1}.
\end{equation} 

Requiring  $\phi$ emission to represent a subdominant 10\% fraction of the total WD luminosity requires $\Lambda \gtrsim 40 \tev$, but as it turns out the actual bound on $\Lambda$ from the white dwarf luminosity function is significantly less constraining. We now compute this bound following the procedure in \cite{raffelt1996}. 

The white dwarf looses internal energy $U$ with time due to emission of photons, neutrinos and (in our case) $\phi$'s, so that $dU/dt = - (L_\gamma + L_\nu + L_\phi)$, where $L_\gamma$ is the total photon luminosity of the star. 
Assuming a constant star formation rate $B$, the number density of white dwarfs in a given magnitude interval is proportional to the time it takes to cool through that interval, so
\begin{equation}
\frac{d N}{d M_\mathrm{bol}} = B \frac{dt}{d M_\mathrm{bol}} = - B \frac{dU/d M_\mathrm{bol}}{L_\gamma + L_\nu + L_\phi}.
\end{equation}
 For a white dwarf, the photon emissivity can be be computed using Kramer's opacity law:
\begin{equation}
\epsilon_\gamma \approx 3.3 \times 10^{-3} \left( \frac{T}{10^7 K}\right)^{7/2} \mathrm{erg} \  \mathrm{s}^{-1} \mathrm{g}^{-1}.
\end{equation}
Bolometric magnitude gives the photon luminosity relative to the sun, $\log_{10}(L_\gamma/L_\mathrm{sun}) = (4.74 - M_\mathrm{bol})/2.5$. Therefore $T \propto 10^{-4 M_\mathrm{bol}/35}$ and we get
\begin{equation}
\label{e.WDLF}
\log_{10} \left[ \frac{d N}{d M_\mathrm{bol}}\right]  = \mathrm{C} + \frac{2}{7} M_\mathrm{bol} + \log_{10}\left[ \frac{\epsilon_\gamma}{\epsilon_\gamma + \epsilon_\nu + \epsilon_\phi}\right].
\end{equation}
where we have absorbed details of the star formation rate and white dwarf heat capacity into the constant $C$. (When comparing to observational data it is conventional to normalize this constant to the data point with the smallest uncertainty.) For pure photon cooling, this reduces to the well-known Mestel's cooling law~\cite{Mestel}, shown as the black dashed line in \fref{WDLF}. This already gives a reasonable fit, but full simulations (thin green curve) are needed to account for the observed data in detail.

\begin{figure}
\begin{center}
\includegraphics[width=8cm]{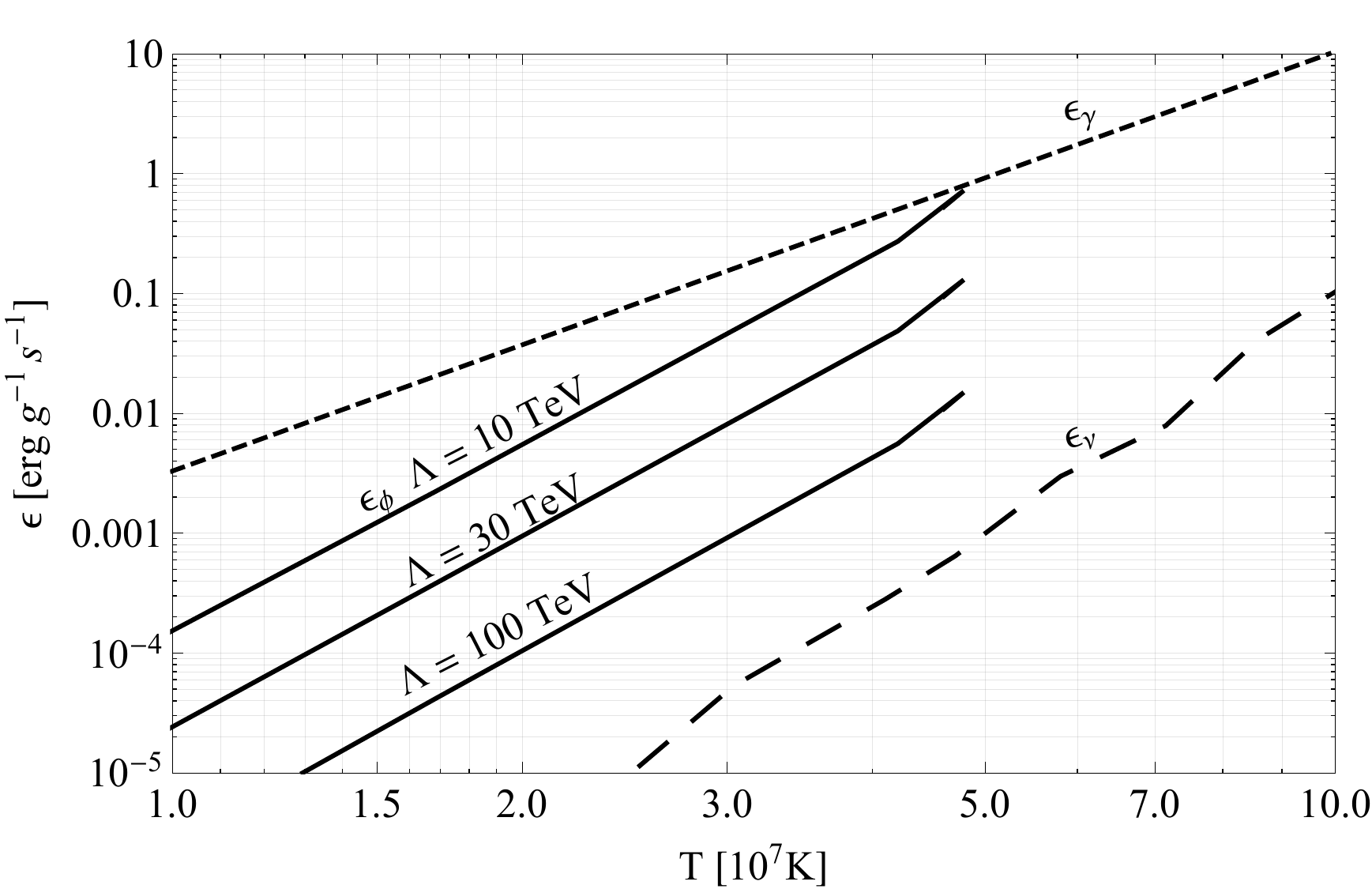}
\end{center}
\caption{
Photon (short-dashed) and neutrino (long-dashed) emissivities for typical white dwarfs with $\rho_\mathrm{core} = 10^6 \ \gram \ \cmmthree$ as a function of core temperature \cite{raffelt1996,Mestel}. Solid lines indicate $\phi$ emissivities obtained for our benchmark dwarf. 
}
\label{f.WDcoolingluminosities}
\end{figure}

\begin{figure}
\begin{center}
\includegraphics[width=7.5cm]{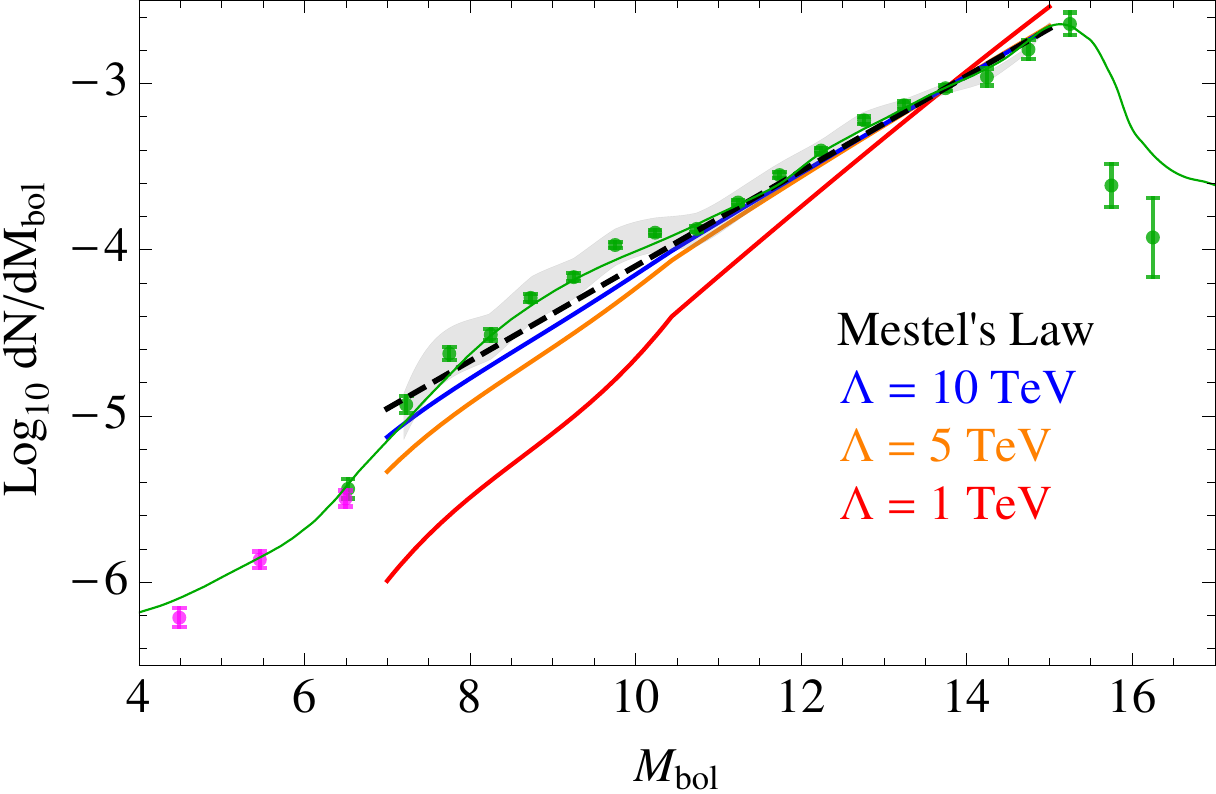}
\end{center}
\caption{
Green and magenta datapoints: measured white dwarf luminosity function from \cite{Harris:2005gd} and \cite{Krzesinski}, showing the number density of observed white dwarfs per bolometric magnitude interval per $\mathrm{pc}^{3}$. The gray band indicates how much the WDLF of \cite{Harris:2005gd} would change by varying the assumed scale height of the galactic disk between 200 and 350pc. 
Green line: WDLF from a full simulation, assuming constant star formation rate, taken from \cite{Dreiner:2013tja}. Black dashed line: Mestel's cooling law (pure photon cooling). Colored solid lines: modification of Mestel's law due to additional $\phi$ cooling for $\Lambda = 1, 5$ and $10 \tev$. All cooling curves have been shifted to pass through the datapoint with the smallest uncertainty.
}
\label{f.WDLF}
\end{figure}

For the range of core temperature we consider, neutrino cooling can be neglected. Using the $\phi$ emissivities of our simulated benchmark dwarf in \eref{WDLF}, we obtain the modifications to Mestel's cooling law shown in \fref{WDLF} for different values of $\Lambda$. Given the crudeness of our estimate a full fit to the data is not appropriate. However, we can estimate the \emph{sensitivity} of a full stellar simulation to $\phi$ cooling by the size of the deviation from Mestel's law. 
Given the scale of astrophysical uncertainty in the WDLF (illustrated by the gray band in \fref{WDLF}), a reasonable rough bound on the allowed modification to standard white dwarf cooling is
\begin{equation}
\Lambda \gtrsim 10 \tev.
\end{equation}

The effect of $\phi$ cooling is more pronounced for young, hot white dwarfs (smaller $M_\mathrm{bol}$). 
A more thorough study, including full simulation of $\phi$ cooling throughout the life of the white dwarf, might therefore give a somewhat more stringent bound on $\Lambda$. However, as we see in the next section, a much stronger constraint is supplied by neutron star cooling. 

Finally, one might worry about $\phi$ being produced in electron collisions or plasmon decay via its loop-induced coupling to the $Z$-boson, see \eref{zphiphi}. However, this coupling is $\sim10^{-3} (10\tev/\Lambda)$ smaller than the equivalent tree-level electroweak coupling. According to the discussion in \cite{Dreiner:2013tja},
$\phi$ emission from the plasmon decay and electron Bremsstrahlung is therefore  $\lsim 10^{-8}(10\tev/\Lambda)^2\mathrm{erg} \ \mathrm{s}^{-1} \mathrm{g}^{-1}$ at $T \sim 4 \times 10^7$ K, which is much lower than the nuclear production discussed above.

\subsection{Neutron Star Cooling}
\label{sss.neutronstarcooling}
Neutron stars are the evolutionary endpoint for heavy stars that do not collapse to a black hole. They are supported by neutron degeneracy pressure and constitute the densest form of matter in the universe. This introduces many subtleties into their cooling processes, which are not yet fully understood even in the Standard Model (see e.g. \cite{Yakovlev:2004iq, 
Yakovlev:2007vs,
Page:2005fq,
Page:2009fu,
Yakovlev:2010ed,
Page:2010aw,
Shternin:2010qi,
Potekhin:2011xe,
Page:2012se,
Elshamouty:2013nfa}). However, neutron stars are such powerful ``$\phi$-factories'' in dmDM that we can still set very strong constraints despite these uncertainties.

Neutron stars are born in hot supernovae explosions with $T \sim 10^{11}\mathrm{K} \simeq 10 \mev$ but quickly cool down and enter the neutrino cooling phase when their internal temperature reaches about $T \sim 10^9 \mathrm{K} \sim 100$ keV (see e.g. \cite{Yakovlev:2004iq} for a review). Neutrino cooling dominates for $\sim 10^5$ years, after which photon cooling takes over. For a given equation of state, the mass of the neutron star fixes both the radius and density profile. The radius is about 10km, while the central density is $\rho \sim 2 - 10 \times \rho_0$, where $\rho_0 \approx 2.8 \times 10^{14} \ \gram \ \cmmthree$ is the density of nuclear matter at saturation. 

The neutron star core extends to about 1km below the surface and is divided into an inner core ($\rho \gtrsim 2 \rho_0$) and an outer core ($\rho \lesssim 2 \rho_0$). (Light neutron stars with $M \lesssim 1.3 M_\mathrm{sun}$ do not have an inner core.) The characteristics of the outer core are well constrained by nuclear  theory and laboratory data, while the inner core is much less well understood, with hypotheses for its composition ranging from normal nuclear matter to hyperions, pion or kaon condensates, or a pure quark fluid (called `strange quark matter' due to the presence of $s$ quarks). However, the recent observation of a 2  solar mass neutron star \cite{Demorest:2010bx} is in conflict with all core hypotheses other than normal nuclear matter, which provides the only equation of state `stiff' enough to support such large masses. We shall therefore only consider neutron stars with nucleon cores.

The neutron star is surrounded by an outer crust of thickness $\sim$ few 100 m, consisting of a \emph{non-degenerate} neutron gas with characteristic density of order $\rho_N \sim 4 \times 10^{11} \ \mathrm{g} \ \cmmthree$. During the neutrino cooling phase the outer crust acts as a heat blanket, thermally insulating the neutron star interior against radiative losses into space. For a nonmagnetic iron envelope the surface temperature of the star can be related to the interior temperature by \cite{Gundmundsson1,Gundmundsson2,Page:2004fy}
\begin{equation}
T_\mathrm{surface} = (0.87 \times 10^{6}\mathrm{K})  \left(\frac{g_s}{10^{14} \mathrm{cm/s}^2}\right)^{1/4} \left( \frac{T_\mathrm{core}}{10^8 \mathrm{K}}\right)^{0.55},
\label{e.Tsurface}
\end{equation}
where $g_s = G M / R_\mathrm{star}^2$ is the surface gravity.
The inner crust has a thickness of $\sim$ 1km and forms the transition between the heat blanket and the core. The thermal conductivity of nuclear matter is so high that the interior  below the blanket is close to isothermal. 

Late-time cooling is constrained by $\sim 20$ observations of neutron stars for which both surface temperature and age could be determined, see green data points in \fref{neutronstarcoolingcurves}. The mass of an individual neutron star, which is not known in the dataset, determines the cooling curve $T_\mathrm{surface}(M; t)$. Different cooling models can be excluded by the requirement that the observed data points fall into the range of allowed cooling curves. For non-superconducting neutron stars with non-magnetic iron heat blankets in the Standard Model this range is illustrated with the two blue dashed lines in \fref{neutronstarcoolingcurves} \cite{Yakovlev:2007vs}. Accretion of light elements in the crust and the presence of strong magnetic fields at the surface would increase the thermal conductivity of the outer neutron star layers, increasing $T_\mathrm{surface}$ by a factor of a few during the neutrino cooling phase. Furthermore, the core may be in different phases of neutron and/or proton superfluidity, which can affect the surface temperature by at least a similar factor. Nevertheless, quite stringent constraints on $\phi$-cooling can be obtained from light, slow-cooling neutron stars.

It is necessary to understand how the standard range of allowed cooling curves changes when $\phi$ emission is included. We will therefore estimate first the emissivity $\epsilon_\phi(T_\mathrm{core})$ and then the cooling curves $T_\mathrm{surface}(t)$ for a light, slow-cooling neutron star and a heavy, fast-cooling neutron star, which bounds the range of allowed cooling curves. The relevant parameters of our benchmark stars are given in \tref{NSbenchmarks}.

\begin{table}
\begin{center}
\begin{tabular}{| c | c | c | c | }
\hline
$M/M_\mathrm{sun}$ & $R$ (km) & $R_\mathrm{core}$ (km) & $\rho_\mathrm{core}/\rho_0$\\
\hline 
1.1 & 13 & 11 & 2\\
2.0 & 11 & 10 & 10\\ \hline
\end{tabular}
\end{center}
\caption{
Parameters of light and heavy neutron stars to determine the range of allowed cooling curves in dmDM. $\rho_\mathrm{core}$ is the central density. Adapted from~\cite{Yakovlev:2004iq}, which assumed nucleon cores.
}
\label{t.NSbenchmarks}
\end{table}

We model the neutron star core as a sphere of constant temperature and density. Assume for the moment that the annihilation process $\phi \phi^* \to \gamma \gamma$ can be ignored, and that $\phi$ is free-streaming in both the core and the crust. In that case, the $\phi$ emissivity is given simply by
\begin{equation}
\epsilon_\phi \sim r_\phi^\mathrm{create} T_\mathrm{core}.
\end{equation}
$\phi$ creation proceeds via the process $n n \to n n \phi \phi^*$.  Here we have to take into account Pauli-blocking: since the neutrons in the core are strongly degenerate, only the subpopulation living on the Fermi surface can participate in reactions, and furthermore the phase space of reactions is suppressed since neutrons cannot scatter into the occupied Fermi volume. The neutron Fermi Energy $E_F = \hbar^2 (3 \pi^2 n_n)^{2/3}/(2 m_n)$ is 95 MeV (280 MeV) for $\rho = 2 \rho_0$ $(10 \rho_0)$. The fraction of neutrons on the Fermi surface is roughly $T/E_F$, so we define the number density of `available neutrons' (with kinetic energy $\approx E_F$) as
\begin{equation}
n_{n_F} \sim n_n \frac{T}{E_F}.
\end{equation}
This gives 
\begin{equation}
r_\phi^\mathrm{create} \ \sim \ 2 \ c \ n_{n_F}^2 \  \sigma_{n n \to n n \phi \phi^*}^F.
\end{equation}
$\sigma_{n n \to n n \phi \phi^*}^F$ is much smaller than $\sigma_{n n \to n n \phi \phi^*}$ from \fref{crosssectionphilowE} due to Pauli Blocking: two neutrons with kinetic energy $\sim E_F$ interact softly to produce two $\phi$'s with energy $\lesssim T$ so that their final energy is still on the Fermi surface. We can roughly estimate this phase space suppression $\zeta(E_F, T)$ using MadGraph, shown in \tref{fermixsectable}. This gives
\begin{eqnarray}
\nonumber \sigma_{n n \to n n \phi \phi^*}^F &\sim& \sigma_{n n \to n n \phi \phi^*}(E_F) \times \zeta(E_F, T)
\\
\nonumber&\sim& \sigma_\mathrm{prod}^0 \ \left(\frac{\tev}{\Lambda}\right)^2 \ \left(\frac{T_\mathrm{core}}{\kev}\right)^2
\\
\label{e.NSsigmaprod}
\end{eqnarray}
where $\sigma_\mathrm{prod}^0 = \mathcal{B}_\mathrm{prod} \ \times \  7 \times 10^{-9}  \ \mathrm{pb}$ and $\mathcal{B}_\mathrm{prod} = 0.3 \ - \ 3$ is a parameter we vary to account for the uncertainty of this estimate.  Interestingly the  cross section is constant up to a factor of $\approx 2$ for $E_F$ in the range of 95 to 280 MeV, so we absorb this $\rho_\mathrm{core}$ dependence into the uncertainty. 

\begin{table}
\begin{center}
\begin{tabular}{| c | c | c | c | }
\hline
$\rho_\mathrm{core}/\rho_0$ & $E_F$ (MeV) & $\sigma_{n n \to n n \phi \phi^*} $ (pb) & $\zeta(E_F, T)$\\
\hline
$2$ & 95 	& $60  \left( \frac{\tev}{\Lambda}\right)^2$	& $1 \times 10^{-10} \left(\frac{T}{\kev}\right)^2$ \\
\hline
$10$ & 280	& $600 \left( \frac{\tev}{\Lambda}\right)^2$	& $5 \times 10^{-12} \left(\frac{T}{\kev}\right)^2 $ \\
\hline
\end{tabular}
\end{center}
\caption{
Cross section of $n n \to n n \phi \phi^*$ for two neutrons both with kinetic energy $E_F$, computed in MadGraph in the one-pion exchange approximation \cite{steigman}. The third column gives the phase space suppression when requiring both  final state neutrons to have kinetic energy in the range $(E_F - T, E_F + T)$. All quantities are understood to be $\sim$ estimates.
}
\label{t.fermixsectable}
\end{table}

\begin{figure}
\begin{center}
\includegraphics[width=7.5cm]{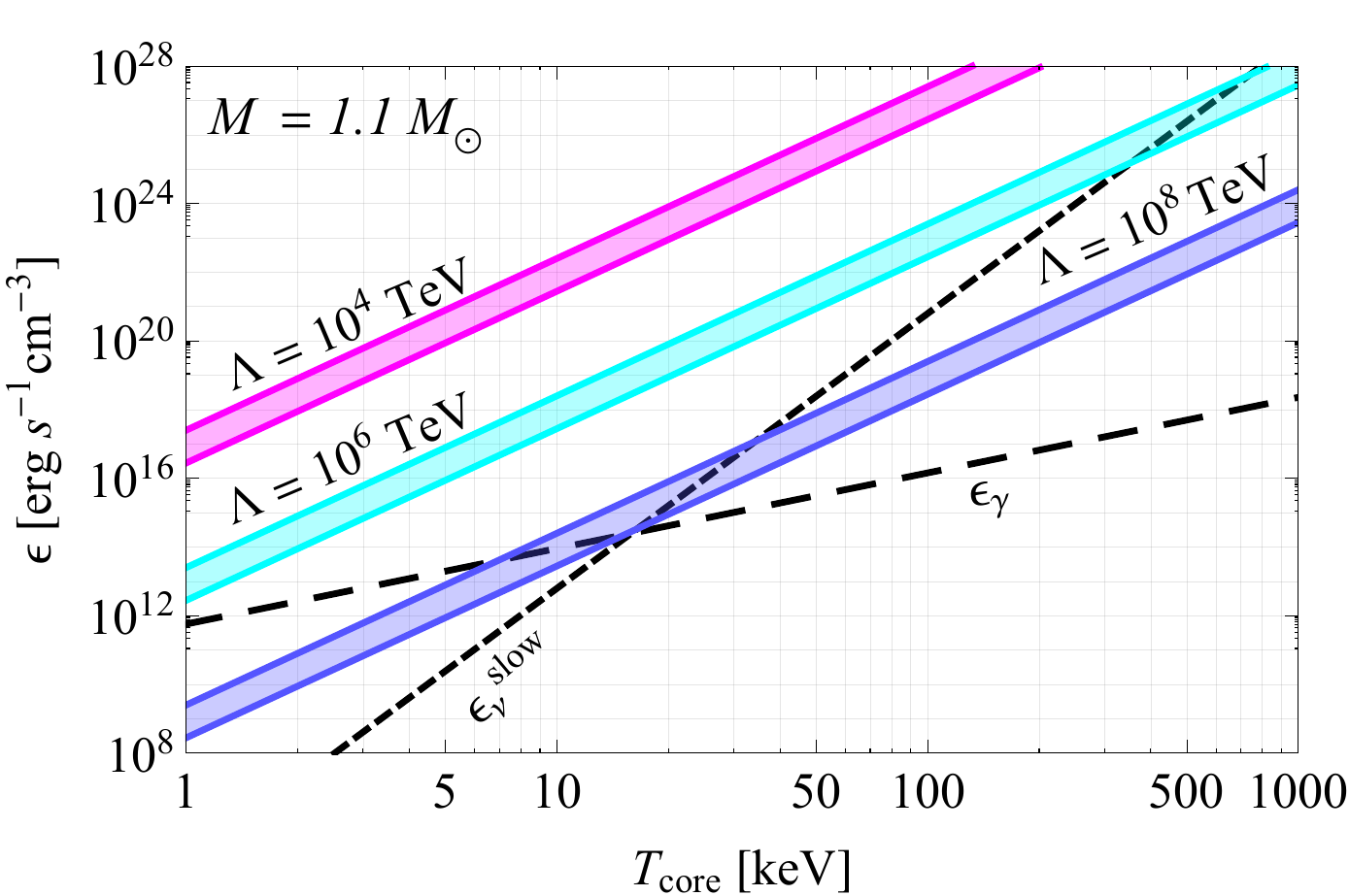}
\end{center}
\caption{
$\phi$ emissivity $\epsilon_\phi(T_\mathrm{core})$ for different $\Lambda$ in the light neutron star defined in \tref{NSbenchmarks}. The allowed range for each $\Lambda$ comes from the production cross section uncertainty in \eref{NSsigmaprod}.
Also shown is slow neutrino emission (dashed) via the modified Urca process, which occurs in all neutron star cores \cite{Shapiro:1983du}, and the effective emissivity from photon emission~\cite{Gundmundsson1,Gundmundsson2,Page:2004fy} (long-dashed).
}
\label{f.neutronstaremissivities}
\end{figure}

Defining $\tilde E_F \approx 95 \mev$ and $n_n^0 \approx 3.3 \times 10^{38} \cmmthree$ to be the Fermi energy and neutron number density when $\rho_\mathrm{core} = 2 \rho_0$, and specifying the actual number density via the dimensionless ratio $\tilde n_n = {n_n}/{n_n^0}$,
we obtain
\begin{eqnarray}
\nonumber 
\epsilon_\phi  &=& 
 \left[ 2 c (n_n^0)^2  \left(\frac{\kev}{\tilde E_F}\right)^2 \right] \sigma_\mathrm{prod}^0 \tilde n_n^{2/3} \left(\frac{\tev}{\Lambda}\right)^2 \frac{T_\mathrm{core}^5}{(\kev)^4} \\
 \label{e.ephiNS}
\end{eqnarray}
This is shown in \fref{neutronstaremissivities} for the light neutron star as a function of core temperature, and compared to the effective emissivity from neutrino and photon emission. Requiring that $\phi$-cooling be subdominant to standard cooling mechanisms in the light neutron star sets the strong constraint $\Lambda \gtrsim 10^8 \tev$. The constraint derived from the heavy neutron star is much weaker, since the powerful direct Urca neutrino emission process is active when the central density is $\rho_\mathrm{core} \gtrsim 2 \rho_0$ \cite{Lattimer:1991ib}.

We have checked that for $\Lambda \gtrsim 10^4 \tev$, the equilibrium $\phi$ density in the neutron star is indeed small enough to render the annihilation process $\phi \phi^* \to \gamma \gamma$ irrelevant. Furthermore, $\phi$ becomes free-streaming in the crust (core) when $\Lambda \gtrsim 5000 \tev$ ($500 \tev$). This validates the assumptions of our estimate, and allows us to circumvent the subtleties of $\phi$-scattering inside the neutron star core and crust (see \cite{Bertoni:2013bsa} for some of the involved issues).

\begin{figure}
\begin{center}
\includegraphics[width=7.5cm]{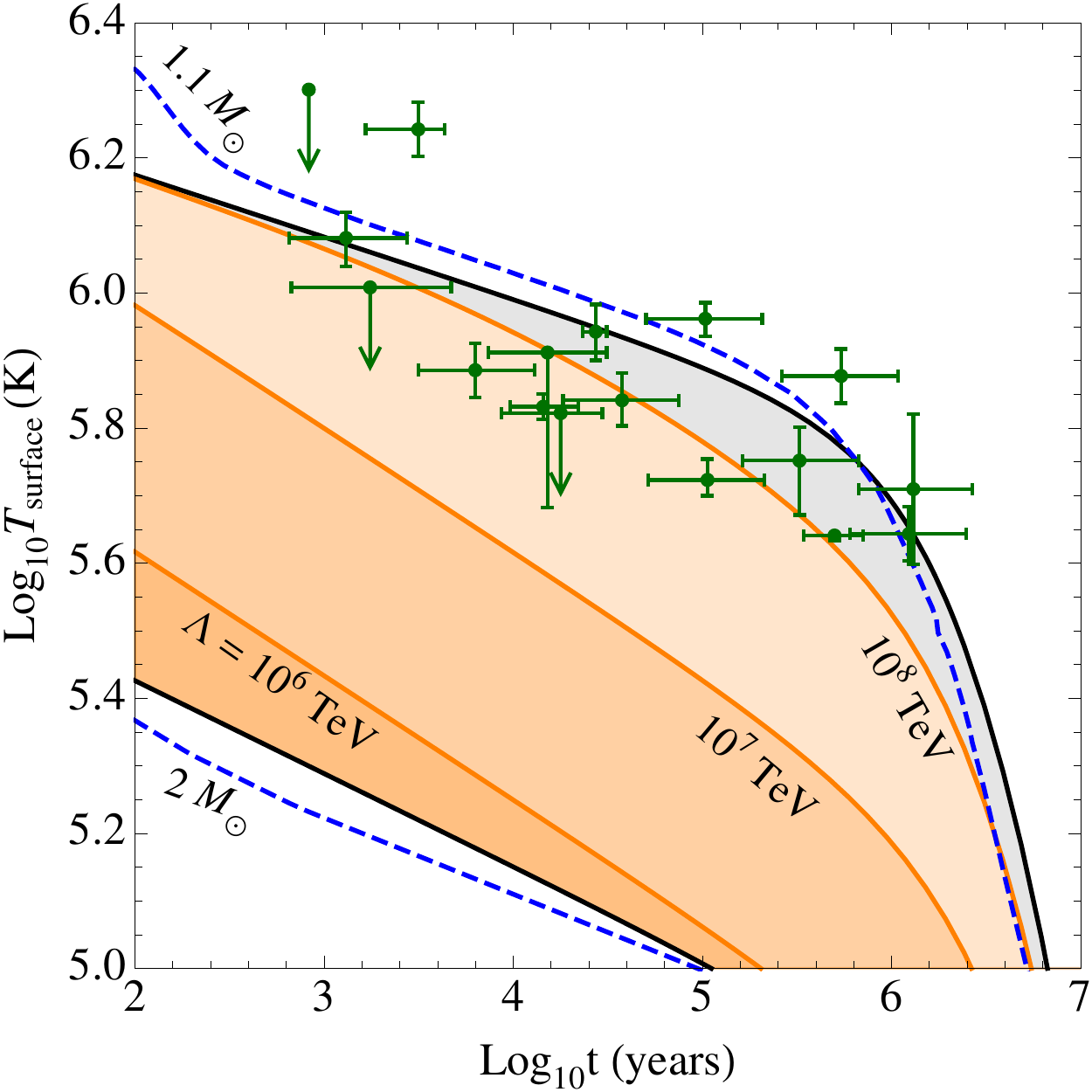}
\end{center}
\caption{
Green data points: surface temperature and age of observed neutron stars \cite{Yakovlev:2007vs}. Blue dotted lines: cooling curves for heavy and light non-superconducting neutron stars with non-magnetic iron heat blankets \cite{Yakovlev:2007vs}. Solid black lines: our corresponding estimate of these cooling curves using \eref{ephiNS} and \eref{coolingcurveNS}. Orange contours: estimate of the light neutron star cooling curve with $\phi$ emission for $\Lambda = 10^{6}, 10^{7}, 10^{8} \tev$ and $\mathcal{B}_\mathrm{prod} = 0.3$. (For $\Lambda >10^6 \tev$, the cooling curve for the heavy NS does not change perceptibly.) In all our estimates we multiplied $T_\mathrm{surface}(t)$ by 0.6 (0.2 units on vertical axis) to bring them into better agreement with the full calculation by \cite{Yakovlev:2007vs}.
}
\label{f.neutronstarcoolingcurves}
\end{figure}

We can explicitly demonstrate the effect of $\phi$ emission on neutron star cooling. Following \cite{Kouvaris:2007ay}, a reasonable estimate of the cooling curve can be obtained by solving the differential equation
\begin{equation}
\label{e.coolingcurveNS}
\frac{d T_\mathrm{core}}{d t} = - \frac{\epsilon_\nu + \epsilon_\gamma + \epsilon_\phi}{c_V},
\end{equation}
where the specific heat for a gas of non-interacting fermions is
\begin{equation}
c_V = \frac{k_B^2 T_\mathrm{core}}{3 \hbar^3 c} \sum_{i = n,p,e} p_F^i \sqrt{m_i^2 c^2 + (p_F^i)^2},
\end{equation}
and the Fermi momenta are $p_F^N = (340 \mev) (2 \tilde n_n)^{1/3}$ and $p_F^{n,e} = (60 \mev) (2 \tilde n_n)^{2/3}$. The surface temperature is then approximately given by \eref{Tsurface}.
The resulting cooling curves for the heavy and light neutron star are shown in \fref{neutronstarcoolingcurves}. (Since we are interested in the effect of introducing $\phi$-cooling \emph{compared to} the standard scenario, we multiply all our $T_\mathrm{surface}(t)$ by 0.6 to bring our estimates into better agreement with complete cooling calculations. This corresponds to a uniform downward shift of 0.2 units on the vertical axis of \fref{neutronstarcoolingcurves}.)

The heavy neutron star cooling curve is not visibly affected for $\Lambda \gtrsim 10^6 \tev$. To avoid altering light neutron star cooling curves by much more than the plausible size of the effects of surface accretion, magnetic fields, and the likely presence of a superfluid component in the core \cite{Yakovlev:2004iq}, requires 
\begin{equation}\label{e.NSbound}
\Lambda \gtrsim 10^8 \tev.
\end{equation}
This confirms our earlier estimate of the constraint. Light neutron stars therefore supply a very strong bound on $\Lambda$ in dmDM for $n_\phi = 1$. However, as we shall see in \sref{phisummary}, this constraint is easily circumvented when $n_\phi = 2$.

\vspace{3mm} 
\section{Other Constraints on $\phi$}
\label{s.constraints}

While stellar astrophysics provides the most impressive constraints on the operator $\bar q q \phi \phi^*/\Lambda$, cosmology and LHC searches bound the dmDM parameter space in complementary directions.

We find that all cosmological constraints are satisfied as long as the only \emph{stable} dark mediator is lighter than $\sim \ev$. LHC searches provide a constraint of $\Lambda \gtrsim$ few TeV that does not depend strongly on $n_\phi$ or $m_\phi$. Flavor physics bound could restrict the allowed SM flavor structure of the coupling in \eref{qqphiphi}, but we avoid those constraints by making the operator diagonal in the SM quark mass basis. 

Dark mediators can also be probed, in principle, using fixed target experiments, precision electroweak measurements or indirect detection of dark matter annihilation. However, as discuss in Appendix~\ref{a.otherconstraints}, these measurements yield no meaningful constraints.

\subsection{LHC Searches}
\label{ss.colliderbounds}

The LHC is sensitive not just to the effective coupling in \eref{qqphiphi} but also to the UV completion of dmDM. We therefore analyze constraints in terms of the dark vector quark model of \ssref{uvcompletion}.

The di-jet + MET search by CMS \cite{CMS:2013gea} is sensitive to on-shell production of two heavy vector-like quarks via the process $p p \rightarrow \psi_Q \bar \psi_Q \rightarrow \phi \phi^* j j$. The constraint is straightforward to apply in our model, since the vector quarks are produced by gauge interactions. The resulting bound is
\begin{equation}
M_Q > 1.5 \tev.
\end{equation}

The 20/fb CMS mono-jet search \cite{ATLAS:2012zim} is sensitive to the processes
\begin{equation}
p\,p\to q^* \to \phi\,\bar{\psi}_{Q,q}\to\phi\phi^* \,j,\quad p\,p\to \phi\,\phi^*+\rm{ISR}
\end{equation}
by doing a counting experiment in different missing energy bins. We simulated the dmDM signal expectation in MG5$+$Pythia 6.420$+$PGS4 \cite{Alwall:2011uj,Sjostrand:2006za} and validated our simulations by reproducing the CMS $jZ(\nu\bar{\nu})$ background prediction with an overall scaling factor of $K = 1.4$. The same scaling factor was also applied to the dmDM signal. The resulting $95\%$ CL lower bound on the $\bar q q \phi \phi^*/\Lambda$ operator depends on whether the intermediate dark vector quark is produced on-shell:
\begin{equation}
\Lambda_\mathrm{eff} \gtrsim 2 \  (6.6) \  \tev \ \ \ \mathrm{for} \ \ \ M_Q = 4 \ (1.5) \  \tev.
\end{equation}
where $\Lambda_\mathrm{eff} = \Lambda$ for $n_\phi = 1$. For $n_\phi > 1$, 
\begin{equation}
\Lambda_\mathrm{eff} = \left(\sum_{i \geq j} \frac{1}{\Lambda_{ij}^2} \right)^{-1/2},
\end{equation}
since the total signal production cross section is given as the sum of all the  $\phi_i \phi_i^*$ cross sections.

\subsection{Cosmological Constraints}
\label{ss.cosmoconstraints}

The mass of the dark mediator should be smaller than about an MeV to allow for sizable direct detection of $\chi$. This means $\phi$ can be thermally produced in the early universe even after $\chi$ freezes out. Such a stable light degree of freedom can overclose the universe and affect Big Bang Nucleosynthesis (BBN) as well as structure formation. In this section, we discuss the thermal history of $\phi$ and derive the relevant cosmological constraints at each step.

\subsubsection{Thermal $\phi$ production}
\label{sss.thermalphi}

The relic density of a light $\phi$ is given by  \cite{Kolb:1990vq}
\begin{equation}\label{e.Omegaphi}
\Omega_{\phi}\,h^2\equiv 7.83\times 10^{-2}\,\frac{g_{\phi}}{g_{*S}}\,\frac{m_{\phi}}{eV},
\end{equation}
where $g_\phi = 2$ is the number of degrees of freedom (d.o.f.) associated with a complex scalar and $g_{*S}$ is the total number of d.o.f. at the temperature at which $\phi$ decouples from the thermal bath. Given the possible size of $g_{*S}$, it is clear that dark mediators with sizable couplings to the SM will overclose the universe unless the heaviest stable scalar has a mass of $m_\phi \lesssim \ev$. This is effectively massless for the purpose of computing all other bounds in this section, which we shall assume from now on.

Assessing the effect of $\phi$ on BBN requires knowing its freeze-out temperature more precisely. For values of $\Lambda$ relevant to direct detection, the hadronic coupling to the SM bath keeps $\phi$ in thermal equilibrium at least until pions decay at $T \sim 100 \mev$. After pion decay, the process $\phi \phi^* \leftrightarrow \gamma \gamma$ maintains thermal equilibrium until the time taken for two photons to annihilate exceeds the hubble time, i.e. 
\begin{equation}
\sigma_{\gamma \gamma \rightarrow \phi \phi*} \ \times \ \frac{2 \zeta(3)}{\pi^2} T^3 \lesssim g_*^{1/2}\,\frac{T^2}{M_\mathrm{pl}}.
\end{equation}
Substituting \eref{gagaphiphiestimate} and the smallest possible $g_* \approx 3$ to slightly underestimate the freeze-out temperature, we obtain
\begin{equation}
T_\phi^\mathrm{freeze-out} \approx (10 \mev) \left(\frac{\Lambda}{\tev} \ \frac{1}{\mathcal{B}}\right)^{2/3}.
\end{equation}
For $\Lambda \gtrsim \tev$, $\phi$ will decouple from the SM bath before BBN.

\subsubsection{Big Bang Nucleosynthesis}
The presence of $\phi$ during the BBN epoch ($T$ around $10$ to $0.1$ MeV) can affect the generation of light elements in two ways. First, even though  $\phi$-nucleon scattering does not change the relative number of neutrons and protons, the presence of an additional light degree of freedom speeds up the expansion of the universe and makes the neutron-proton ratio freeze out at a larger value. This leads to an over production of $^4$He, an effect that can be constrained by measuring the effective number of neutrino flavors, $N_\mathrm{eff}$ during  BBN. Current observation gives $N_\mathrm{eff}=3.3\pm0.6$ \cite{Ade:2013zuv} at 95\% CL from Plank$+$WMAP$+$HighL CMB observations. Since $\phi$ is relativistic it will contribute to an effective number of light neutrino flavors,
\begin{equation}\label{eq:deltaNeff}
\delta N =\frac{8}{7}\times \left(\frac{g^*_{\rm{BBN}}}{g^*_{\phi\,\,\rm{decouple}}}\right)^{4/3},
\end{equation}
to the SM value of $N_\mathrm{eff} = N_\nu = 3$. Assuming all $\phi_i$ are in thermal contact with the SM bath during BBN, this restricts $n_\phi < 2$ (1) if $\phi$ is real (complex).\footnote{$\phi_i$ are colder than photons during the era of Baryon Acoustic Oscillations (BAO), so that $N_\mathrm{eff}$ measurement provides a weaker constraint.} Note that this constraint is weaker if $\phi_i$ decouples earlier. 

When light elements are formed around $0.1$ MeV, $\phi$ can dissociate the nuclei if it gives a recoil energy larger than the nuclear binding energy.  However, due to the lightness of $\phi$, the energy that $\phi$ can give to a nuclei is rather small. For example, in the $^2$H rest-frame, the maximum recoil energy of the $^2$H nucleus being hit by a $\phi$ is $E_R^{max} = {2\,E_{\phi}^2}/{m_{^2\rm{H}}}$. This is only larger than the $^2$H binding energy of $2.2$ MeV if $E_\phi > 47 \mev$, which is much higher than the $\phi$ temperature at the same time.  The effect on nuclear number densities is  negligible.

\subsubsection{Structure Formation}
During the structure formation era (around a temperature of $10$ eV), the scattering length between $\phi$ and $^4\mathrm{He}$ was about $3\times 10^4$ Mpc for $\Lambda = 10 \tev$, so we can treat $\phi$ as a collisionless particle. $\phi$ therefore generates a Landau damping to the primordial density fluctuations, with a free-streaming length that can be estimated as \cite{Kolb:1990vq} 
\begin{equation}
\lambda_{FS,\,\phi}\simeq 20\,\rm{Mpc} \left(\frac{m_{\phi}}{10\,\rm{eV}}\right)^{-1}.
\end{equation} 
This is close to free-streaming neutrinos with $\lambda_{FS,\,\nu}\simeq 20\,\rm{Mpc} \left(\frac{m_{\nu}}{30\,\rm{eV}}\right)^{-1}$, and $\phi$ should satisfy similar constraints as a thermally produced sterile neutrino, with cold dark matter still dominating relic density.  As discussed in \sssref{thermalphi}, this latter requirement of a sub-dominant hot dark matter $\phi$ component requires $m_\phi \lesssim$ eV. The scenario is then similar to the case studied by  \cite{Wyman:2013lza}. The existence of sterile neutrinos at sub-eV scale can relax the tension between Planck result and the local measurements of galaxy clusters on matter perturbation and the expansion rate of the universe. Similar conclusions apply to scalars, meaning sub-eV scale $\phi$'s are compatible with structure formation bounds.

\subsubsection{Dark Acoustic Oscillations}

When DM particles $\chi$ couple to a bath of nearly massless $\phi$ scalars, we expect the DM-$\phi$ system to give rise to dark acoustic oscillations (DAO), similar to the acoustic oscillations of baryons. 
The temperature and polarization spectra obtained from CMB data strongly constrain this effect, which translates to an upper bound on  $\chi$-$\phi$ scattering.

In dmDM, the only tree-level $\chi$-$\phi$ interaction that is not suppressed by  $m_\chi$ is the process  $\chi\phi\to\chi^c\phi\phi$. This is mediated by a $t$-channel $\phi$ and is generated both by the DM yukawa coupling to $\phi$ and the quartic coupling $\lambda \phi^4$ in eq.~(\ref{e.dmdm}).
The transverse cross section of this process is only suppressed by the energy transfer $m_{\chi}^2v_{\chi}^4$ and decouples at a very late time for light $\phi$. Therefore, for scalars with mass below about 10 eV (the temperature of structure formation), CMB data sets stringent upper bounds on the coupling combination $\lambda y_\chi$ to ensure DAO do not generate a sizable effect. 

Although the coupling between four of the lightest scalars has no direct  implication for direct detection signals, it is still useful to understand this constraint for completeness. A detailed analysis using CMB data is beyond the scope of this work (for an example of a analysis, see \cite{Cyr-Racine:2013fsa}). 
However, we can estimate a conservative bound on $y_\chi \lambda$ by requiring the scattering to decouple before structure formation ($T \sim 10 \ev$).


There are two ways for the $\chi\phi\to\chi^c\phi\phi$ process to freeze out before structure formation. 
\begin{enumerate}

\item The lightest scalar could have a mass above 10 eV. To avoid overclosure, \eref{Omegaphi} then requires $\phi$ to freeze out when $g_{*S}\simeq 10^2$, i.e. before the electroweak scale. This can be the case if $\Lambda$ is very large. Indeed, for $n_\phi = 1$, this is required by neutron star cooling, see \eref{NSbound}. However, such a scenario would be sterile with respect to direct detection.

\item In the next section we will define a $n_\phi = 2$ model which avoids neutron star bounds while allowing for direct detection. In that case, the lightest scalar must have a mass below $\sim \ev$ to avoid overclosure. The light scalar therefore remains in thermal contact with dark matter, and remains relativistic during structure formation, which translates to a strong constraint on its quartic coupling. 

The extent to which DM motion is influenced by $\phi$ scattering is given by the transverse cross section for $\chi\phi\to\chi^c\phi\phi$, which we can estimate as
\begin{equation}
\sigma_T\sim\frac{(y_{\chi}^L)^2\,\lambda^2}{16\pi^2\times 16\pi\,m_{\chi}^2v_{\chi}^4}\ln\left(\frac{4\pi m_{\chi}v^2}{y_{\chi}^L\lambda m_{\phi}}\right).
\end{equation}
The logarithmic factor comes from the Coulomb potential of the long-range $\phi$ interaction, and the additional phase space suppression of emitting an additional scalar is approximated by a factor of $(16\pi^2)^{-1}$ . Assuming the scattering rate to be smaller than Hubble before $10$ eV, the upper bound on the couplings translates to
\begin{equation}
\label{e.DAObound}
y_{\chi}^L\,\lambda\lsim 10^{-12},
\end{equation}
for benchmark parameters $m_{\chi}=10$ GeV, $m_{\phi_L}=1$ eV, and DM with kinetic energy $\sim 10$ eV.
\end{enumerate}
A more detailed study may relax the rather conservative bound, but a sizable tuning with $\lambda\sim 10^{-8}$ is expected for $y_{\chi}^L \sim 10^{-4}$. However, this bound has no bearing on direct detection. 

\section{A Realistic dmDM Scenario for Direct Detection}
\label{s.phisummary}

\renewcommand{\arraystretch}{2}
\begin{table*}
\begin{center}
\begin{tabular}{| m{4cm} | m{9cm} |}
\hline 
Avoiding $\phi$-overclosure & Heaviest stable $\phi$ must have $m_\phi \lesssim \ev$\\
\hline
$N_\mathrm{eff}$ during BBN & At most two real light scalars: $n_\phi \leq 2$\\
\hline \hline
LHC direct searches & $\Lambda_\mathrm{eff} = \left(\sum_{i \geq j} \Lambda_{ij}^{-2} 
\right)^{-2} > 2 \tev$.
\\
\hline \hline
Solar Heat Transfer & $\Lambda_{ij} \gtrsim 10 \tev$ if $m_{\phi_{i,j}} \lesssim \kev$\\
\hline
White Dwarf Cooling & $\Lambda_{ij} \gtrsim 10 \tev$ if $m_{\phi_{i,j}} \lesssim$ few keV\\
\hline
Neutron Star Cooling & $\Lambda_{ij} \gtrsim 10^8 \tev$ if $m_{\phi_{i,j}} \lesssim 100 \kev$\\
\hline
Supernovae & $\Lambda_{ij} \lesssim 10^6 \tev$ or $\Lambda_{ij} \gtrsim 10^{11} \tev$ if $m_{\phi_{i,j}} \lesssim 10 \mev$\\
\hline
\end{tabular}
\end{center}
\caption{Bounds on light scalars coupling to SM via operators $\bar q\,q \,\phi_i \phi_j^*/\Lambda_{ij}$. $m_{\phi_{i,j}}$ refers to \emph{both} scalars, not either. Indirect detection via DM annihilation, fixed target experiments and precision measurement bounds yield no relevant constraints if the coupling is SM flavor-blind, see \aref{otherconstraints}.
}
\label{t.phiconstraints}
\end{table*}

A summary of our derived constraints on light dark mediators and their coupling to the SM, formulated for $n_\phi \geq 1$, is shown in \tref{phiconstraints}. To place these in context, recall  from \sref{yXbounds} that the dominant Yukawa coupling should be $y_\chi \sim 10^{-3} - 10^{-2}$. Furthermore, as we review in \sref{directdetection}, direct detection of dmDM in the $2\to3$ regime is feasible if $\Lambda_{ij} \lesssim 10^4 \tev$ with $m_{\phi_i} \lesssim \kev$ and $m_{\phi_j} \lesssim \mev$, so that one scalar can be emitted while the other acts as a light mediator.

With this in mind it is clear that any $n_\phi = 1$ scenario of dmDM with realistic direct detection prospects is \emph{completely excluded} neutron star bounds. In fact, the minimum value of $\Lambda$ required by neutron star cooling truncates the length of the supernova neutrino burst, so the actual lower bound on $\Lambda$ becomes $10^{11} \tev$.

However, there is a very simple $n_\phi = 2$ scenario which behaves almost identically to the minimal $n_\phi = 1$ model for purposes of direct detection, yet is not excluded by any of the bounds in \tref{phiconstraints}.

Consider a dmDM setup like \eref{dmdm} with two light dark mediators, real scalars $\phi_L$ and $\phi_H$ having masses $m_{\phi_L} \lesssim \ev$ and $m_{\phi_H} \sim \mev$. 
We also add a quartic coupling to allow $\phi_H$ to decay into $\phi_L$:
\begin{eqnarray}
\nonumber 
\mathcal{L}_\mathrm{DM} & \supset &
\bar q q \left(\frac{1}{\Lambda_{HH}}  \phi_H \phi_H  + \frac{1}{\Lambda_{LL}}  \phi_L \phi_L  + \frac{1}{\Lambda_{LH}}  \phi_H \phi_L \right)\\
\nonumber 
&& +\  \overline{\chi^c} \chi \left( y_\chi^H \phi_H + y_\chi^L \phi_L \right) + h.c.\\
\label{e.HLmodel}
&& +\  \lambda \phi_H \phi_L^3
\end{eqnarray}
Other quartic couplings are omitted for simplicity\footnote{The quartic coupling for $\phi_L^4$ would have to obey the constraint from dark acoustic oscillations, see \eref{DAObound}.}. When $\lambda > 10^{-9}$, $\phi_H \to \phi_L \phi_L \phi_L$ is instantaneous when the temperature drops below one MeV, leaving $\phi_L$ with a similar relic density to \eref{Omegaphi}.

Now let $\Lambda_{LL} > 10^8 \tev$ to comply with neutron star bounds, while $\Lambda_{HH}, \Lambda_{LH} < 10^6 \tev$ avoids supernova bounds by trapping both $\phi_L$ and $\phi_H$ in the stellar medium. In that case, all the bounds in \tref{phiconstraints} are satisfied. Importantly, $\Lambda_{LH}$, which can be relatively small, now controls direct detection.  This can give a large rate for the process $\bar \chi N \to \bar \chi N \phi_L$.  Since $\phi_H$ is much lighter than the typical momentum exchange of $\gtrsim 10 \mev$ for ambient DM scattering off nuclei, the nuclear recoil spectrum is nearly identical to the $n_\phi = 1$ case with $m_\phi \sim \ev$.

A small $\Lambda_{LH}$ will generate an effective $\Lambda_{LL}$ coupling through a loop of constituent quarks and $\phi_H$. The size of this effective operator is 
\begin{equation}
\Lambda_{LL}^\mathrm{eff} \approx 80 \pi^2 \frac{\Lambda_{LH}^2}{m_q},
\end{equation}
where $m_q \approx 263 \mev$ is the constituent quark mass. The neutron star bound on $\Lambda_{LL}$ then translates to a bound on $\Lambda_{LH}$:
\begin{equation}
\label{e.NSboundHL}
\Lambda_{LH} \gtrsim 10 \tev
\end{equation}
which is the bound we adopt when discussing direct detection in the next section.

Finally, the presence of two Yukawa couplings to dark matter gives additional freedom. A very modest hierarchy $y_\chi^H/y_\chi^L \gtrsim 10$ would suppress the $\chi N \to \chi N$ loop process. This makes it possible for $y_\chi^H$ to be large enough for a thermal relic $\chi$ and  ameliorate the inconsistencies between dwarf galaxy simulations and observation, all while being in the $2\to3$ regime of direct detection (see \fref{yXboundplot}).

This scenario can be realized in the UV completion of \ssref{uvcompletion} by assuming hierarchical Yukawa couplings between dark mediators, dark vector quarks and different chiralities of the SM quarks.

\vspace{3mm} 
\section{Direct Detection of \lowercase{dm}DM}
\label{s.directdetection}

In this section we outline in detail our computation of nuclear recoil spectra and direct detection constraints on dmDM, first summarized in \cite{Curtin:2013qsa}. We work with the minimal $n_\phi = 1$ scenario with effectively massless $\phi$ for simplicity, with the understanding that this phenomenology can be replicated by the unexcluded $n_\phi = 2$ scenario defined by \eref{HLmodel}.

The discussion of the previous two sections derived constraints on the Yukawa couplings between dark mediators and dark matter, and the coupling between dark mediators and SM quarks. Direct detection is sensitive to a combination of the two. We predict the dmDM signal at XENON100 \cite{Aprile:2012nq}, LUX \cite{Akerib:2013tjd}, CDMS-Si \cite{Agnese:2013rvf} and CDMSlite \cite{Agnese:2013lua} and demonstrate that large regions of the direct detection plane are not yet excluded. 

It is instructive to compare the dmDM interaction with nuclei to the contact operator 
\begin{equation}
\label{e.contactoperator}
\frac{\bar q q \bar \chi \chi}{\tilde \Lambda^2},
\end{equation}
since it is the standard choice for showing constraints from different direct detection experiments in the same $(m_\chi, \sigma_\mathrm{SI}^n)$-plane. Referring to the above interaction model as the ``standard-WIMP'', we find that $\mathcal{O}(100 \gev)$ dmDM will fake a different lighter $\mathcal{O}(10 \gev)$ WIMP at different experiments. This is due to  energy loss from the outgoing $\phi$, which leads to an underestimate of the DM energy when assuming the above contact operator. We study this interesting phenomenon by first examining at the parton level cross section, understanding the parametric dependence of the recoil spectrum, and then produce the full experimental recoil prediction including form factors and the velocity distribution.

\subsection{Differential cross section calculation}

\begin{figure*}
\begin{center}
\includegraphics[width=4.9cm]{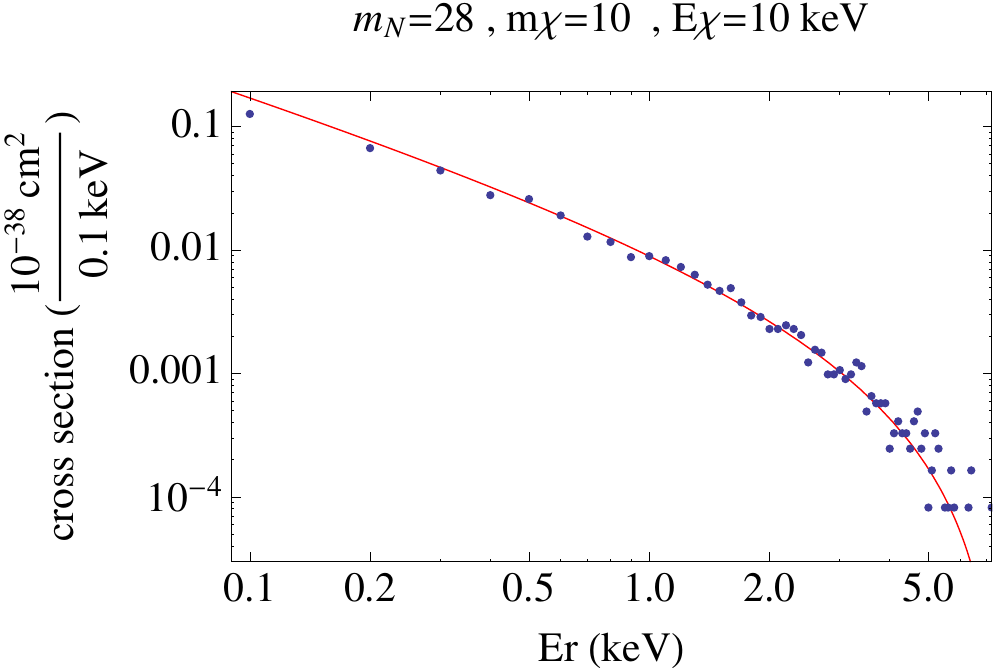}\quad\includegraphics[width=4.9cm]{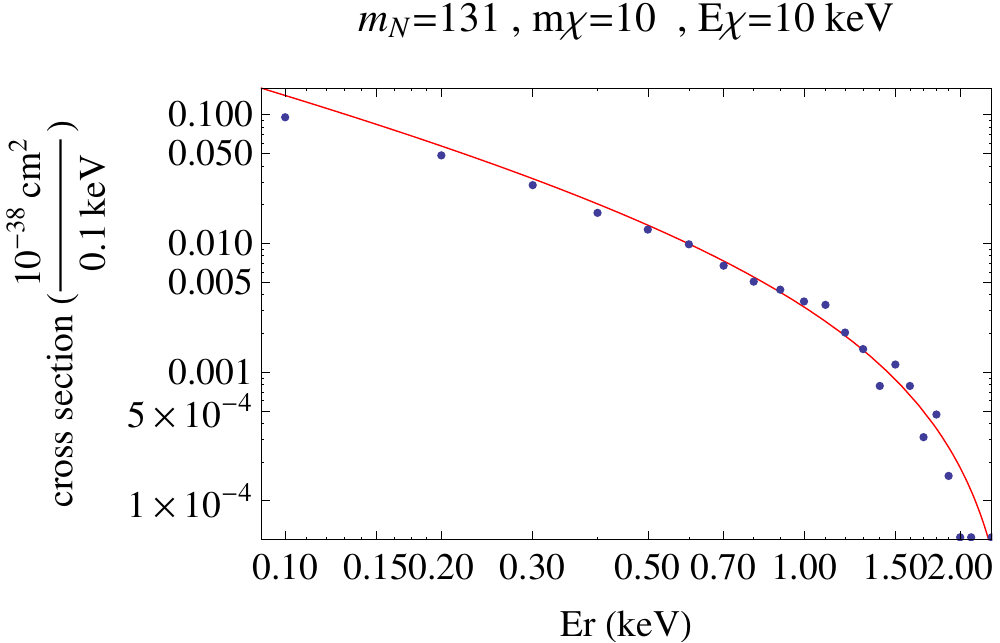}\quad\includegraphics[width=4.9cm]{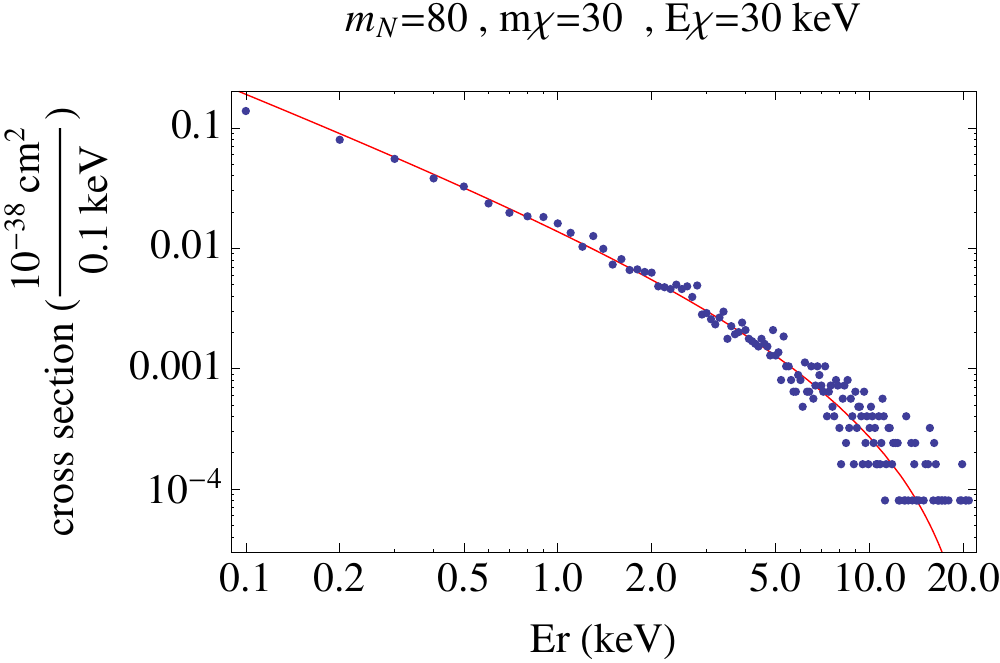}\\
\includegraphics[width=4.9cm]{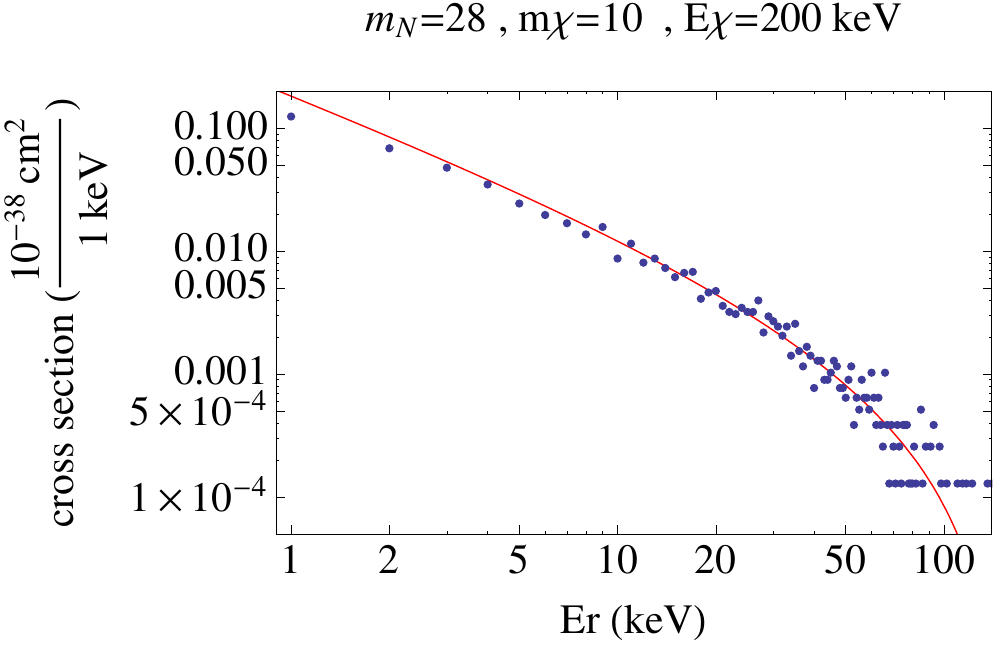}\quad\includegraphics[width=4.9cm]{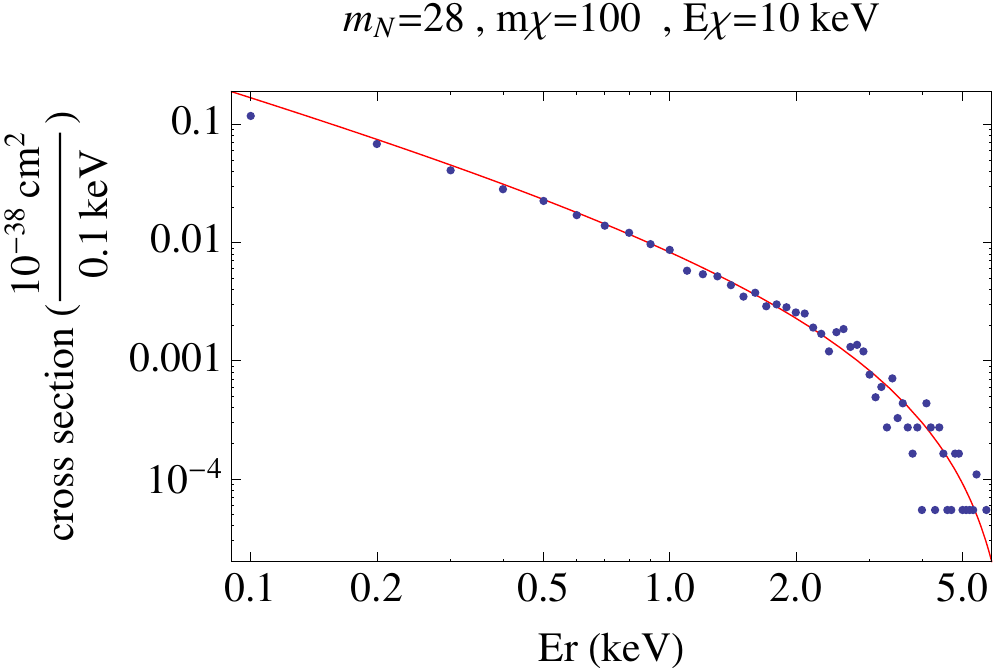}\quad\includegraphics[width=4.9cm]{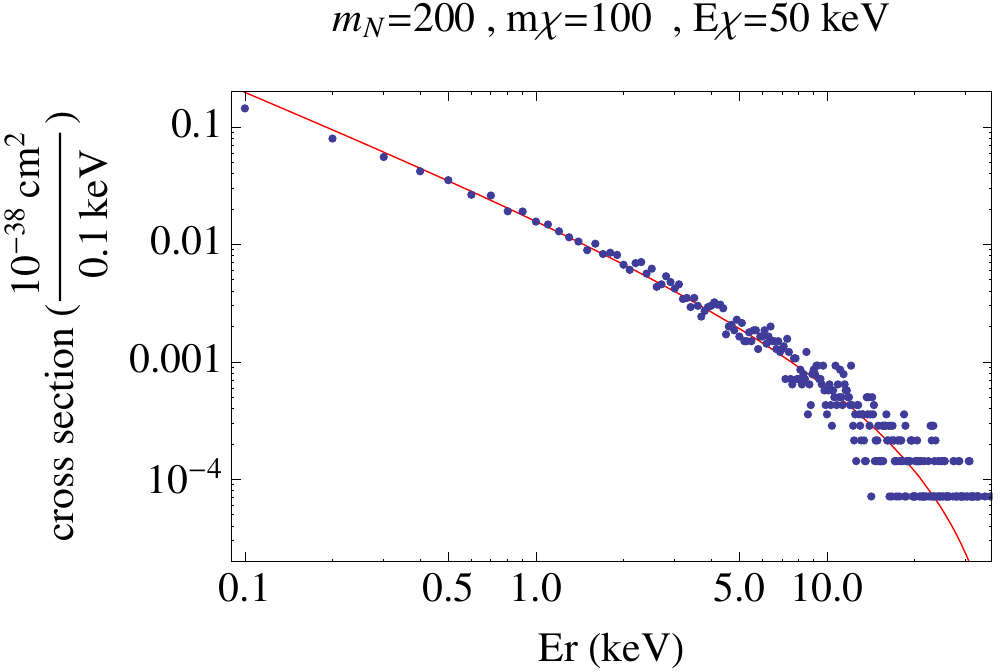}
\end{center}
\caption{Examples of nuclear recoil spectra with dmDM at `parton-level' (without nuclear/nucleus form factors and coherent scattering enhancement) for different $m_N, m_{\chi}$ and a given incoming energy $E_{\chi}$. The blue datapoints are given by the MG5 simulation, and the red curve are the analytical approximation of the spectrum in Eq.~\eref{Erfit2}.
}
\label{f.partonxsection}
\end{figure*}

We are interested in the differential cross section for a dark matter particle hitting a stationary nucleus which then recoils with kinetic energy $E_r$. This is given by
\begin{equation}
\frac{d \sigma_N}{d E_r} \  = \ F^2(E_r) \  A^2  \  \left( \Sigma B \right)^2 \  \frac{d \sigma_N^\mathrm{bare}}{d E_r}.
\end{equation}

$d \sigma_N^\mathrm{bare}/d E_r$ is the `parton-level' differential cross section evaluated for the process $q \chi \to \bar q \chi \phi$ or $q \chi \to q \chi$ with the substitution of $m_q \to m_N$. This is because ambient dark matter is extremely non-relativistic with velocites of order a few $100$ km/s, interacting with the entire nucleus coherently. $d \sigma_N^\mathrm{bare}/d E_r$ is easily evaluated analytically for the $2\to2$ loop process using \eref{2to2operator}, reproducing the result of a standard WIMP with an additional suppression at high momentum transfer. For $2\to3$ scattering we adopt a Monte-Carlo approach\footnote{This was more practical than the analytical approach for evaluating different possible models that realize $2\rightarrow3$ scattering. It is unlikely that fully analytical cross section expressions would have been extremely illuminating.
We do discuss analytical approximations below.} by defining  a \texttt{MadGraph5} \cite{Alwall:2011uj} model containing the DM Yukawa coupling and the $\bar N N \phi \phi^*/\Lambda$ effective operator using \texttt{FeynRules1.4}~\cite{Ask:2012sm}. We discuss the resulting spectrum below.

The factor $\left( \Sigma B \right)^2$ is a quark-nucleon form-factor to convert the amplitude from quark- to nucleon-level by taking into account the values of quark currents inside the proton or neutron (see \cite{Belanger:2008sj,Crivellin:2013ipa} for a review). Since the momentum transfer $q^2$ is much less than the QCD confinement scale we can take this form factor to be constant. The relevant case for dmDM is the scalar operator  $\langle N | m_q \bar q q |N\rangle = f_q^N m_N$, which is interpreted as the contribution of quark $q$ to the nucleon mass $m_N$. Importantly, the contribution of all sea quarks is \emph{additive}, giving a large matrix element enhancement. $f_q^N < 1$, since each sea quark contributes more than its bare mass to the proton mass, and can be computed from lattice techniques. This gives matrix elements $\langle N|\bar{q}q|N\rangle\equiv B^N_q$, where 
\begin{eqnarray*}
& B_u^p=8.6,\quad B_d^p=6.3,\quad B_s^p=2.4,\\
& B_u^n=6.8,\quad B_d^n=8.0,\quad B_s^n=2.4.
\end{eqnarray*}
Assuming equal coupling of $\phi$ to all SM quarks, the $|\mathcal{M}|^2$ enhancement is therefore 
\begin{equation}
\label{e.Bsqfactor}
\bigg( \sum_{q = u,d,s} B_q^{n,p} \bigg)^2 \approx 300.
\end{equation}

Going from nucleon- to nucleus-level, the cross section is enhanced by $A^2$ (assuming equal $\phi$ coupling to protons and neutrons) and must be convolved with the \emph{Helm Form Factor} \cite{Engel:1991wq, Lewin:1995rx}. This is just the Fourier transform of the radial nuclear density distribution,
\begin{equation}
\label{e.Helm}
F^2(E_r) = \left( \frac{3 j_1(q r_0) }{q r_0}\right)^2 e^{-s^2 q^2},
\end{equation}
where $j_1$ is a Bessel Function, $q = \sqrt{2 m_N E_r}$, $s = 1 \ \mathrm{fm}$, $r_0 = \sqrt{r^2 - 5 s^2}$ and $r = 1.2 A^{1/3}$.

It is instructive to compare $d \sigma^\mathrm{bare}_N/dE_r$ for the $2\to3$ scenario to the simple WIMP case generated by the contact operator \eref{contactoperator}. We examine the case of massless emitted $\phi$. $m_\phi \sim $ few keV could be interesting to introduce shape features into the recoil spectrum but is cosmologically disfavored, see \tref{phiconstraints}.
As shown in \fref{partonxsection}, the recoil spectrum of dmDM can be well described by the function\begin{eqnarray}
\frac{d\,\sigma_{2\to3}^{bare}}{d\,E_r}&\simeq& \, \frac{\mathcal{C}}{E_r} \,\left(1-\sqrt{\frac{E_r}{E_r^\mathrm{max}}}\right)^2,
\label{e.Erfit2}
\end{eqnarray}
where $\mathcal{C}=1.3\times 10^{-42}\,(\tev/\Lambda)^2$ cm$^2$. $E_r^{max}\simeq 2\,\frac{\mu_{\chi N}^2}{m_N}\,v^2$ is the maximum allowed nuclear recoil energy for a given incoming DM velocity, same as for the standard WIMP. The above approximation holds for massive dark mediators as well, provided the intermediate $t$-channel $\phi$ has mass $\lesssim \mev$ and the emitted $\phi$ has mass $\lesssim \kev$. 

Eq.~\ref{e.Erfit2} can be decomposed into the phase space part of $\chi\,N\to\chi\,N\,\phi$ scattering via a contact interaction, times the propagator of the light mediator $\phi$. The phase space can be approximated by 
\begin{equation}
\frac{d\,\sigma_{2\to3}^{contact}}{d\,E_R}\propto m_N^2\,E_R\left(1-\sqrt{\frac{E_R}{E_R^{max}}}\right)^2,
\end{equation}
which vanishes when $E_R$ reaches its maximum value, or when $E_{R}\to 0$, since a relativistic $\phi$ itself cannot compensate both the energy and momentum of a non-relativistic DM particle. The non-relativistic scattering also requires the spatial momentum exchange to be much larger than the kinetic energy, which makes the spatial momentum of the relativistic $\phi$ negligible in energy-momentum conservation. Because of this, the propagator of $\phi$ in the dmDM scattering is $(2m_N\,E_R)^{-2}$, which is dominated by the spatial momentum exchange between $N$ and $\chi$ and gives the spectrum in \eref{Erfit2}.

The contact operator \eref{contactoperator} produces a flat parton-level nuclear recoil spectrum for $E_r  < E_r^\mathrm{max}$. On the other hand, \eref{Erfit2} features a suppression at large recoil. 
The functional form of this recoil suppression is different than for $2\to2$ scattering with light mediators and/or derivative couplings. Furthermore, the scaling of total cross section with $m_N, m_\chi$ is unique. This necessitates a full re-interpretation of all direct detection bounds to understand how a heavy dmDM candidate fakes different light WIMPs at different detectors. We expect the recoil suppression to  increase the sensitivity advantage enjoyed by low-threshold Xenon detectors over CDMS.

\begin{figure}
\begin{center}
\includegraphics[width=8cm]{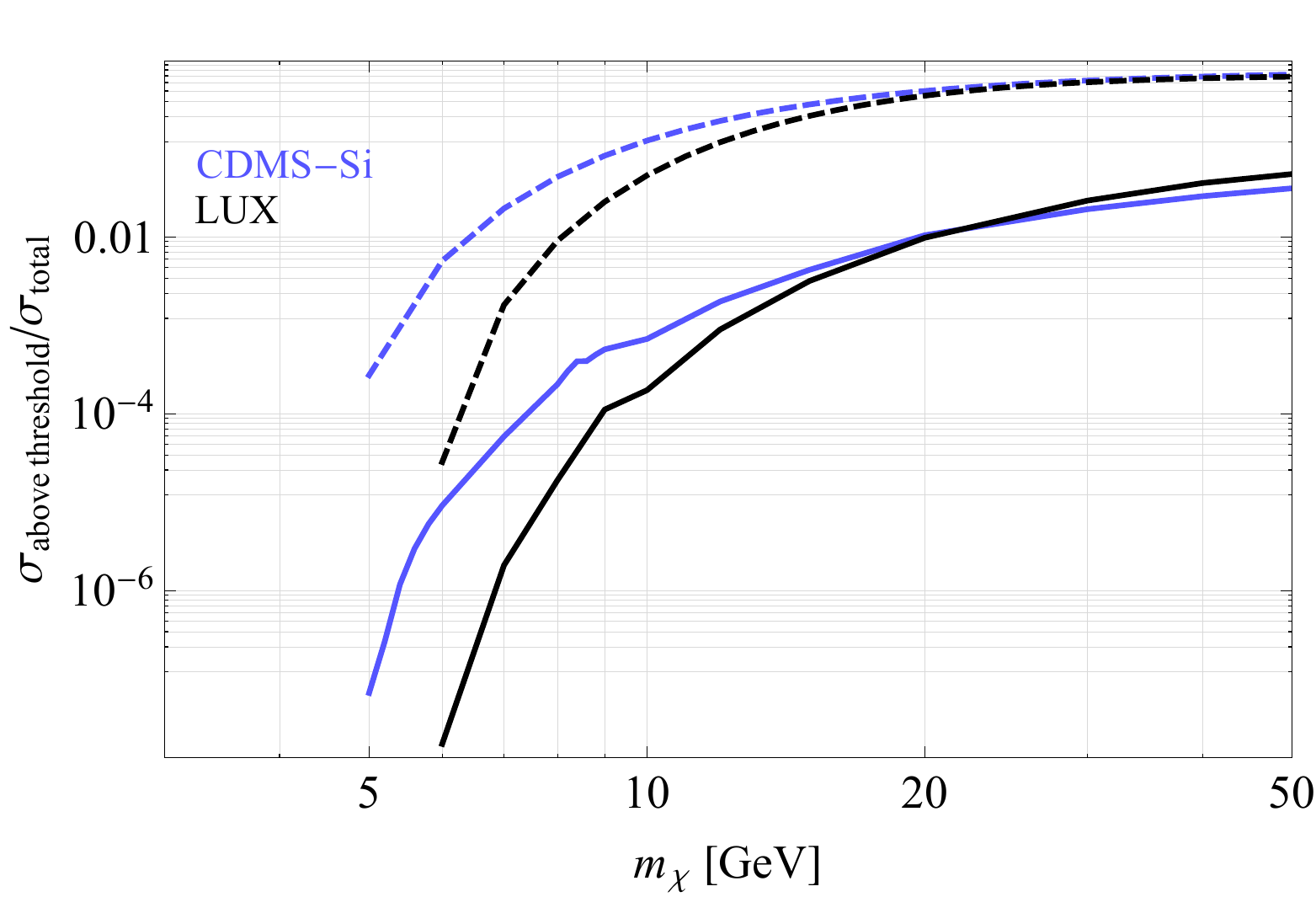}
\end{center}
\caption{
The fraction of the dmDM (solid) and WIMP DM (dashed) direct detection cross section above experimental threshold (blue for CDMS II Si $E_r > 7 \kev$, black for LUX $S1 > 2$) as a fraction of the total cross section.  $m_\phi < \kev$. $S1$ light collection efficiency is taken into account but signal selection cuts have not been applied. 
}
\label{f.ddefficiency}
\end{figure}

\begin{figure*}
\begin{center}
\includegraphics[width=15cm]{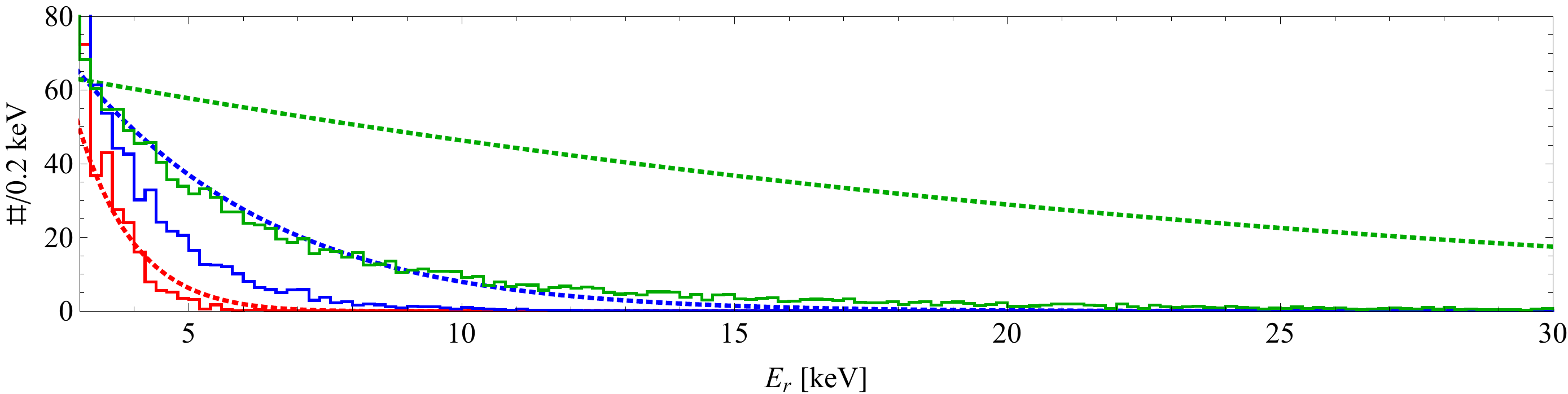}
\end{center}\vspace{-3mm}
\caption{
Nuclear recoil spectra at CDMS II Silicon ($m_N = 28 \gev$) with 140.2 kg$\cdot$days exposure for dmDM (solid) and WIMP DM (dotted) of mass 5 (red), 10 (blue) and 50 (green) GeV. Experimental efficiencies are not included, and the recoil spectrum is shown only for $E_r > 3 \kev$ because the dmDM spectrum is so sharply peaked at the origin that no other features would be visible if it were included. The shown WIMP-nucleon cross sections for (5, 10, 50) GeV are $(4, 2, 6) \times 10^{-40} \ \cmtwo$, while the dmDM parameters are $y_\chi = 0.02$, $\Lambda = (29, 91, 91) \tev$ and $m_\phi < \kev$
}
\label{f.DDspectraCDMS}
\end{figure*}

\begin{figure*}
\begin{center}
\includegraphics[width=15cm]{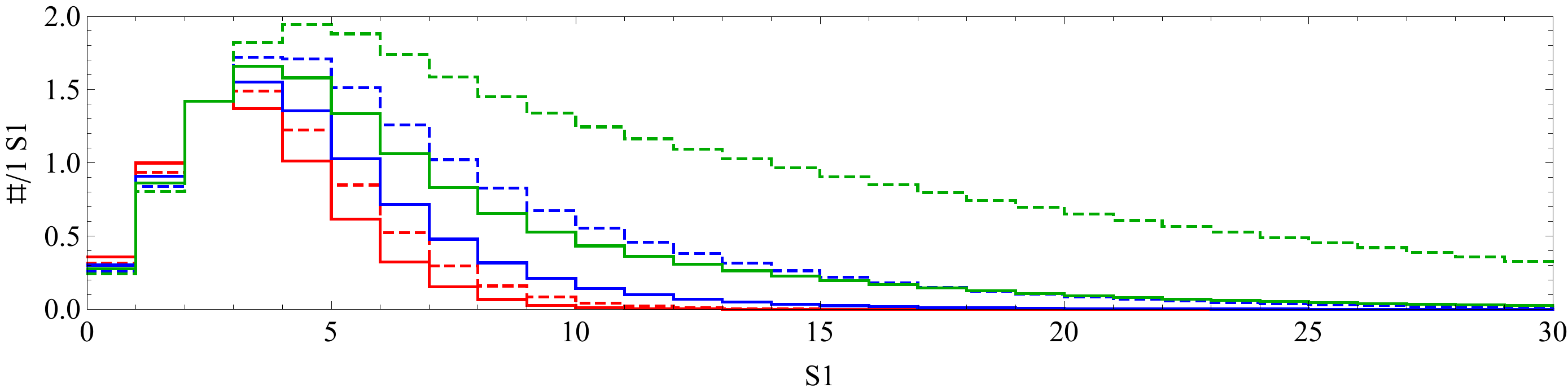}
\end{center} \vspace{-3mm}
\caption{
S1 spectra at LUX ($m_N = 131 \gev$) with 10065.4 kg$\cdot$days exposure for dmDM (solid) and WIMP DM (dotted) of mass 10 (red), 20 (blue) and 50 (green) GeV. The 14\% S1 light gathering efficiency is included but selection cuts are not. No DM signal below $E_r = 3 \kev$ is included due to limitations of the measured $\mathcal{L}_{eff}$, in accordance with the collaboration's analysis. The shown WIMP-nucleon cross sections for (10, 20, 50) GeV are $(18.5, 3.6, 4.9) \times 10^{-45} \ \cmtwo$, while the dmDM parameters are $y_\chi = 0.02$ and $\Lambda = (1900, 9700, 13000) \tev$ and $m_\phi < \kev$.
}
\label{f.DDspectraLUX}
\end{figure*}

\begin{figure}
\begin{center}
\includegraphics[width=7.5cm]{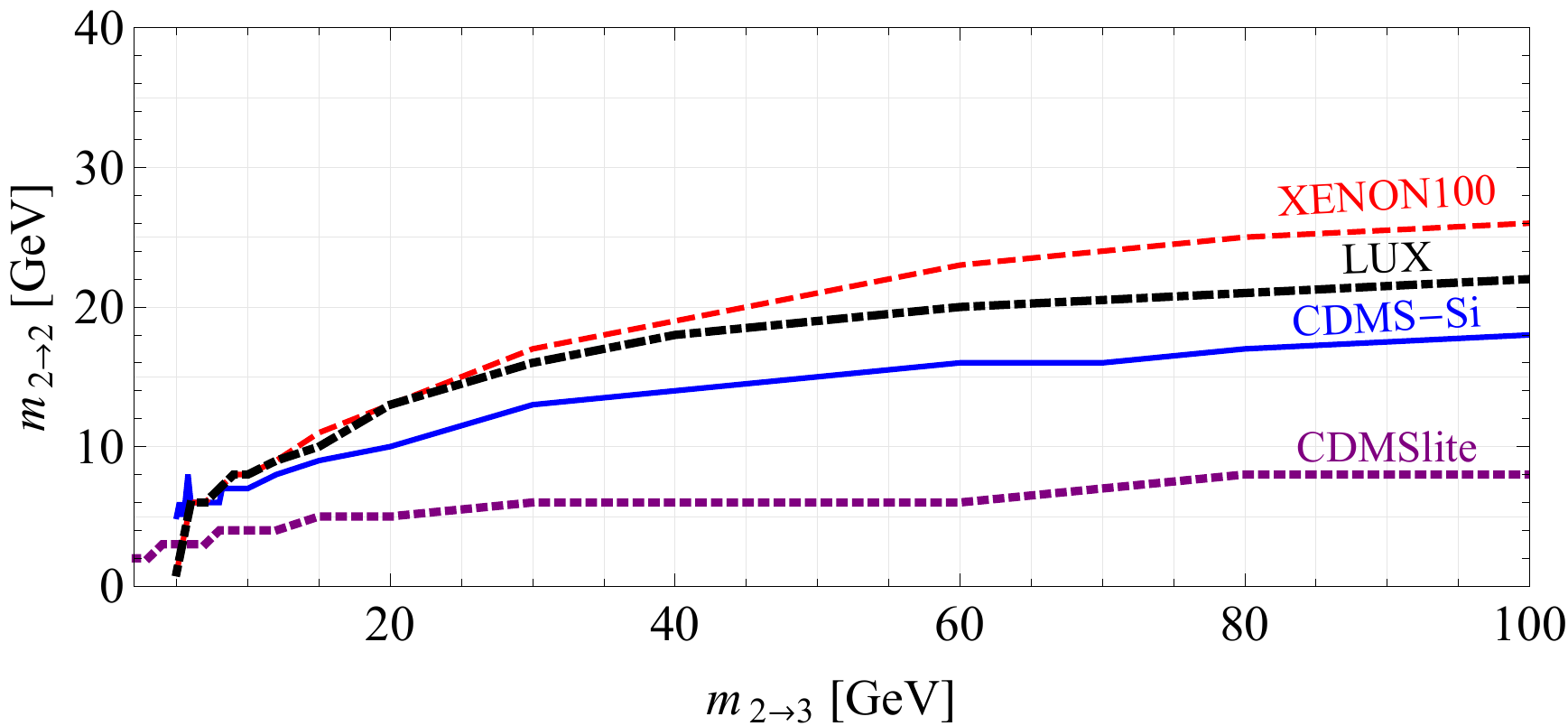}
\end{center}
\caption{
The measured nuclear recoil spectrum produced by a dmDM candidate with mass $m_\chi = m_{2\rightarrow3}$ is very similar to that of a WIMP with mass $m_{2\to2} < m_{2\to3}$, interacting with nuclei via the contact operator \eref{contactoperator}. $m_{2\to2}(m_{2\to3})$ 
is shown for XENON100 ($S1 > 3$ with 6\% light gathering efficiency, dashed red line), LUX ($S1 > 2$ with 14\% light gathering efficiency, dash-dotted black line), CDMS II Silicon ($E_r > 7 \kev$, solid blue line), and CDMSlite (Germanium, $Er > 0.2 \kev$, dotted purple line) before selection cuts.
}
\label{f.compare2to2to2to3}
\end{figure}

\subsection{Nuclear Recoil Spectra}

To compute the expected nuclear recoil spectrum at a direct detection experiment, the differential scattering cross section must be convolved with the dark matter speed distribution in the earth frame,
\begin{equation}
\frac{d R}{d E_r} = N_T \frac{\rho_\chi}{m_\chi} \int dv \ v f(v) \frac{d \sigma_N}{d E_r},
\end{equation}
The speed distribution is given in \aref{ddcomputation}.
In our Monte Carlo calculation for $2\to3$ scattering, we simulate $\chi N \rightarrow \chi^c N \phi$ for different $m_N, m_\chi, m_\phi$ and incoming DM velocities $v$ in bins of $20$ km/s to build up a table of the various required $\frac{d \sigma_N}{d E_r}$ and perform this convolution numerically. For verification, we applied our pipeline to WIMP-nucleus scattering, reproducing the expected analytical results. 

The spectrum of nuclear recoil events that occurred in the detector must  be translated to actual experimental observables. This involves folding in efficiencies, as well as converting the nuclear recoil signal to a scintillation light signal in the case of liquid Xenon detectors. These details are also given in \aref{ddcomputation}.

The detection efficiency for $2\to3$ scattering in dmDM is about $100 - 1000$ times smaller  compared to the standard WIMP (and also $2\to2$ scattering in dmDM), see \fref{ddefficiency}. This is expected, given the additional $E_r$-suppression. In the next subsection we will take care to understand the parameter regions where $2\to3$ scattering dominates over the $2\to2$ process in dmDM

 \fref{DDspectraCDMS} shows some $2\to3$ nuclear recoil spectra at CDMS II Si before taking detection efficiency into account. dmDM is compared to WIMPs for different DM masses, and the principal experimental feature of our model is apparent: a $\sim 50 \gev$ dmDM candidate looks like a $\sim 10 \gev$ WIMP. \fref{DDspectraLUX} shows different $S1$ spectra at LUX, where a $\sim 50 \gev$ dmDM candidate looks more like a $\sim 20 \gev$ WIMP. This mass remapping compared to the standard contact operator interpretation is shown for different experiments in \fref{compare2to2to2to3}. This dependence of recoil suppression on the detector and DM parameters is unique to dmDM, and could be added to other DM models by including the emission of a light particle..

\subsection{Direct Detection Constraints}
\label{ss.ddconstraints}

We compute direct detection bounds on dmDM in two ways. The first is by remapping the bounds provided by the respective experimental collaborations using the remapping of dmDM to standard WIMP parameters \cite{Curtin:2013qsa}, part of which is shown in \fref{compare2to2to2to3}. These results are then reproduced, for verification, by using a full modified maximum likelihood analysis \cite{Barlow:1990vc} for each experiment. The resulting bounds in the direct detection plane for dominant $2\to3$ and $2\to2$ scattering in dmDM are shown in \fref{mappingbounds} and \fref{mappingbounds2to2}. To provide a lower boundary on the relevant parameter space we indicate where in the direct detection plane the dmDM signal gets drowned out by the irreducible neutrino background \cite{Billard:2013qya}. 

\begin{figure}
\begin{center}
\hspace*{-7mm}
\includegraphics[width=8cm]{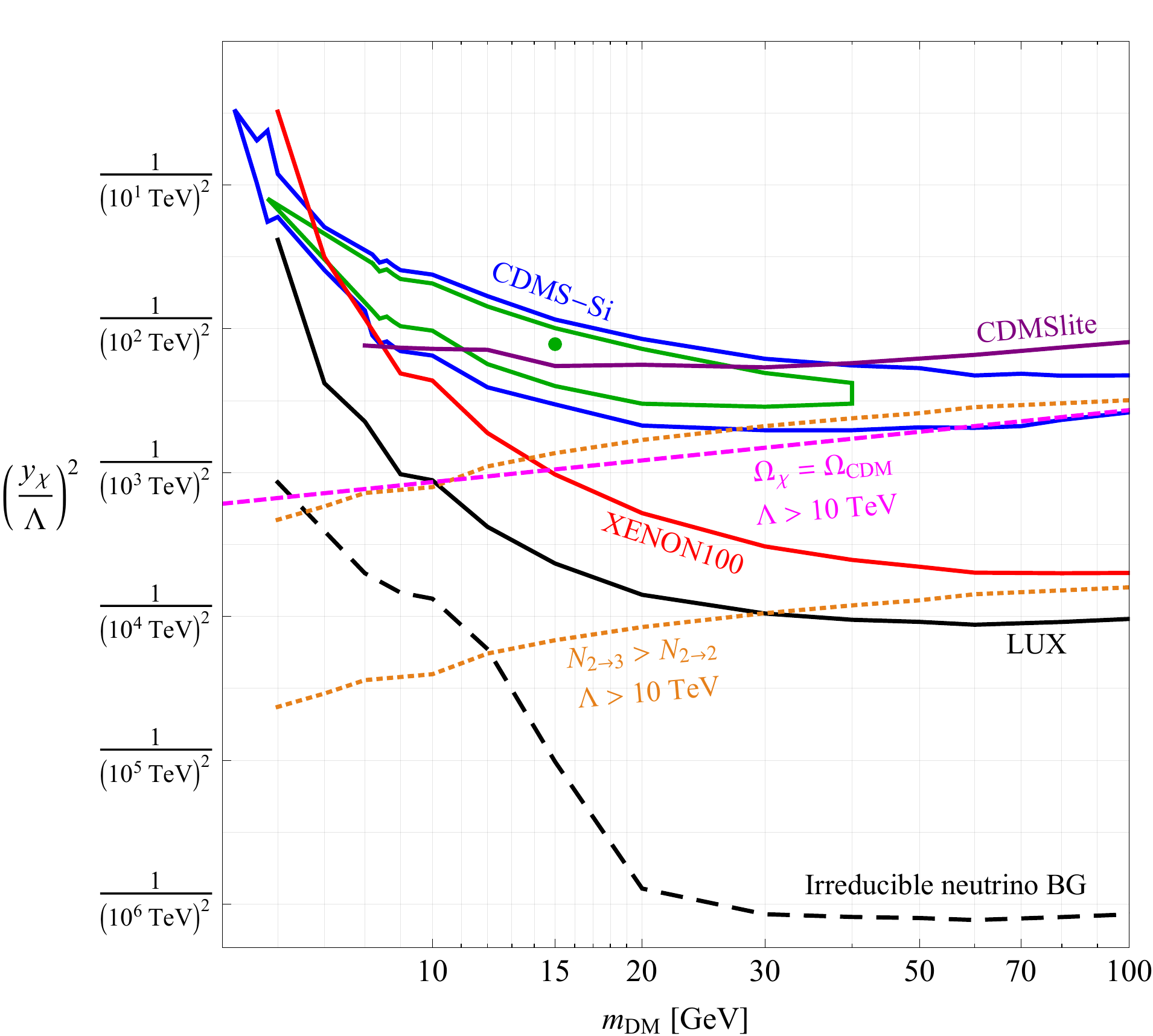}
\end{center}
\caption{Direct detection bounds on the $2\rightarrow3$ regime of $n_\phi = 1$ dmDM. The vertical axis is proportional to $\sigma_{\chi N \to \bar \chi N \phi}$, and is understood to be $(y_\chi^H/\Lambda)^2$ for the $n_\phi = 2$ model of \sref{phisummary}. \emph{Solid lines}: 90\% CL bounds by XENON100 (red), LUX (black) and \mbox{CDMSlite} (purple), as well as the best-fit regions by CDMS II Si (blue, green). The large-dashed black line indicates where the dmDM signal starts being drowned out by the irreducible neutrino background \cite{Billard:2013qya}. 
\emph{Small-dashed magenta line}: $y_\chi = y_\chi^\mathrm{relic}(m_\chi)$ and $\Lambda = 10 \tev$. Need to be below this line for a thermal relic to be compatible with the neutron star cooling bound \eref{NSboundHL}.
\emph{Lower dotted orange line}: for $n_\phi = 1$, below this line $y_\chi$ is small enough to ensure the $2\to3$ process dominates direct detection while also satisfying the neutron star cooling bound. This line can be arbitrarily moved when $n_\phi = 2$.}
\label{f.mappingbounds}
\end{figure}

\begin{figure}
\begin{center}
\hspace*{-7mm}
\includegraphics[width=8cm]{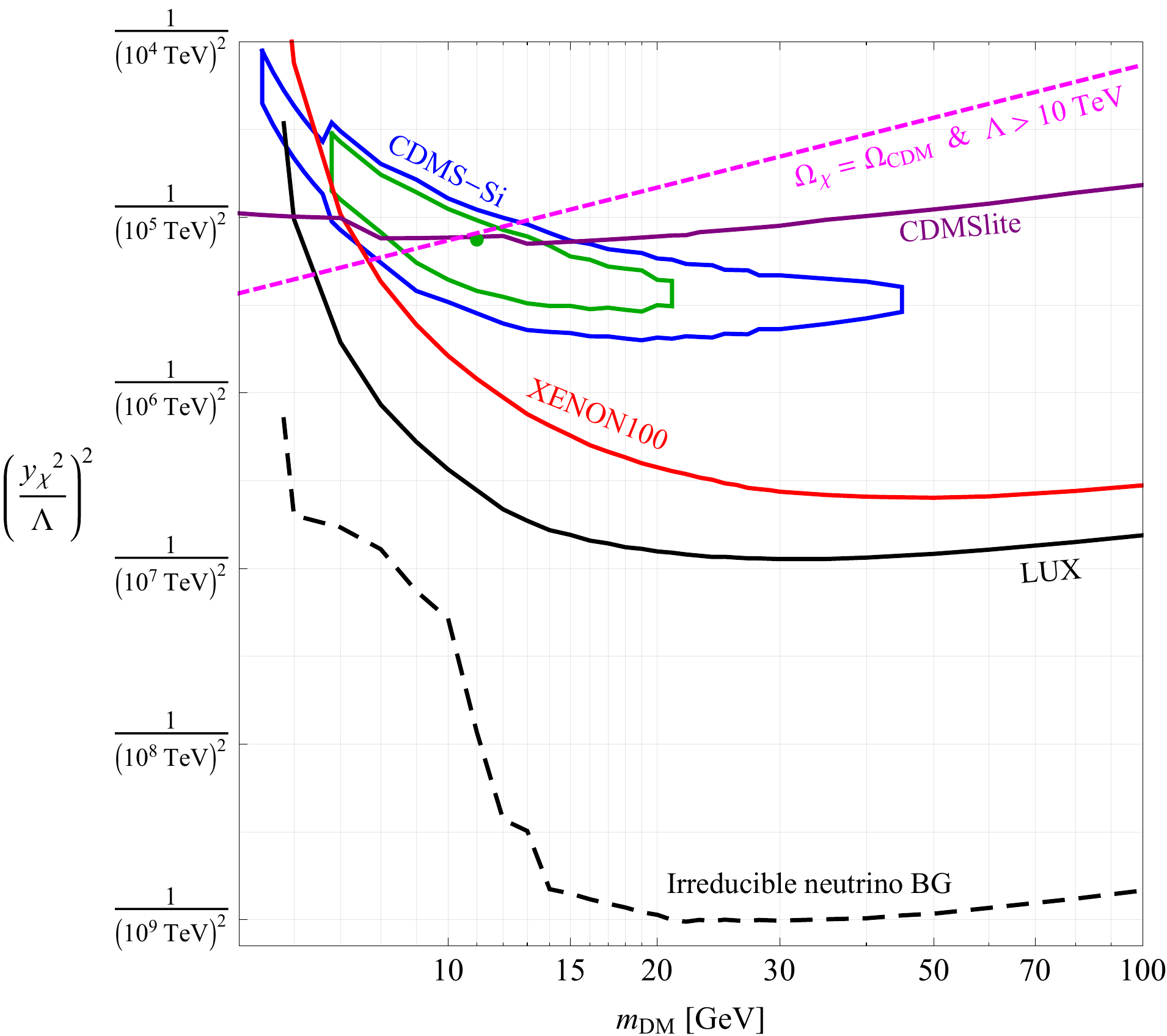}
\end{center}
\caption{Direct detection bounds on the $2\rightarrow2$ regime of $n_\phi = 1$ dmDM. The vertical axis is proportional to $\sigma_{\chi N \to \bar \chi N}$, and is understood to be $(y_\chi^H y_\chi^L/\Lambda)^2$ for the $n_\phi = 2$ model of \sref{phisummary}. 
Same labeling as \fref{mappingbounds}. }
\label{f.mappingbounds2to2}
\end{figure}

For the $2\to3$ and $2\to2$ scattering regimes, direct detection probes $y_\chi/\Lambda$ and $y_\chi^2/\Lambda$ respectively. The neutron star cooling bound \eref{NSboundHL} for the $n_\phi = 2$ model $\Lambda \gtrsim 10 \tev$ and the bounds on dark matter Yukawa coupling $y_\chi$ can be combined to be shown in the direct detection planes of Figs. \ref{f.mappingbounds} and \ref{f.mappingbounds2to2}. The assumption of a thermal relic then sets bounds which supersede the liquid Xenon experiments for $m_\chi \lesssim 10 \gev$.

For $n_\phi = 1$, the $2\to2$ loop process in \fref{feynmandiagram} dominates if $y_\chi \gtrsim 10^{-3}$. This is indicated, together with the neutron star bound, by the dashed orange line in  \fref{mappingbounds}. However, in the $n_\phi = 2$ model of \sref{phisummary}, the vertical axis of Figs.~\ref{f.mappingbounds} and \ref{f.mappingbounds2to2} is $(y_\chi^{H}/\Lambda)^2$ and $(y_\chi^{H} y_\chi^L/\Lambda)^2$ respectively, so this orange line can be moved arbitrarily upwards. This means the $n_\phi = 2$ model can realize $2\to3$ dominated direct detection while being consistent with a thermal relic, as well as the SIDM solution to the inconsistencies between dwarf galaxy simulations and observation.

\vspace{3mm} 
\section{Conclusion}
\label{s.conclusion}

Previous theoretical investigations have shown that direct detection can proceed very differently from the standard WIMP scenario. Investigating all possibilities is of vital importance. For one, the current list of experimental anomalies naively conflicting with other collaborations' bounds motivates the search for alternative interpretations of the data. Another general reason for achieving `full theoretical coverage' is the looming irreducible neutrino background \cite{Billard:2013qya}  that direct detection could become sensitive to in about a decade. Hitting this neutrino floor without a clear dark matter signal is an undesirable scenario, but being left without alternative options to explore would be an even more dire situation.

\emph{Dark Mediator Dark Matter} is the first example of a slightly non-minimal dark sector where the mediators connecting dark matter to the Standard Model are themselves charged under the same symmetry that makes dark matter stable. Phenomenologically, this closes a long-standing gap in the list of investigated scenarios by realizing $2\to3$ nuclear scattering at direct detection experiments. 

We carry out the first systematic exploration of light scalar mediators coupling to the SM quarks via operators of the form $\bar q q \phi \phi^*/\Lambda$. Their existence and coupling can be strongly constrained by cosmological bounds, LHC direct searches and stellar astrophysics, see \tref{phiconstraints}. Neutron star cooling excludes detectable dmDM scenarios with a single dark mediator completely, but an $n_\phi = 2$ scenario can easily evade all bounds while giving identical direct detection phenomenology. 

The presence of a light mediator and additional particle emission means that the nuclear recoil spectrum of dmDM at direct detection experiments is strongly peaked towards the origin. The functional form of this recoil suppression and the overall cross section dependence on nucleus and DM mass is unique. As a consequence of this suppression, a $\sim 100 \gev$ dmDM candidate fakes different $\mathcal{O}(10 \gev)$ standard WIMPs at different experiments. We compute direct detection bounds on dmDM for both nuclear scattering processes, $\chi N \to \chi N \phi$ and the loop suppressed $\chi N \to \chi N$ and find large regions that are not excluded but discoverable in the future. The abovementioned $n_\phi = 2$ scenario can realize $2\to3$ direct detection while being compatible with a thermal relic and the SIDM solution for the inconsistencies between dwarf galaxy simulations and observation.

Our model represents an interesting combination of light mediator and inelastic scattering ideas, since the latter is realized by having a light scalar $\phi$ from a direct-detection point of view. This allows us to smoothly map dmDM spectra to similar WIMP spectra, and the resulting map of dmDM parameters to WIMP parameters makes transparent how the direct detection bounds are re-interpreted (see also \cite{Curtin:2013qsa}). While dmDM does not reconcile the conflicting signals and constraints, it may point the way towards another model that does. For example, it might be interesting to explore how this new scattering process changes models with non-standard form factors or exothermic down-scattering.

\vspace{4mm}
\textbf{Acknowledgements}

The authors would like to gratefully acknowledge the contributions of Yue Zhao and Ze'ev Surujon during  early stages of this collaboration. 
We thank Patrick Meade for valuable comments on an early draft  of this paper. We are very grateful to 
Haipeng An, 
Brian Batell, 
Joseph Bramante,
Rouven Essig, 
Greg Gabadadze, 
Roni Harnik, 
Jasper Hasenkamp, 
Patrick Meade,
Matthew McCullough,
Olivier Mattelaer,
Ann Nelson,
Matthew Reece, 
Philip Schuster, 
Natalia Toro, 
Sean Tulin,
Neal Weiner, 
Itay Yavin and
Hai-Bo Yu
for valuable discussions. D.C. is supported in part by the National Science Foundation under Grant PHY-0969739. Y.T. is supported in part by the Department of Energy under Grant DE-FG02-91ER40674. The work of Y.T. was also supported in part by the National Science Foundation under Grant No. PHYS-1066293 and the hospitality of the Aspen Center for Physics. The work of D.C. and Y.T. was also supported by the hospitality of the Center for Future High Energy Physics in Beijing, China.

\appendix

\section{Radial Profiles for Stellar Cooling Calculation}
\label{a.radialprofiles}

\begin{figure}
\begin{center}
\includegraphics[width=6cm]{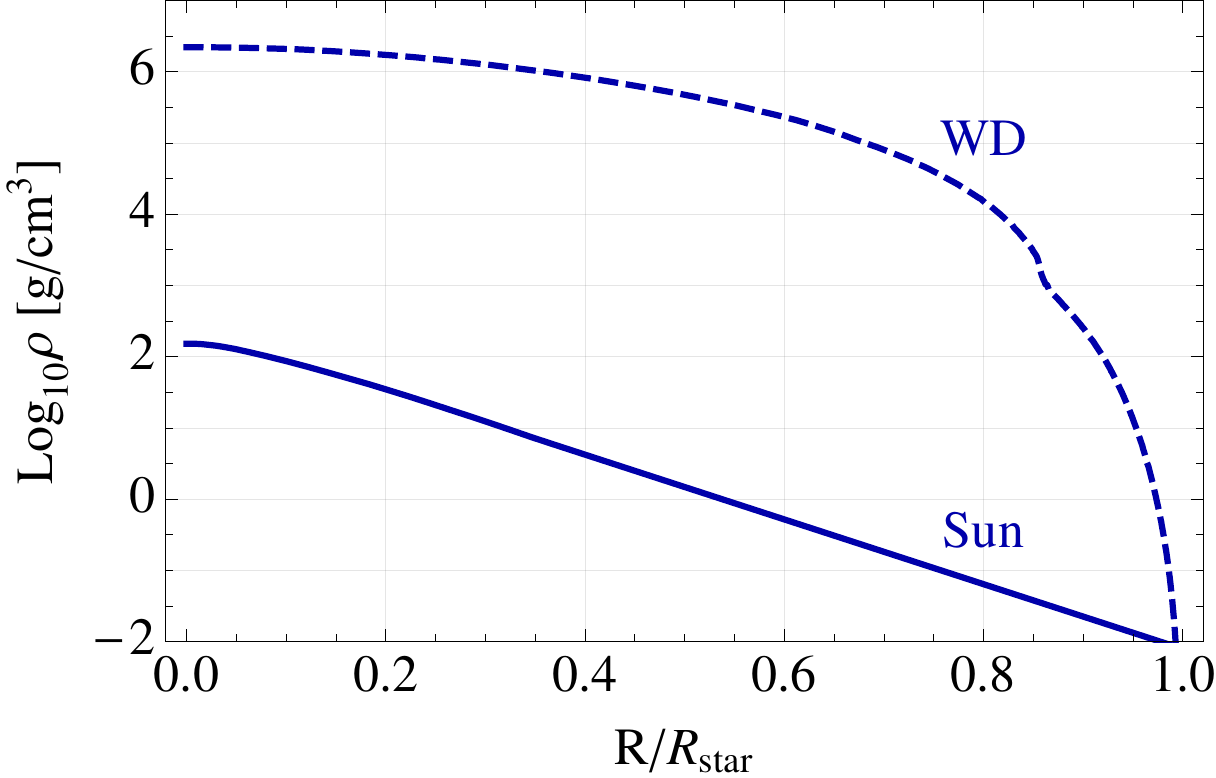}\quad
\includegraphics[width=6cm]{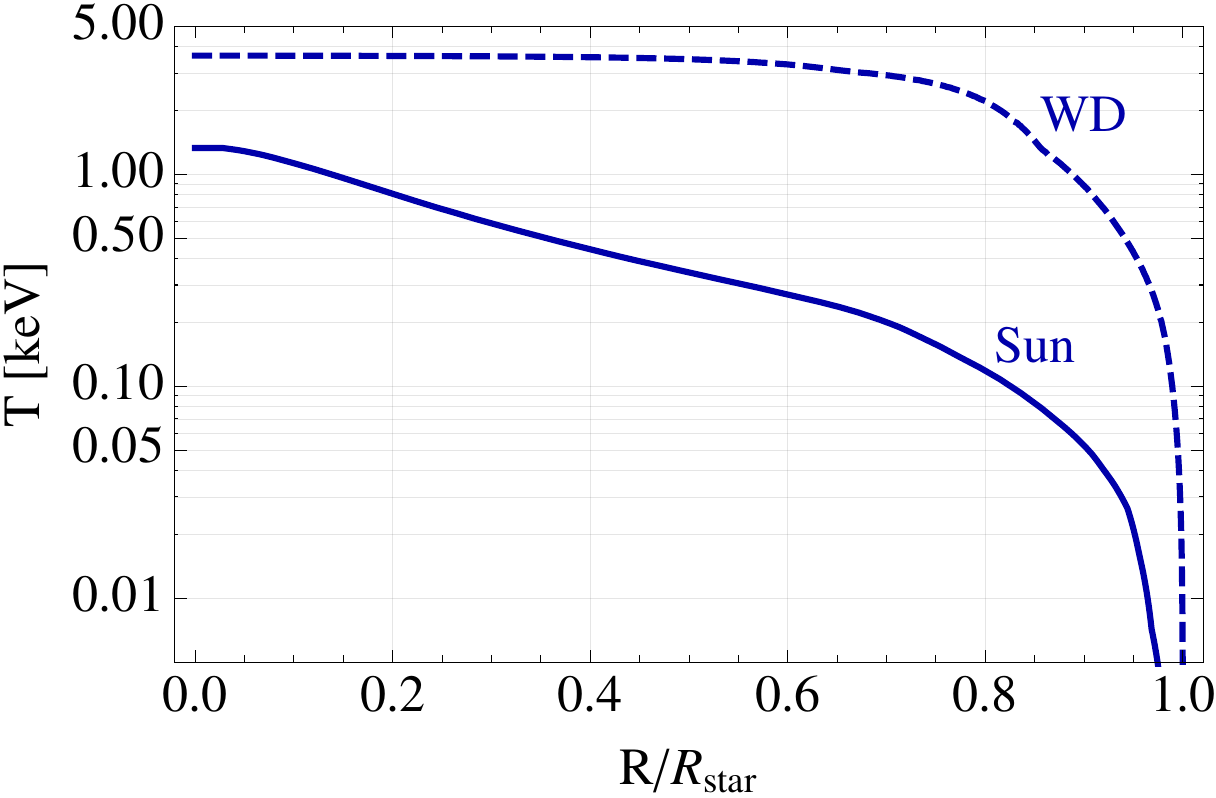}\\
\includegraphics[width=6cm]{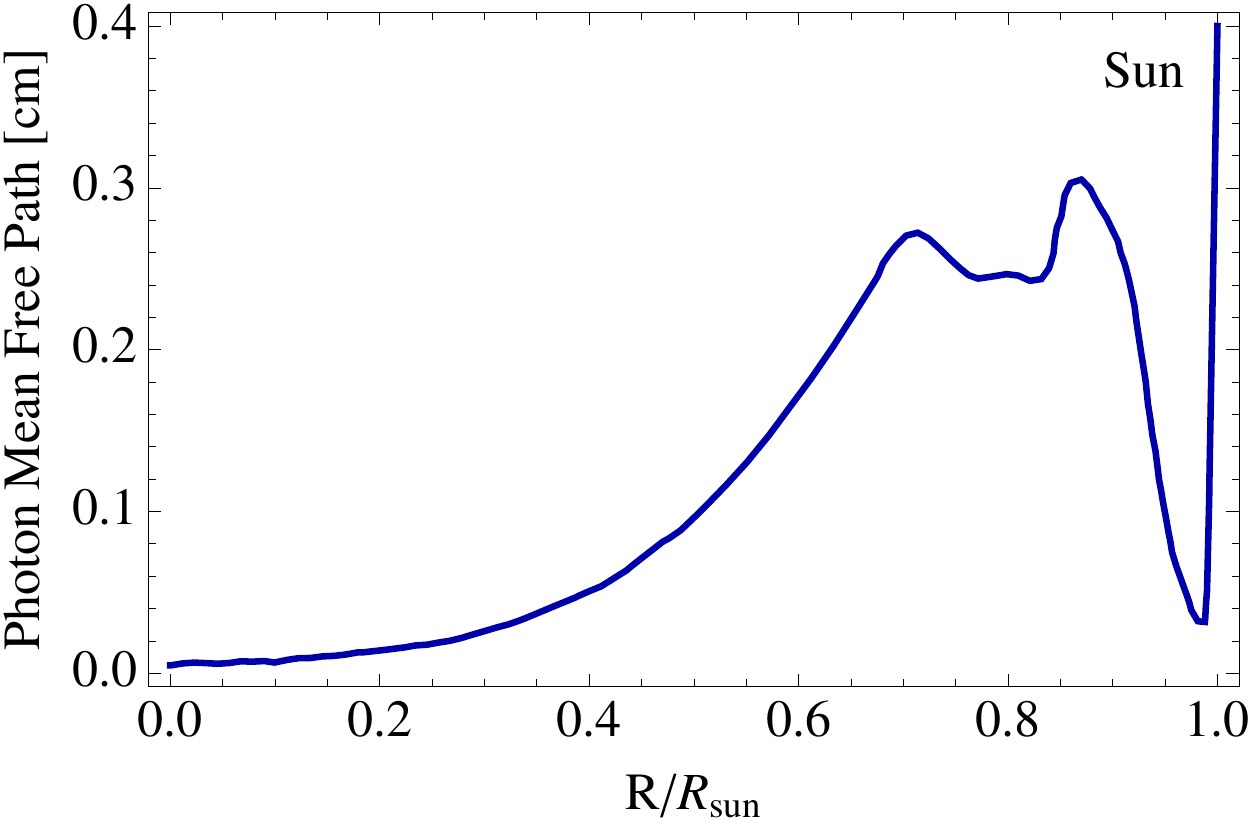}\quad
\includegraphics[width=6cm]{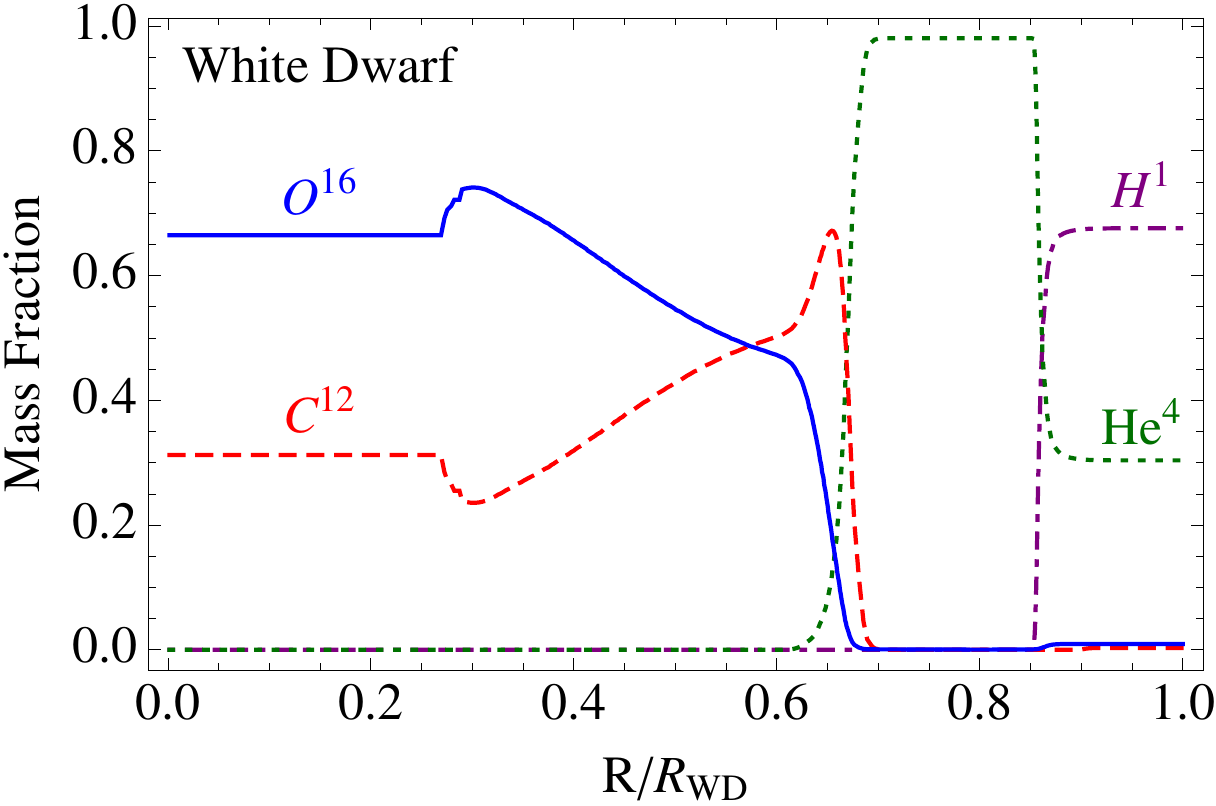}\\
\includegraphics[width=6cm]{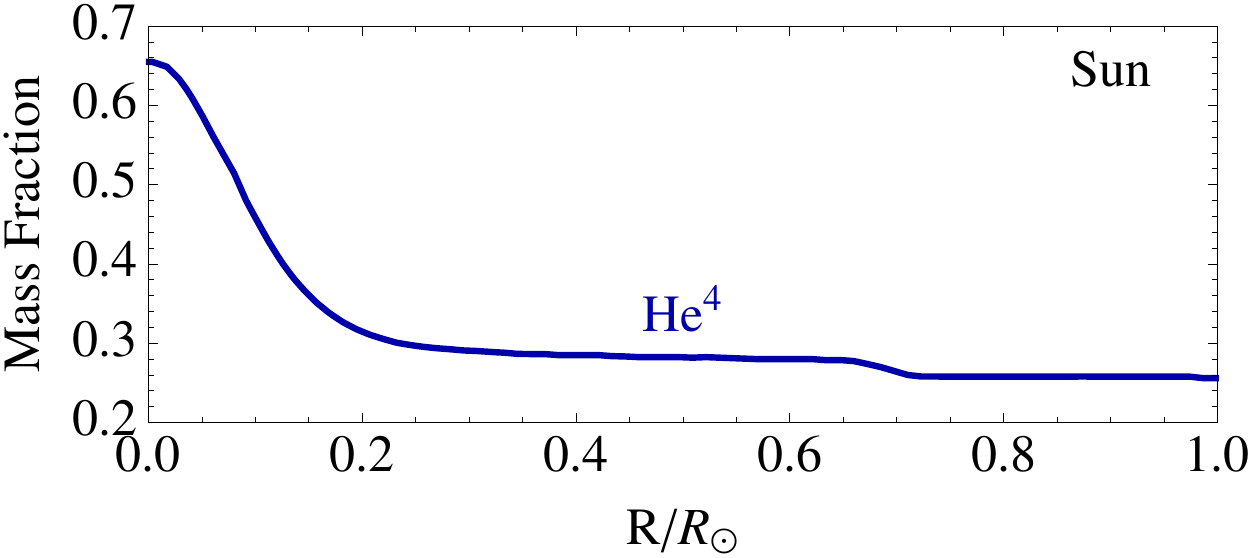}\\
\end{center}
\caption{
Radial profiles used in our solar and white dwarf cooling calculations. Source: $T_\mathrm{sun}(R), \rho_\mathrm{sun}(R)$, solar mass fraction \cite{carrollostlie}; Photon mean free path in sun from standard solar model, Guenther et al. (1992) \cite{photonMFPsun}. The profiles for our benchmark white dwarf were produced with the \texttt{MESA} code \cite{mesa} by Max Katz.}
\label{f.solarprofiles}
\end{figure}

The solar energy loss and radiative heat transfer calculation in  \sssref{solarcooling} makes use of standard radial profiles for temperature, density and composition of the sun, shown in \fref{solarprofiles}. These can be found in basic astrophysics textbooks like  \cite{carrollostlie}. The radius and power output of the sun are $R_\mathrm{sun} \approx 3.85 \times 10^{26}$ cm and $P_\mathrm{sun} \approx 3.85 \times 10^{26}$ Watts.

For the white dwarf cooling calculation in \sssref{whitedwarfcooling} we gratefully acknowledge the help of Max Katz, who simulated the evolution of an approximately one solar mass sun-like star from the main sequence to a very old white dwarf using the \texttt{MESA} stellar evolution code \cite{mesa}. The mass of this white dwarf, $\approx 0.5$ solar masses, is representative of the majority of white dwarfs in the luminosity function dataset \cite{raffelt1996, DeGennaro:2007yw}. The luminosity (power output in solar units) and core temperature of this dwarf over time are shown in \fref{WDevolution}, along with the relationship between core temperature and power output.

Radial density, temperature and composition profiles for this white dwarf when its power output was $P_\mathrm{WD}/P_\mathrm{sun} \approx 0.1$ are shown in \fref{solarprofiles}. This point in the evolution of the dwarf marks the start of dominant photon cooling \cite{Raffelt:1988rx}, which we compare to $\phi$ emission in \sssref{whitedwarfcooling}.

\begin{figure}
\begin{center}
\includegraphics[width=4.9cm]{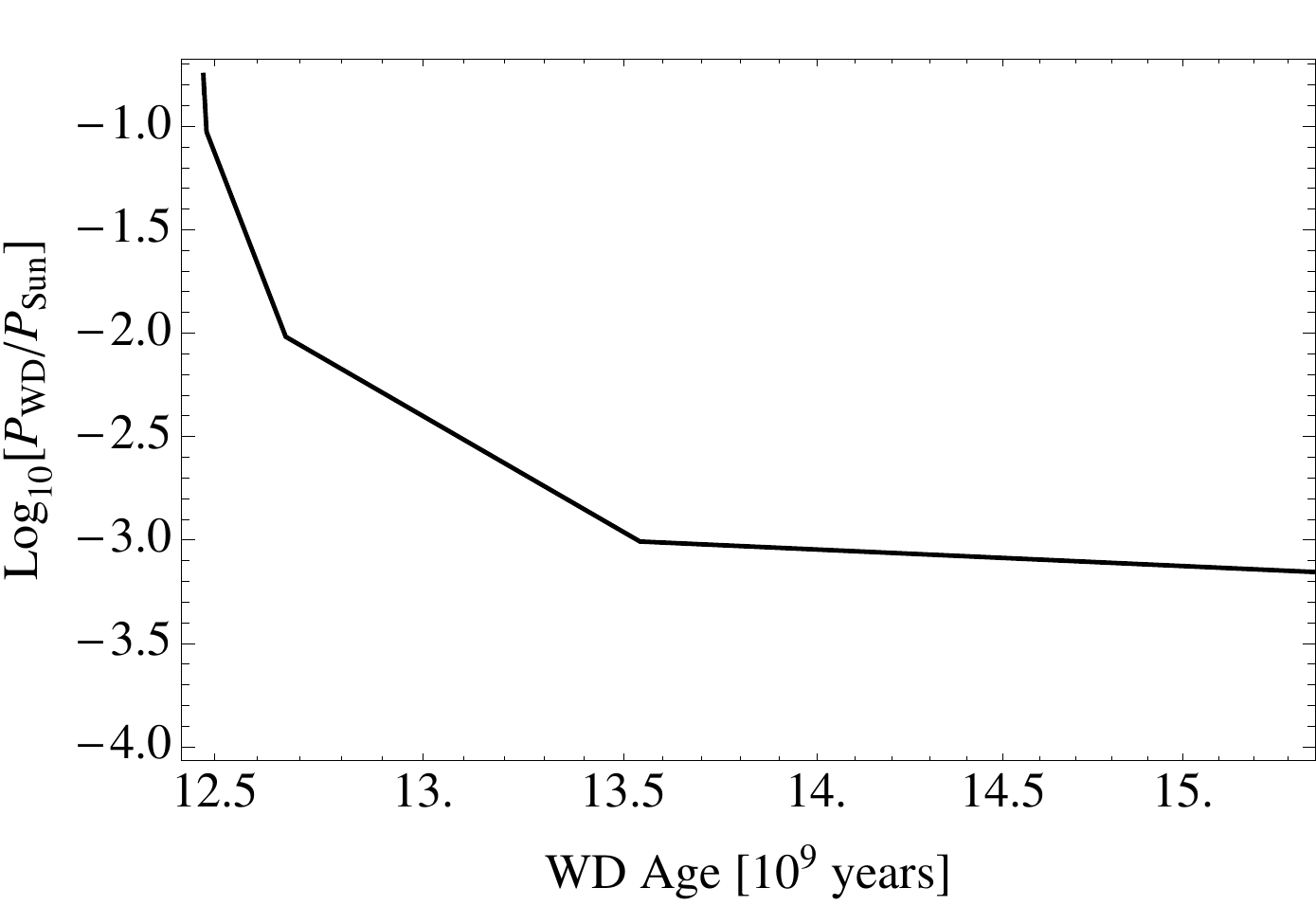}\quad
\includegraphics[width=4.9cm]{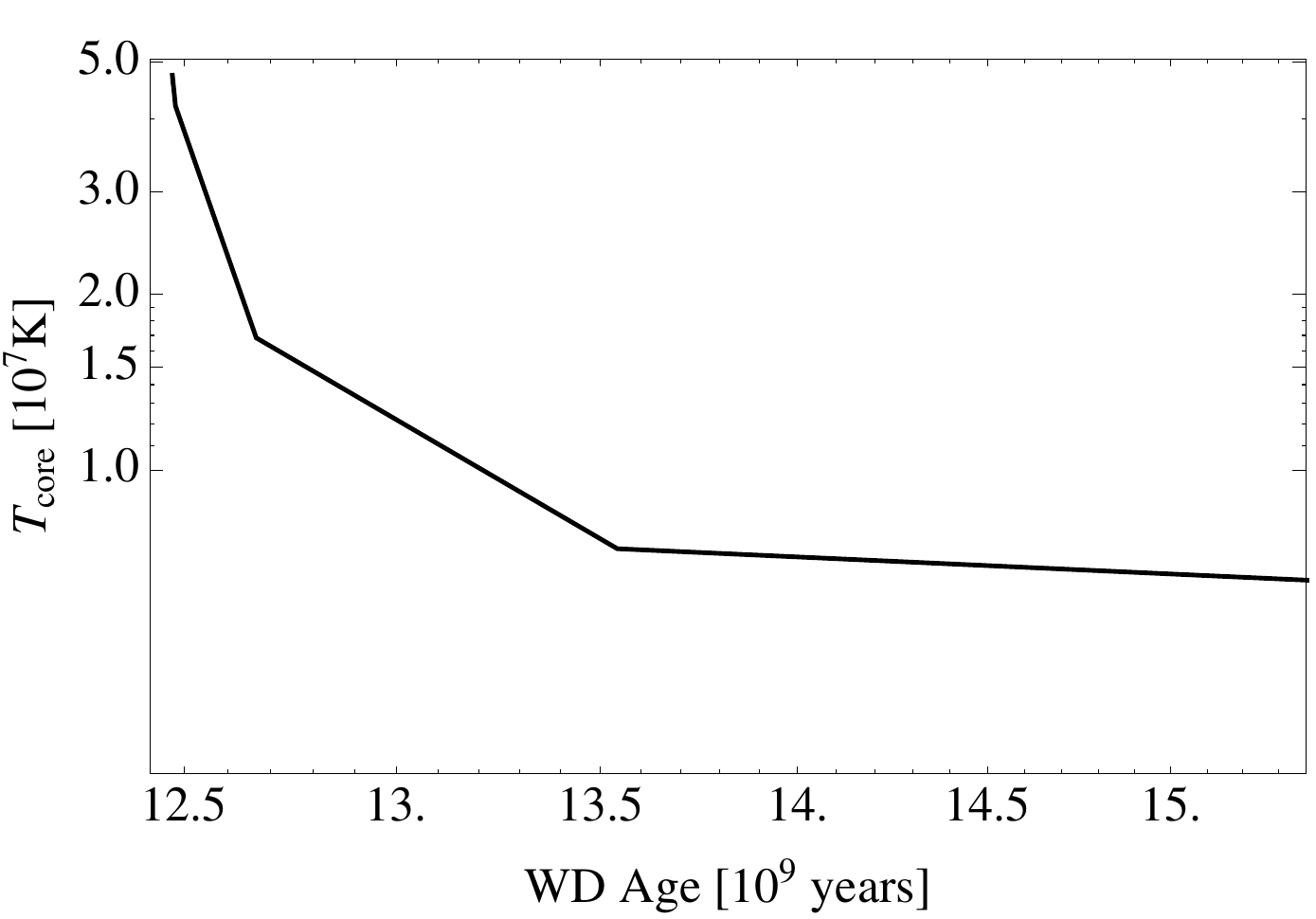}\quad
\includegraphics[width=4.9cm]{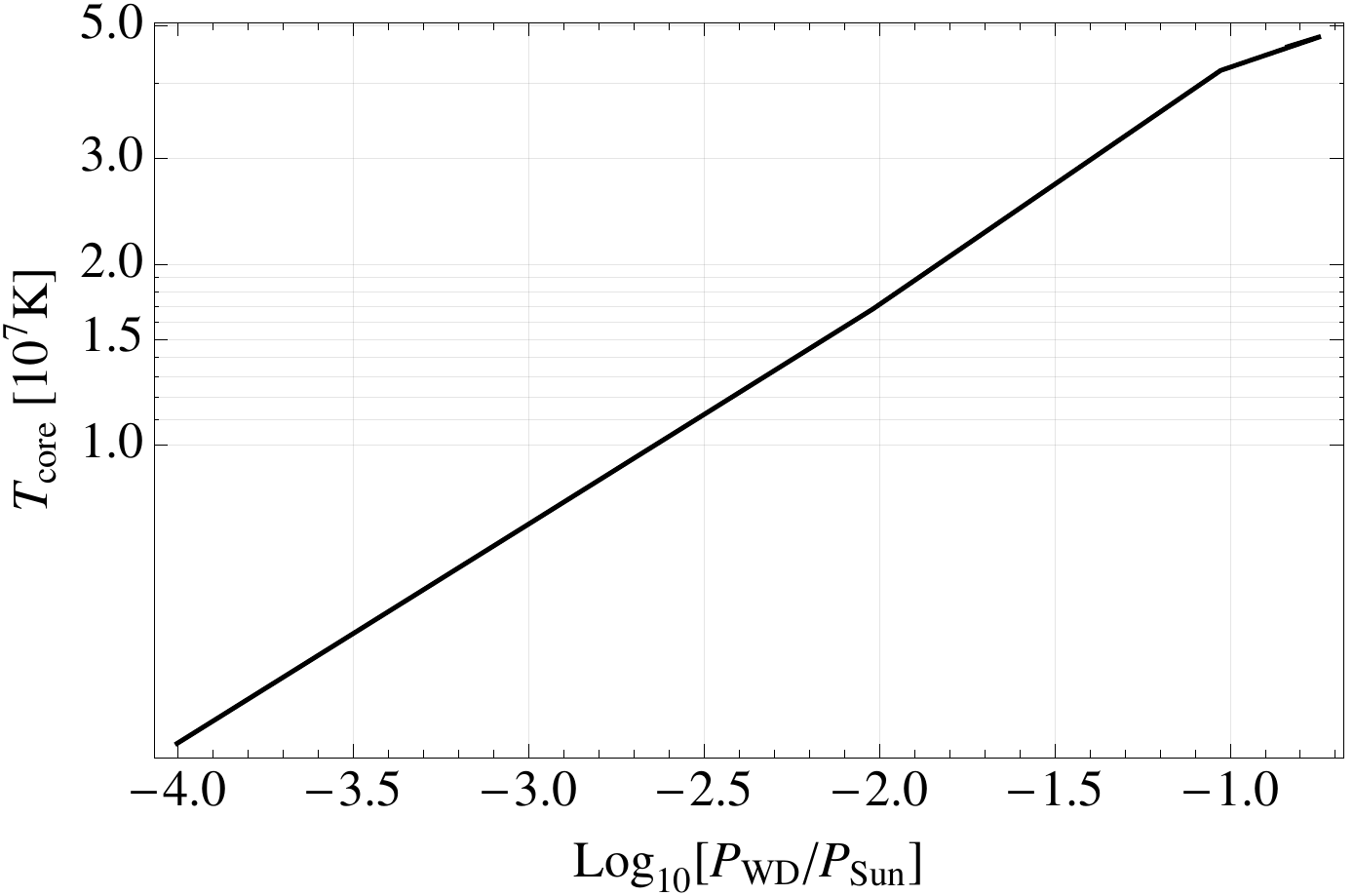}
\end{center}
\caption{
Top two plots: Evolution of power output and core temperature for the one solar mass white dwarf simulated by Max Katz using \texttt{MESA} \cite{mesa}. Bottom: relationship between power output an core temperature.
}
\label{f.WDevolution}
\end{figure}

\section{Checking other constraints on $\phi$}
\label{a.otherconstraints}

Here we briefly demonstrate that fixed target experiments, precision measurements bounds, and indirect detection do not constrain dark mediators. 

\subsection{Fixed target experiments}

The fixed target experiments 
MINOS ($E_p = 120 \gev$) \cite{Barr:2013wta, Adamson:2012gt}, T2K ($E_p = 30 \gev$) \cite{Ishida:2013kba, Abe:2011ks, Fukuda:2002uc}, MiniBooNE ($E_p = 8.9 \gev$) \cite{Aguilar-Arevalo:2013pmq} and LSND ($E_p = 0.8 \gev$) \cite{Mills:2001tq} bombarded graphite or beryllium targets with $\sim 10^{20} - 10^{23}$ protons-on-target ($N_\mathrm{pot}$).
Dark mediators can be produced in these collisions via the process $p p \rightarrow p p \phi \phi$, which has cross section $\sigma_\mathrm{produce} \sim 10^{-5}$ to $10^{-3}$ pb for $\Lambda = 10 \tev$, depending on $E_p$. (This was computed in \texttt{MadGraph5}.) We can estimate whether these experiments are sensitive to $\phi$ production with this cross section.

Hitting a target of some length and proton number density $L_\mathrm{target}, n_\mathrm{target}$ with $N_\mathrm{POT}$ protons produces the following number of $\phi$'s:
\begin{equation}
N_\phi^\mathrm{produced} = N_\mathrm{POT} n_\mathrm{target} \sigma_\mathrm{produce} L_\mathrm{target}.
\end{equation}
For $\Lambda = 10 \tev$, the interaction cross section of a high-energy $\phi$ with stationary protons is $\sigma_{p \phi \rightarrow p \phi} \sim 0.1 $ pb over the relevant range of proton energy $E_p$. This makes the mean free path of a high-energy $\phi$ in a typical material with densities $\sim \ \gram \ \cmmthree$ about  $\sim 10^{10}$ meters, so we can assume all $\phi$'s leave the target. The chance that a single $\phi$ is detected is
\begin{equation}
P_\phi^\mathrm{detect} = \epsilon n_\mathrm{detector} \sigma_\mathrm{detect} L_\mathrm{detector}
\end{equation}
where $\epsilon$ (likely $\ll1$) is some efficiency factor to account for the finite $\phi$-beam width as well as likelihood of detecting the target nuclear recoil. $n_\mathrm{detect}$ and $L_\mathrm{detect}$ are the number densities and lengths of the detector material, and $\sigma_\mathrm{detect} \sim \sigma_{p \phi \rightarrow p \phi}$ is the interaction cross section of $\phi$ with the target material. Therefore, the number of $\phi$'s detected by this experiment is
\begin{eqnarray}
\nonumber N_\phi^\mathrm{detected} &\sim& N_\mathrm{POT} n_\mathrm{target} \sigma_\mathrm{produce} L_\mathrm{target} \\ 
&&  \epsilon n_\mathrm{detector} \sigma_\mathrm{detect} L_\mathrm{detector}
\end{eqnarray}
Substituting graphite and iron densities for target and detector respectively, as well as the production cross section at  $E_p = 150 \gev$ and $\epsilon = 1$ to \emph{strongly overestimate} $\phi$ detection, we get
\begin{equation}
\frac{N_\phi^\mathrm{detected}}{10^{-6}} \sim \left(\frac{N_\mathrm{POT}}{10^{21}}\right) \left(\frac{10 \tev}{\Lambda}\right)^4 \left(\frac{L_\mathrm{target} L_\mathrm{detector}}{\mathrm{meter}^2}\right) 
\end{equation}
Typical physical dimensions for target and detector are $\mathcal{O}(1 - 10 \mathrm{m})$. Therefore, fixed target experiments have no sensitivity to $\phi$ production for $\Lambda \gtrsim 10 \tev$.


\subsection{Precision Measurement Bounds}
\label{ss.ewpobounds}
Since  $\phi \phi^*$ has potentially sizable coupling to the scalar quark current it could contribute to meson decays and the invisible $Z$-width. We show below that no meaningful constraints are derived from these processes. However, the heavy vector-like quarks in the UV-completion do contribute to the $S$-parameter, which bounds their Yukawa coupling to the Higgs.

\begin{figure}
\centering
		\begin{tikzpicture}[line width=1.5 pt, scale=1.2] 
		\draw[color=white] (-1.7,-1.7) rectangle (1.7,1.7);
		\draw[fermion] (-1.5,0) -- (-1.75,0);
		\draw[fermionnoarrow] (-1.5,0) -- (-1.0,0);
		\draw[fermionbar] (1,0) -- (1.75,0);
		\draw[fermionnoarrow] (-1,0) arc (180:135:1);
		\draw[fermionnoarrow] (135:1) arc (135:45:1);
		\draw[fermionnoarrow] (45:1) arc (45:0:1); 
		\draw[vector] (1,0) arc (0:-180:1);
		\draw[vector] (40:1) -- (40:2);
		\draw[fermion] (30:2.5) -- (40:2);
		\draw[fermion] (40:2) -- (50:2.5);
		%
		\begin{scope}[rotate=180]
		\begin{scope}[shift={(1.3,0)}] 
			\clip (0,0) circle (.175cm);
			\draw[fermionnoarrow] (-1,1) -- (1,-1);
			\draw[fermionnoarrow] (1,1) -- (-1,-1);
		\end{scope}	
		\end{scope}
		\begin{scope}[rotate=135]
		\begin{scope}[shift={(1,0)}] 
			\clip (0,0) circle (.175cm);
			\draw[fermionnoarrow] (-1,1) -- (1,-1);
			\draw[fermionnoarrow] (1,1) -- (-1,-1);
		\end{scope}	
		\end{scope}
		\begin{scope}[rotate=70]
		\begin{scope}[shift={(1,0)}] 
			\clip (0,0) circle (.175cm);
			\draw[fermionnoarrow] (-1,1) -- (1,-1);
			\draw[fermionnoarrow] (1,1) -- (-1,-1);
		\end{scope}	
		\end{scope}
				\node at (0,-1.4) {$W$};
		\node at (1.8,2.0) {$\nu$};
		\node at (2.2,1.5) {$\bar{\nu}$};
		\node at (22:1.2) {$t_L$};
		\node at (100:1.2) {$t_R$};
		\node at (160:1.20) {$t_L$};
		\node at (-1.5,-.35) {$s_R$};
		\node at (1.5,-.35) {$d_L$};
		\node at (135:.6) {$m_t$};
		\node at (60:.6) {$m_t$};
	\end{tikzpicture}
	\qquad\qquad
		\begin{tikzpicture}[line width=1.5 pt, scale=1.2] 
		\draw[color=white] (-1.7,-1.7) rectangle (1.7,1.7);
		\draw[fermion] (-1.5,0) -- (-1.75,0);
		\draw[fermionnoarrow] (-1.5,0) -- (-1.0,0);
		\draw[fermionbar] (1,0) -- (1.75,0);
		\draw[fermionnoarrow] (-1,0) arc (180:135:1);
		\draw[fermionnoarrow] (135:1) arc (135:45:1);
		\draw[fermionnoarrow] (45:1) arc (45:0:1); 
		\draw[vector] (1,0) arc (0:-180:1);
		\draw[scalar] (30:2.5) -- (40:1);
		\draw[scalar] (40:1) -- (50:2.5);
		%
		\begin{scope}[rotate=180]
		\begin{scope}[shift={(1.3,0)}] 
			\clip (0,0) circle (.175cm);
			\draw[fermionnoarrow] (-1,1) -- (1,-1);
			\draw[fermionnoarrow] (1,1) -- (-1,-1);
		\end{scope}	
		\end{scope}
		\begin{scope}[rotate=135]
		\begin{scope}[shift={(1,0)}] 
			\clip (0,0) circle (.175cm);
			\draw[fermionnoarrow] (-1,1) -- (1,-1);
			\draw[fermionnoarrow] (1,1) -- (-1,-1);
		\end{scope}	
		\end{scope}
				\node at (0,-1.4) {$W$};
		\node at (1.8,2.0) {$\phi$};
		\node at (2.2,1.5) {$\phi^*$};
		\node at (22:1.2) {$t_L$};
		\node at (100:1.2) {$t_R$};
		\node at (160:1.2) {$t_L$};
		\node at (-1.5,-.35) {$s_R$};
		\node at (1.5,-.35) {$d_L$};
		\node at (135:.6) {$m_t$};
		\node at (40:.6) {$\Lambda^{-1}$};
	\end{tikzpicture}
  \caption{One of the penguins in the $K^+\to\pi^+\nu\,\nu$ and the corresponding $K^+\to\pi^+\phi\,\phi$ process.}
  \label{f.kaondecay} 
\end{figure}
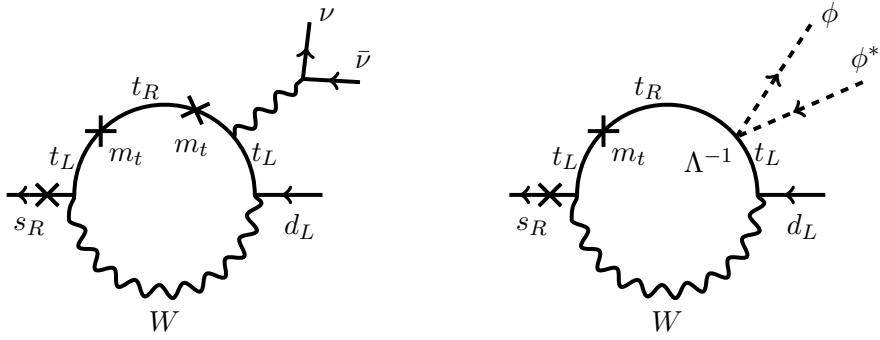

\subsubsection{Bound from the meson decays} 
Pions are the lightest QCD pseudoscalars  and cannot decay to $\phi \phi^*$, but Kaon decay with $\phi \phi^*$ in the final state is possible. The measurement \cite{Beringer:1900zz} 
\begin{equation} 
Br(K^+\to\pi^+\nu\bar{\nu})_{\rm{exp}}=17.3^{+11.5}_{-10.5}\cdot 10^{-11},
\end{equation}
is quite close to the SM value \cite{Brod:2010mj}
\begin{equation} 
Br(K^+\to\pi^+\nu\bar{\nu})_{\rm{SM}}=(8.5\pm 0.7)\cdot 10^{-11}.
\end{equation}
Since $K^+\to \pi^+\phi\phi$ gives the same signal as the neutrino process, the same measurement sets a lower bound on the suppression scale $\Lambda$. The dominant diagrams involve a $W$-up-quark loop which suffers from a GIM mechanism and requires up-type mass, dominated by the top. 
The leading order SM result is given by  \cite{Buras:2004uu} and we can use it to estimate the dmDM matrix element:
\begin{equation}
\frac{|\mathcal{M}_{\phi\phi}|}{|\mathcal{M}_{\rm{SM}}|}\sim\frac{m_Z^2}{g_{Zq}\,g_{Z\nu}\,m_t\,\Lambda}.
\end{equation} 
This is clear from \fref{kaondecay} by chirality and dimensionality arguments. For $\Lambda = 10 \tev$ the ratio is about $0.03$, much less than the current experimental precision. Kaon decay therefore supplies no meaningful dmDM bounds.

\subsubsection{Electroweak Precision Measurement} 
As explained in \ssref{uvcompletion}, the $\bar q q \phi \phi^*/\Lambda$ operator is most plausibly generated by a single generation of vector-like quarks. Since these quarks have $SU(2)_L \times U(1)_Y$ charge they contribute to the oblique parameters $S, T, U$ \cite{Peskin:1991sw}. The resulting constraints on such a Top Partner Doublet model have been computed by \cite{Dawson:2012di}. For $M_Q \approx 1 \ \tev$, the mass splitting between the up and down type vector-like quarks must be less than about $10 \ \gev$, which sets a strong bound on the flavor structure. However, for flavor-diagonal couplings there are no constraints.

\subsubsection{The $Z\,\phi\,\phi$ coupling} 

The properties of the $Z$ boson are extremely well measured. The $|\phi|^2\bar{Q}q/\Lambda$ operator contributes to the invisible $Z$-width. Of all the electroweak precision constraints, this contribution gives the strongest bound on dmDM.

Assuming flavor-diagonal $|\phi|^2$ coupling to quarks in the SM mass basis, the dominant contribution to $Z\to\phi \phi^*$ comes from a top loop with a single mass insertion. This can be expressed as an effective $Z\phi\phi^*$ coupling
\begin{eqnarray}
\nonumber g_{\phi}&=&\frac{3\, g_{L-R}\,m_t}{8\,\pi^2\,\Lambda}\left(\ln\frac{\Lambda^2}{m_t^2}+\mathcal{O}\left(\frac{m_t^2}{\Lambda^2}\right)\right)\\
&\simeq& 10^{-3}\,\left(\frac{10\,\rm{TeV}}{\Lambda}\right),\label{e.zphiphi}
\end{eqnarray}
where $g_{L-R}\simeq 0.2$ is the difference between the $Z$ coupling of the left- and right-handed tops. The resulting partial width of $Z\to \phi\,\phi$ is 
\begin{eqnarray}
\Gamma_{Z\to\phi\phi}&\simeq& \frac{m_Z}{8\,\pi}\left[\frac{3\,g\,m_t}{8\,\pi^2\,\Lambda}\,\ln\frac{\Lambda^2}{m_t^2}\right]^2\\
\nonumber &\simeq& 4\times10^{-3}\left(\frac{10\,\rm{TeV}}{\Lambda}\right)^2\,\rm{MeV}.
\end{eqnarray}
This is much smaller than the current precision of $\Gamma(Z\to \mathrm{invisible})=499.0\pm 1.5$ MeV \cite{Beringer:1900zz}, meaning electroweak precision constraints supply no meaningful bounds on dmDM.

\subsection{Indirect Detection}
\label{ss.indirectdetection}
A potential indirect detection signal may arise from the annihilation process $\overline{\chi^c}\chi\to q\bar q\phi^*$ (assuming the $2\rightarrow2$ loop process $\bar \chi \chi \rightarrow \bar q q$ is suppressed like in direct detection). The total annihilation cross section is found using \texttt{MadGraph5} to be
\begin{eqnarray}
   (\sigma v)_\mathrm{ann} &=&  \left(1.5 \times 10^{-40} \, {\rm cm}^3 \mathrm{s}^{-1} \right) \ \times\\
   \nonumber && \left(\frac{y_\chi}{0.05}  \frac{9 \tev}{\Lambda}\right)^2 \times \left( \frac{v}{35 \mathrm{km/s}}\right)^2
\end{eqnarray}
This result is independent of the DM mass and similar to the behavior of the usual $2\to2$ annihilation of Dirac DM via a scalar in the $s$-channel. 

The Fermi-LAT collaboration~\cite{Ackermann:2013yva} reports dwarf galaxy bounds\footnote{For these objects there is an independent handle on the local DM densities by means of measuring the peculiar velocity, thereby significantly reducing dependence on any particular halo model compared to the galactic center.} on $(\sigma_{\chi \chi \rightarrow q q} v)_\mathrm{ann}$  between $\sim10^{-26}$ and 
$10^{-25}$ ${\rm cm}^3 \mathrm{s}^{-1}$ for $m_\chi$ between 2 GeV and 100 GeV. The dmDM annihilation cross section is many orders of magnitude smaller. Furthermore, the Fermi bounds assume $2\rightarrow2$ annihilation, resulting in a monochromatic quark spectrum $E_q = m_\chi$ before hadronization and decay. In dmDM the spectrum is triangularly rising towards  $E_q = m_\chi$, which is harder to detect. It is therefore clear that dmDM leaves no detectable signal in the gamma ray sky.

\section{Computing direct detection signals}
\label{a.ddcomputation}

\subsection{Dark matter speed distribution}
The velocity of thermalized cold dark matter in the \emph{halo frame} has an approximate Maxwell-Bolzmann distribution:
\begin{equation}
f(\vec v_H) = \left\{ 
\begin{array}{ll}
\left( \frac{1}{\pi v_0}\right)^{3/2} e^{- |v_H|^2/v_0^2} & \mathrm{for} \  |\vec v_H| \leq v_{esc}
\\
0 &\mathrm{for} \  |\vec v_H| > v_{esc},
\end{array}
\right.
\end{equation}
where $v_0 \approx 220$ km/s and $v_{esc} \approx 544$ km/s are the temperature of the distribution and the galactic escape speed\footnote{We implement the galactic escape speed with a hard cutoff, which is not entirely realistic, but the effect of $v_{esc} < \infty$ is very small in the earth frame so this is sufficient for our purposes.}  \cite{Smith:2006ym}. Transforming this distribution to the earth frame moving at an average velocity of $v_e \approx 233$ km/s through the galaxy gives the relevant speed distribution for our calculations.
\begin{eqnarray}
\label{e.dmspeeddistrib}
f(v) &=& \frac{1}{\eta} \ \frac{v}{\sqrt{\pi} v_0 v_e} \ e^{-(v^2 + v_e^2)/v_0^2} \\
\nonumber && \left( e^{2 v v_e/v_0^2} - e^{2 \cos \phi_{max} v v_e/v_0^2} \right),
\end{eqnarray}
where $\eta$ is a normalization factor,
\begin{equation}
\eta = \mathrm{Erf}\left( \frac{v_{esc}}{v_0}\right) - \frac{2 v_{esc}}{\sqrt{\pi} v_0} e^{-v_{esc}^2/v_0^2},
\end{equation}
and $\cos \phi_{max}$ as a function of $v$ is given by
\begin{equation}
\cos \phi_{max} = \left\{
\begin{array}{ll}
1 & \mathrm{for} \ v \leq v_{esc} - v_e \\
-1 & \mathrm{for} \ v \geq v_{esc} + v_e\\
\displaystyle \frac{v_{esc}^2 - v_e^2 - v^2}{2 v v_e} & \mathrm{otherwise}
\end{array}
 \right. .
\end{equation}

\subsection{Computing experimental observables}

\subsubsection{CDMS II Silicon}

The CDMS II Silicon detectors ($m_N = 38 \gev$) have accumulated 140.2 kg$\cdot$days of expsure and use simultaneous measurement of ionization and non-equilibrium phonons to measure nuclear recoil and distinguish from electron recoil background. The recoil cutoff is 7 KeV (higher for some sub-detectors), which together with the fiducial volume and phonon timing cuts (to eliminate background) results in the WIMP-nucleon scattering efficiency curve shown in \cite{Agnese:2013rvf}. To obtain experimental predictions for recoil spectra we simply multiply each recoil bin by this efficiency, which is about 20\% for $E_r = 10 \kev$ and asymptotes to about $40\%$ at $30 \kev$.

\subsubsection{XENON100 and LUX}

Xenon100 \cite{Aprile:2012nq} and LUX \cite{Akerib:2013tjd} ($m_N = 131 \gev$) have accumulated exposures of 7636 and 10065 kg$\cdot$days and detect nuclear recoil with two experimental signals: the number of produced scintillation photons ($S1$) and the charge signal once the ionization travels up to the gaseous phase ($S2$). Nuclear recoils can be distinguished from electron recoil backgrounds using the time difference between the $S1$ and $S2$ signals, as well as their ratio $S2/S1$. 

XENON100 has a light gathering efficiency of about 6\% for S1 photons. The \emph{expected} number of scintillation photons for a given nuclear recoil event is  \cite{Aprile:2012nq} 
\begin{equation}
\label{e.expectedS1}
\langle S1 \rangle = E_r  \ \times \  \frac{S_{nr} L_y}{S_{ee}} \mathcal{L}_{eff}(E_r),
\end{equation}
where $S_{ee} = 0.58$, $S_{nr} = 0.95$ and $L_y = (2.28 \pm 0.04)$ (photo electrons)/(keV$_{ee}$). The relative scintillation efficiency $\mathcal{L}_{eff}$ must be extracted from experiment. The fit used by the collaboration can be found in \cite{Aprile:2011hi}. Due to limitations of the $\mathcal{L}_{eff}$ measurement, no DM signal below $E_r = 3 \kev$ is included. This makes the bounds conservative.
The same applies for LUX, appropriately rescaled to account for the highter 14\% light gathering efficiency.

The expected S1 spectrum can then be computed from the expected nuclear recoil spectrum,
\begin{equation}
\frac{dN}{d S1} = \int d E_r  \ \frac{dN}{d E_r}  \ \mathrm{Poi}(S1,\langle S1 \rangle),
\end{equation}
where $\langle S1 \rangle$ is a function of $E_r$ as per \eref{expectedS1}.
Once an S1 signal has been detected it must pass selection cuts. The probability of a WIMP signal passing these cuts is S1-dependent and asymptotes to about 0.8  for $S1 \gtrsim 5$ at XENON100 and about 1 for $S1 \gtrsim 2$ at LUX. The XENON100 signal region is $S1 \in (3, 30)$, while for LUX it is $S1 \in (2, 30)$. 

\vspace{2mm}
\subsubsection{CDMSlite}

CDMSlite \cite{Agnese:2013lua} was a light DM search using a single Super-CDMS iZIP detector operated at higher bias voltage to lower the nuclear recoil detection threshold to 0.84 keV at the cost of giving up background discrimination. The accumulated exposure is 6.18 kg$\cdot$days. Instead of computing the dmDM CDMSlite signal with experimental efficiencies, we transform the collaboration's DM bounds using the dmDM$\rightarrow$WIMP parameter map discussed in \ssref{ddconstraints}, having validated the method on XENON and CDMS-Si data.


\begin{thebibliography}{99}

\bibitem{Ade:2013zuv} 
  P.~A.~R.~Ade {\it et al.}  [Planck Collaboration],
  arXiv:1303.5076 [astro-ph.CO].

\bibitem{Jungman:1995df} 
  See, for example, G.~Jungman, M.~Kamionkowski and K.~Griest,
  Phys.\ Rept.\  {\bf 267}, 195 (1996)
  [hep-ph/9506380]. G.~Bertone and D.~Merritt,
  Mod.\ Phys.\ Lett.\ A {\bf 20}, 1021 (2005)
  [astro-ph/0504422].
 
\bibitem{Goodman:1984dc} 
  M.~W.~Goodman and E.~Witten,
  Phys.\ Rev.\ D {\bf 31}, 3059 (1985).

  
\bibitem{Akerib:2013tjd} 
  D.~S.~Akerib {\it et al.}  [LUX Collaboration],
  arXiv:1310.8214 [astro-ph.CO].


\bibitem{Agnese:2013rvf} 
  R.~Agnese {\it et al.}  [CDMS Collaboration],
  [arXiv:1304.4279 [hep-ex]].


\bibitem{Aprile:2012nq} 
  E.~Aprile {\it et al.}  [XENON100 Collaboration],
  Phys.\ Rev.\ Lett.\  {\bf 109}, 181301 (2012)
  [arXiv:1207.5988 [astro-ph.CO]].


\bibitem{Bernabei:2013cfa} 
  R.~Bernabei, P.~Belli, S.~d'Angelo, A.~Di Marco, F.~Montecchia, F.~Cappella, A.~d'Angelo and A.~Incicchitti {\it et al.},
  Int.\ J.\ Mod.\ Phys.\ A {\bf 28}, 1330022 (2013)
  [arXiv:1306.1411 [astro-ph.GA]].


\bibitem{Aalseth:2012if} 
  C.~E.~Aalseth {\it et al.}  [CoGeNT Collaboration],
  Phys.\ Rev.\ D {\bf 88}, 012002 (2013)
  [arXiv:1208.5737 [astro-ph.CO]].


\bibitem{Angloher:2011uu} 
  G.~Angloher, M.~Bauer, I.~Bavykina, A.~Bento, C.~Bucci, C.~Ciemniak, G.~Deuter and F.~von Feilitzsch {\it et al.},
  Eur.\ Phys.\ J.\ C {\bf 72}, 1971 (2012)
  [arXiv:1109.0702 [astro-ph.CO]].


\bibitem{Curtin:2013qsa} 
  D.~Curtin, Z.~Surujon and Y.~Tsai,
  arXiv:1312.2618 [hep-ph].
  
  
\bibitem{TuckerSmith:2001hy} 
  D.~Tucker-Smith and N.~Weiner,
  Phys.\ Rev.\ D {\bf 64}, 043502 (2001)
  [hep-ph/0101138].


\bibitem{Graham:2010ca} 
  P.~W.~Graham, R.~Harnik, S.~Rajendran and P.~Saraswat,
  Phys.\ Rev.\ D {\bf 82}, 063512 (2010)
  [arXiv:1004.0937 [hep-ph]].

\bibitem{Essig:2010ye} 
  R.~Essig, J.~Kaplan, P.~Schuster and N.~Toro,
  [arXiv:1004.0691 [hep-ph]].




\bibitem{MarchRussell:2012hi} 
  J.~March-Russell, J.~Unwin and S.~M.~West,
  JHEP {\bf 1208}, 029 (2012)
  [arXiv:1203.4854 [hep-ph]].
  
  
\bibitem{Chang:2009yt} 
  S.~Chang, A.~Pierce and N.~Weiner,
  JCAP {\bf 1001}, 006 (2010)
  [arXiv:0908.3192 [hep-ph]].


  \bibitem{Essig:2013lka} 
  R.~Essig, J.~A.~Jaros, W.~Wester, P.~H.~Adrian, S.~Andreas, T.~Averett, O.~Baker and B.~Batell {\it et al.},
  arXiv:1311.0029 [hep-ph].



\bibitem{Alves:2009nf} 
  D.~S.~M.~Alves, S.~R.~Behbahani, P.~Schuster and J.~G.~Wacker,
  Phys.\ Lett.\ B {\bf 692}, 323 (2010)
  [arXiv:0903.3945 [hep-ph]].


\bibitem{Kribs:2009fy} 
  G.~D.~Kribs, T.~S.~Roy, J.~Terning and K.~M.~Zurek,
  Phys.\ Rev.\ D {\bf 81}, 095001 (2010)
  [arXiv:0909.2034 [hep-ph]].
  
  
\bibitem{Lisanti:2009am} 
  M.~Lisanti and J.~G.~Wacker,
  Phys.\ Rev.\ D {\bf 82}, 055023 (2010)
  [arXiv:0911.4483 [hep-ph]].
  
  
\bibitem{Cline:2012bz} 
  J.~M.~Cline, A.~R.~Frey and G.~D.~Moore,
  Phys.\ Rev.\ D {\bf 86}, 115013 (2012)
  [arXiv:1208.2685 [hep-ph]].
  
  


\bibitem{Feldstein:2009tr} 
  B.~Feldstein, A.~L.~Fitzpatrick and E.~Katz,
  JCAP {\bf 1001}, 020 (2010)
  [arXiv:0908.2991 [hep-ph]].

\bibitem{Bai:2009cd} 
  Y.~Bai and P.~J.~Fox,
  JHEP {\bf 0911}, 052 (2009)
  [arXiv:0909.2900 [hep-ph]].


\bibitem{Carlson:1992fn} 
  E.~D.~Carlson, M.~E.~Machacek and L.~J.~Hall,
  Astrophys.\ J. {\bf 398}, 43 (1992)

\bibitem{Spergel:1999mh} 
  D.~N.~Spergel and P.~J.~Steinhardt,
  Phys.\ Rev.\ Lett.\  {\bf 84}, 3760 (2000)
  [astro-ph/9909386].
  
 
\bibitem{Tulin:2013teo} 
  S.~Tulin, H.~-B.~Yu and K.~M.~Zurek,
  Phys.\ Rev.\ D {\bf 87}, no. 11, 115007 (2013)
  [arXiv:1302.3898 [hep-ph]].
 
\bibitem{Feng:2011vu} 
  J.~L.~Feng, J.~Kumar, D.~Marfatia and D.~Sanford,
  Phys.\ Lett.\ B {\bf 703}, 124 (2011)
  [arXiv:1102.4331 [hep-ph]].


\bibitem{Kolb:1990vq} 
  E.~W.~Kolb and M.~S.~Turner,
  Front.\ Phys.\  {\bf 69}, 1 (1990).


\bibitem{Feng:2009mn} 
  J.~L.~Feng, M.~Kaplinghat, H.~Tu and H.~-B.~Yu,
  JCAP {\bf 0907}, 004 (2009)
  [arXiv:0905.3039 [hep-ph]].


\bibitem{Peter:2012jh} 
  A.~H.~G.~Peter, M.~Rocha, J.~S.~Bullock and M.~Kaplinghat,
  arXiv:1208.3026 [astro-ph.CO].
  
  
\bibitem{Brice:2013fwa} 
  S.~J.~Brice, R.~L.~Cooper, F.~DeJongh, A.~Empl, L.~M.~Garrison, A.~Hime, E.~Hungerford and T.~Kobilarcik {\it et al.},
  arXiv:1311.5958 [physics.ins-det].

  \bibitem{steigman}
  M.~S.~Turner, H.-S.~Kang, and G.~Steigman, (1989).
  
  
  
\bibitem{Volkov:2009mz} 
  M.~K.~Volkov, Y.~.M.~Bystritskiy and E.~A.~Kuraev,
  arXiv:0901.1981 [hep-ph].

\bibitem{CMS:2013gea} 
  CMS Collaboration [CMS Collaboration],
  CMS-PAS-SUS-13-012.

\bibitem{ATLAS:2012zim} 
  [ATLAS Collaboration],
  ATLAS-CONF-2012-147.  [CMS Collaboration],
  CMS-PAS-EXO-12-048. 
  
\bibitem{Alwall:2011uj} 
  J.~Alwall, M.~Herquet, F.~Maltoni, O.~Mattelaer and T.~Stelzer,
  JHEP {\bf 1106}, 128 (2011)
  [arXiv:1106.0522 [hep-ph]].

\bibitem{Sjostrand:2006za} 
  T.~Sjostrand, S.~Mrenna and P.~Z.~Skands,
  JHEP {\bf 0605}, 026 (2006)
  [hep-ph/0603175].

\bibitem{Wyman:2013lza} 
  M.~Wyman, D.~H.~Rudd, R.~A.~Vanderveld and W.~Hu,
  arXiv:1307.7715 [astro-ph.CO].

\bibitem{Cyr-Racine:2013fsa} 
  F.~Y.~Cyr-Racine, R.~de Putter, A.~Raccanelli and K.~Sigurdson,
  Phys.\ Rev.\ D {\bf 89}, 063517 (2014)
  [arXiv:1310.3278 [astro-ph.CO]].
 
\bibitem{Raffelt:2006cw} 
  G.~G.~Raffelt,
  Lect.\ Notes Phys.\  {\bf 741}, 51 (2008)
  [hep-ph/0611350].
  
  \bibitem{raffelt1996}
  Georg.~G.~Raffelt, "Stars as Laboratories for Fundamental Physics", University Chicago Press, 1996.
  
  
\bibitem{Raffelt:1988rx} 
  G.~G.~Raffelt and G.~D.~Starkman,
  Phys.\ Rev.\ D {\bf 40}, 942 (1989).
  

  \bibitem{Dreiner:2013tja} 
  H.~K.~Dreiner, J.~-F.~�o.~Fortin, J.~Isern and L.~Ubaldi,
  Phys.\ Rev.\ D {\bf 88}, 043517 (2013)
  [arXiv:1303.7232 [hep-ph]].
  
  
  
\bibitem{DeGennaro:2007yw} 
  S.~DeGennaro, T.~von Hippel, D.~E.~Winget, S.~O.~Kepler, A.~Nitta, D.~Koester and L.~Althaus,
  Astron.\ J.\  {\bf 135}, 1 (2008)
  [arXiv:0709.2190 [astro-ph]].
  
  
  
  
\bibitem{mesa} 
  B.~Paxton, L.~Bildsten, A.~Dotter, F.~Herwig, P.~Lesaffre and F.~Timmes,
  Astrophys.\ J.\ Suppl.\  {\bf 192}, 3 (2011)
  [arXiv:1009.1622 [astro-ph.SR]].
  
  
  
  
  
\bibitem{Mestel} 
  L.~Mestel,
  Mon Not R Astron Soc {\bf 112}, 583 (1952).



\bibitem{Harris:2005gd} 
  H.~C.~Harris, J.~A.~Munn, M.~Kilic, J.~Liebert, K.~A.~Williams, T.~von Hippel, S.~E.~Levine and D.~G.~Monet {\it et al.},
  Astron.\ J.\  {\bf 131}, 571 (2006)
  [astro-ph/0510820].
  
  \bibitem{Krzesinski}
  J.~Krzesinski, S.~J.~Kleinman, A.~Nitta, S.~ H\"ugelmeyer, S.~Dreizler, J.~Liebert and H.~Harris, A\&A \textbf{508}, 339 (2009), http://dx.doi.org/10.1051/0004-6361/200912094.
\bibitem{Yakovlev:2004iq} 
  D.~G.~Yakovlev and C.~J.~Pethick,
  Ann.\ Rev.\ Astron.\ Astrophys.\  {\bf 42}, 169 (2004)
  [astro-ph/0402143].
  
  
\bibitem{Yakovlev:2007vs} 
  D.~G.~Yakovlev, O.~Y.~Gnedin, A.~D.~Kaminker and A.~Y.~Potekhin,
  AIP Conf.\ Proc.\  {\bf 983}, 379 (2008)
  [arXiv:0710.2047 [astro-ph]].
  
  
\bibitem{Page:2005fq} 
  D.~Page, U.~Geppert and F.~Weber,
  Nucl.\ Phys.\ A {\bf 777}, 497 (2006)
  [astro-ph/0508056].
  
  
  
\bibitem{Page:2009fu} 
  D.~Page, J.~M.~Lattimer, M.~Prakash and A.~W.~Steiner,
  Astrophys.\ J.\  {\bf 707}, 1131 (2009)
  [arXiv:0906.1621 [astro-ph.SR]].
  
  
  
\bibitem{Yakovlev:2010ed} 
  D.~G.~Yakovlev, W.~C.~G.~Ho, P.~S.~Shternin, C.~O.~Heinke and A.~Y.~Potekhin,
  Mon.\ Not.\ Roy.\ Astron.\ Soc.\  {\bf 411}, 1977 (2011)
  [arXiv:1010.1154 [astro-ph.HE]].
  
  
  
  
\bibitem{Page:2010aw} 
  D.~Page, M.~Prakash, J.~M.~Lattimer and A.~W.~Steiner,
  Phys.\ Rev.\ Lett.\  {\bf 106}, 081101 (2011)
  [arXiv:1011.6142 [astro-ph.HE]].



\bibitem{Shternin:2010qi} 
  P.~S.~Shternin, D.~G.~Yakovlev, C.~O.~Heinke, W.~C.~G.~Ho and D.~J.~Patnaude,
  Mon.\ Not.\ Roy.\ Astron.\ Soc.\  {\bf 412}, L108 (2011)
  [arXiv:1012.0045 [astro-ph.SR]].
  
  
  
  
\bibitem{Potekhin:2011xe} 
  A.~Y.~Potekhin,
  Phys.\ Usp.\  {\bf 53}, 1235 (2010)
  [Usp.\ Fiz.\ Nauk {\bf 180}, 1279 (2010)]
  [arXiv:1102.5735 [astro-ph.SR]].
  
  
  
  
\bibitem{Page:2012se} 
  D.~Page,
  arXiv:1206.5011 [astro-ph.HE].
  
  
  
\bibitem{Elshamouty:2013nfa} 
  K.~G.~Elshamouty, C.~O.~Heinke, G.~R.~Sivakoff, W.~C.~G.~Ho, P.~S.~Shternin, D.~G.~Yakovlev, D.~J.~Patnaude and L.~David,
  Astrophys.\ J.\  {\bf 777}, 22 (2013)
  [arXiv:1306.3387 [astro-ph.HE]].
  
  
  
  

\bibitem{Demorest:2010bx} 
  P.~Demorest, T.~Pennucci, S.~Ransom, M.~Roberts and J.~Hessels,
  Nature {\bf 467}, 1081 (2010)
  [arXiv:1010.5788 [astro-ph.HE]].
  
  
\bibitem{Gundmundsson1}
  E.~H.~Gundmundsson, C.~J.~Pethick and R.~I.~Epstein,
  Astrophys.\  J. {\bf 259}, L19 (1982).


\bibitem{Gundmundsson2}
  E.~H.~Gundmundsson, C.~J.~Pethick and R.~I.~Epstein,
  Astrophys.\  J. {\bf 272}, 286 (1983).
  
  
  
  
  
\bibitem{Page:2004fy} 
  D.~Page, J.~M.~Lattimer, M.~Prakash and A.~W.~Steiner,
  Astrophys.\ J.\ Suppl.\  {\bf 155}, 623 (2004)
  [astro-ph/0403657].
  
  
  
  
  

\bibitem{Shapiro:1983du}
  S.~L.~Shapiro and S.~A.~Teukolsky,
{\it  New York, USA: Wiley (1983) 645 p}




\bibitem{Lattimer:1991ib} 
  J.~M.~Lattimer, M.~Prakash, C.~J.~Pethick and P.~Haensel,
  Phys.\ Rev.\ Lett.\  {\bf 66}, 2701 (1991).
  
  
  
  
\bibitem{Bertoni:2013bsa} 
  B.~Bertoni, A.~E.~Nelson and S.~Reddy,
  Phys.\ Rev.\ D {\bf 88}, 123505 (2013)
  [arXiv:1309.1721 [hep-ph]].
  
  
  
\bibitem{Kouvaris:2007ay} 
  C.~Kouvaris,
  Phys.\ Rev.\ D {\bf 77}, 023006 (2008)
  [arXiv:0708.2362 [astro-ph]].
  
  
  
  
\bibitem{Agnese:2013lua} 
  R.~Agnese, A.~J.~Anderson, M.~Asai, D.~Balakishiyeva, R.~B.~Thakur, D.~A.~Bauer, J.~Billard and A.~Borgland {\it et al.},
  arXiv:1309.3259 [physics.ins-det].
  
  
  
  
\bibitem{Ask:2012sm} 
  S.~Ask, N.~D.~Christensen, C.~Duhr, C.~Grojean, S.~Hoeche, K.~Matchev, O.~Mattelaer and S.~Mrenna {\it et al.},
  arXiv:1209.0297 [hep-ph].


\bibitem{Belanger:2008sj} 
  G.~Belanger, F.~Boudjema, A.~Pukhov and A.~Semenov,
  Comput.\ Phys.\ Commun.\  {\bf 180}, 747 (2009)
  [arXiv:0803.2360 [hep-ph]].

\bibitem{Crivellin:2013ipa} 
  A.~Crivellin, M.~Hoferichter and M.~Procura,
  Phys.\ Rev.\ D {\bf 89}, 054021 (2014)
  [arXiv:1312.4951 [hep-ph]].

\bibitem{Engel:1991wq} 
  J.~Engel,
  Phys.\ Lett.\ B {\bf 264}, 114 (1991).
  
  
\bibitem{Lewin:1995rx} 
  J.~D.~Lewin and P.~F.~Smith,
  Astropart.\ Phys.\  {\bf 6}, 87 (1996).
  
  
\bibitem{Barlow:1990vc} 
  R.~J.~Barlow,
  Nucl.\ Instrum.\ Meth.\ A {\bf 297}, 496 (1990).
  
  
  

\bibitem{Billard:2013qya} 
  J.~Billard, L.~Strigari and E.~Figueroa-Feliciano,
  arXiv:1307.5458 [hep-ph].

\bibitem{carrollostlie}
B.~Carroll and D.~Ostlie, 2006, ``An Introduction to Modern Astrophysics'' (2nd Edition), Boston, MA: Addison-Wesley.


\bibitem{photonMFPsun}
R.~Mitalas and K.R.~Sills, ApJ 401:759-760, 1992 December 20. 

  
  

\bibitem{Barr:2013wta} 
  G.~Barr [MINOS Collaboration],
  PoS ICHEP {\bf 2012}, 398 (2013).
  
\bibitem{Adamson:2012gt} 
  P.~Adamson {\it et al.}  [MINOS Collaboration],
  Phys.\ Rev.\ D {\bf 86}, 052007 (2012)
  [arXiv:1208.2915 [hep-ex]].
  
  
  
\bibitem{Ishida:2013kba} 
  T.~Ishida [group for the Hyper-Kamiokande working Collaboration],
  arXiv:1311.5287 [hep-ex].
  
\bibitem{Abe:2011ks} 
  K.~Abe {\it et al.}  [T2K Collaboration],
  Nucl.\ Instrum.\ Meth.\ A {\bf 659}, 106 (2011)
  [arXiv:1106.1238 [physics.ins-det]].
  







  \bibitem{Fukuda:2002uc} 
  Y.~Fukuda {\it et al.}  [Super-Kamiokande Collaboration],
  Nucl.\ Instrum.\ Meth.\ A {\bf 501}, 418 (2003).
  
  
  
\bibitem{Aguilar-Arevalo:2013pmq} 
  A.~A.~Aguilar-Arevalo {\it et al.}  [MiniBooNE Collaboration],
  Phys.\ Rev.\ Lett.\  {\bf 110}, 161801 (2013)
  [arXiv:1207.4809 [hep-ex], arXiv:1303.2588 [hep-ex]].
  
  
  
  
\bibitem{Mills:2001tq} 
  G.~B.~Mills [LSND Collaboration],
  Nucl.\ Phys.\ Proc.\ Suppl.\  {\bf 91}, 198 (2001).
  
  

\bibitem{Beringer:1900zz} 
  J.~Beringer {\it et al.}  [Particle Data Group Collaboration],
  Phys.\ Rev.\ D {\bf 86}, 010001 (2012).

\bibitem{Brod:2010mj} 
  J.~Brod and M.~Gorbahn,
  Phys.\ Rev.\ D {\bf 82}, 094026 (2010)
  [arXiv:1007.0684 [hep-ph]]. J.~Brod and M.~Gorbahn,
  Phys.\ Rev.\ Lett.\  {\bf 108}, 121801 (2012)
  [arXiv:1108.2036 [hep-ph]].

  \bibitem{Buras:2004uu} 
  A.~J.~Buras, F.~Schwab and S.~Uhlig,
  Rev.\ Mod.\ Phys.\  {\bf 80}, 965 (2008)
  [hep-ph/0405132].
  
\bibitem{Peskin:1991sw} 
  M.~E.~Peskin and T.~Takeuchi,
  Phys.\ Rev.\ D {\bf 46}, 381 (1992).
  
  
  \bibitem{Dawson:2012di} 
  S.~Dawson and E.~Furlan,
  Phys.\ Rev.\ D {\bf 86}, 015021 (2012)
  [arXiv:1205.4733 [hep-ph]].
  
  
  \bibitem{Ackermann:2013yva} 
  M.~Ackermann {\it et al.}  [Fermi-LAT Collaboration],
  arXiv:1310.0828 [astro-ph.HE].



\bibitem{Smith:2006ym} 
  M.~C.~Smith, G.~R.~Ruchti, A.~Helmi, R.~F.~G.~Wyse, J.~P.~Fulbright, K.~C.~Freeman, J.~F.~Navarro and G.~M.~Seabroke {\it et al.},
  Mon.\ Not.\ Roy.\ Astron.\ Soc.\  {\bf 379}, 755 (2007)
  [astro-ph/0611671].


  \bibitem{Aprile:2011hi} 
  E.~Aprile {\it et al.}  [XENON100 Collaboration],
  Phys.\ Rev.\ Lett.\  {\bf 107}, 131302 (2011)
  [arXiv:1104.2549 [astro-ph.CO]].
    
    
\bibitem{Feng:2010zp} 
  J.~L.~Feng, M.~Kaplinghat and H.~B.~Yu,
  Phys.\ Rev.\ D {\bf 82}, 083525 (2010)
  [arXiv:1005.4678 [hep-ph]].
    
    
 
\end{thebibliography}
\end{document}